# TOWARDS PRINCIPLED ANALYSIS AND MITIGATION OF SPACE CYBER RISKS

By

EKZHIN EAR

B.S., Long Beach State University, 2006

M.A., Liberty Baptist Theological Seminary, 2013

M.S., Western Governors University, 2019

M.Div., Southern Baptist Theological Seminary, 2025

A dissertation submitted to the Graduate Faculty of the

University of Colorado Colorado Springs

in partial fulfillment of the

requirements for the degree of

Doctor of Philosophy

Department of Computer Science

2025

This Dissertation for the Doctor of Philosophy degree by

Ekzhin Ear

has been approved for the

Department of Computer Science

by

Shouhuai Xu, Chair

Terrance Boult

Xi Tan

Gregory J. Falco

Wayne C. Henry

May 5, 2025

Date

Ear, Ekzhin (Ph.D., Cybersecurity)

Towards Principled Analysis and Mitigation of Space Cyber Risks

Dissertation directed by Professor Shouhuai Xu

## ABSTRACT

Space infrastructures have become an underpinning of modern society, but their associated cyber risks are little understood. This Dissertation advances the state-of-the-art via four contributions. (i) It introduces an innovative framework for characterizing real-world cyber attacks against space infrastructures, or *space cyber attacks*, including a novel methodology for coping with missing data and three novel metrics. A case study demonstrates the usefulness of the framework on 108 real-world space cyber attacks. (ii) This Dissertation characterizes the state-of-the-practice in space cyber risk analysis and mitigation, namely the Notional Risk Scores (NRS) within the Space Attack Research and Tactic Analysis (SPARTA) framework. (iii) We propose a set of desired properties that should be satisfied by any competent space cyber risk analysis and mitigation tool and applies them to assess two industrial space cyber risk analysis and mitigation tools. (iv) The study introduces a novel framework to analyze and mitigate space cyber risks by explicitly modeling space cyber attack cascading effects and presenting algorithms for *mission risk analysis* and *mission hardening*. We demonstrate the usefulness of the framework by applying it to analyze and mitigate space cyber risks, with testbed-based validation.

## ACKNOWLEDGMENTS

This Dissertation was made possible with the immense mentoring and support of my advisor, Dr. Shouhuai Xu. I am indebted to his continuous encouragement, regimented work ethic, and forward-thinking vision toward space cybersecurity dynamics. I thank the committee members, Dr. Terrance Boult, Dr. Xi Tan, Dr. Gregory J. Falco, and Dr. Wayne C. Henry, for their invaluable feedback. I am also grateful to my labmates in the Laboratory for Cybersecurity Dynamics, and especially Jose Remy and Caleb Chang, for their countless late nights of collaboration in our research.

The encouragements of my family, friends, and colleagues, including Dennis Lam, Darith Lee, Dara Lee, Tammi Huynh, Mike Dill, Victor Padilla, Thor Munoz, Stephanie Tan, Lora Gorsky, Roger Sutton, Amy Mai Phan, Nancy Wong, Andrew Tran, Christophe Schleypen, François-Xavier Stellamans, Ken Lew, Eric Ong, Qiren Que, Kora Gwartney, Alice Tan, Dr. Danielle Tan, Dan Collins, Dr. Todd Gray, Dr. David Highfield, Sorina Highfield, Michael Robbins, and Eric Walters, were indispensable.

I thank Sagecreek Bible Church, my parents, Andrew and Susanna Ear, parents-in-law, Albert and Lisa Wong, brother, Ek T. Ear, sister-in-law, Danie So, pastors, Terry Pautzke, Jose Navarro, and Jerry Briggs, and deacons, Herbert Klumpe and Gus Keiley, for saturating my research with fervent prayers for success that reflects the LORD's graciousness and goodness. Thank you, LORD, for making my research fruitful.

This Dissertation is supported in part by the Army ACS program, the DoD UC2 program, and NSF Grant #2308142. The opinions expressed herein are solely that of the author, and do not reflect any government entities or other organizations, whatsoever.

# DEDICATION

*This Dissertation is lovingly dedicated*

*To God the Father, the Son, and the Holy Spirit; may this work be to Your glory.*

*To my perfect and lovely wife, Ling Sze; you are far more valuable than pearls.*

*To my children, Hanna, Eliza, and Xander; you are treasured gifts from the LORD.*

*To my nephew, Darren, who has suffered much; may the LORD turn your afflictions into inexpressible joy.*

# TABLE OF CONTENTS

CHAPTER

# LIST OF FIGURES

FIGURE

# CHAPTER I

## INTRODUCTION

Space infrastructures have become an underpinning of modern society and economy because their missions include services that support many land, air, maritime, and cyber operations, such as Positioning, Navigation, and Timing (PNT) for the global stock market [88], Satellite Communications (SATCOM) for global beyond-line-of-sight (BLOS) terrestrial voice and data requirements, and remote sensing for space domain awareness and planetary defense (e.g., detecting and deflecting large debris from hitting Earth). For instance, the Space Foundation projects that the volume of Space Economy will reach $1.8T by 2035 [145]. These applications and impacts attest to the importance of securing and defending space infrastructures from malicious attacks, which include cyber attacks, to benefit the greater good of the society and economy.

One characteristic feature of the space cybersecurity problem is that the space domain is interwoven with, and enabled by, the cyber domain, with real-time cyber-physical systems comprising the space segment. Although many of the vulnerability-containing space infrastructures have been in operation for decades without reported security breaches, a study shows that the actual percentage of exploited versus discovered vulnerabilities is quite low, at less than 35% [104]. However, space infrastructures are not less susceptible to cyber attacks. The lower percentage may be attributed to cyber risks associated with space infrastructures and systems, or *space cyber risks* for short, likely being different from their counterpart in enterprise IT or even OT networks, for at least two reasons. One reason is that vulnerabilities in space infrastructures may been less known to attackers or malicious actors, perhaps because of their highly proprietary and unique nature in terms of hardware stacks (e.g., radiation-hardened flight computers, constrained on-board compute and storage capabilities), software stacks (e.g., processing of data transported through RF and laser communications), and other space infrastructure properties (e.g., possible long periodic network disconnections of satellites from ground stations). However, this should never be treated as a source of security, which is a big lesson the cybersecurity community has learned from the last decades, namely that cybersecurity should rely on open design and principled analysis [127]. Another reason is that vulnerabilities in space infrastructures may be more difficult to exploit than their counterparts in IT or terrestrial OT networks. Again, this should never be treated as a leverage that benefits the defender because attackers sooner or later will learn to exploit them.

In fact, cyber attacks can and have affected space infrastructures, as evidenced by numerous space-related security incidents [32,43,53,110]. Space incidents have occurred as early as 1977, with the hijacking of a satellite's audio transmission to broadcast the attacker's own message [43]. In 1998, a U.S.-German RoSat (sensing satellite) experienced a malfunction that led to the satellite turning its x-ray sensor towards the sun, causing permanent damage. While it is debatable whether this incident was caused by cyber attacks, the confirmed cyber attack at the Goddard Space Flight Centre, where the RoSat is controlled, shows that cyber attacks can cause physical damage to space systems [151]. Understanding real-world attacks equips us to cope with the different dimensions of space cyber risks.

Despite their importance, there is no systematic understanding of space cyber risks, likely due to the lack of data. Correspondingly, space cybersecurity practitioners also lack tools to effectively analyze and mitigate space cyber risks. In a major effort to support space cybersecurity practitioners, The Aerospace Corporation developed and incorporated space cyber *Notional Risk Scores* (NRS) [139] into their Space Attack Research and Tactic Analysis (SPARTA) framework [141], by associating a notional evaluation of cyber risks to attack techniques. NRS and SPARTA are founded on a wealth of industry research and precedent, including traditional enterprise Information Technology (IT) cybersecurity (e.g., security controls [40–42]). The intention of NRS is to provide practitioners with a starting point for space cyber risk analysis and mitigation, whereby they can tailor NRS to meet their specific space cyber risk analysis and mitigation needs. In order to harden space infrastructures and systems against cyber threats, we need to identify, quantify, and reduce the cyber risks against them. Although NRS offers practitioners with relatively specific guidance on conducting space cyber risk analysis and mitigation, our research shows that it is just a starting point towards ultimately tackling the space cyber risk analysis and mitigation problem.

## 1.1 Dissertation Research Motivation

The state of the art is that space cybersecurity is an under-investigated topic, meaning that space cyber risks are poorly understood. There are many open problems that have yet to be investigated. This Dissertation addresses the following problems:

- There is a lack of data describing real-world space cyber attacks. As a consequence, there is a lack of deep understanding of them, which hinders space cybersecurity research. *What are the characteristics that are exhibited by real-world space cyber attacks?*
- Analyzing space cyber threats and risks is a necessary step before we can adequately mitigate them. *What space cyber threat and risk analysis methodologies may have been used by practitioners in the real world?*
- There is a lack of understanding on what would construe a competent solution to adequately characterizing and mitigating cyber threats. As a result, we do not know what are the gaps between the status quo solutions and the ideal solutions. This highlights the importance of the following problem: *What are the properties of ideal space cyber risk analysis and mitigation tools?*
- Understanding of the gap between existing tools and the ideal tools paves a way to design the next generation solutions. This prompts the following question: *How can we design a new solution for analyzing and mitigating space cyber threats and risks towards the ideal solutions?*

The preceding open questions motivates this Dissertation research.

## 1.2 Dissertation Research Contributions

This Dissertation makes four contributions, which respectively address the preceding open questions as follows:

- We propose an innovative framework to characterize real-world space cyber attacks. Salient features of the framework include: a principled *extrapolation* methodology for coping with the missing data that is often encountered in real-world space cyber attack reports; and three metrics we introduce, dubbed *attack consequence*, *attack sophistication*, and *attack likelihood.* We demonstrate the usefulness of the framework by applying it to analyze 108 real-world space cyber attacks and draw a number of insights. (**Chapter III**)
- We characterize the state of the practice in space cyber risk analysis and mitigation, by analyzing the Notional Risk Scores (NRS) within the Space Attack Research and Tactic Analysis (SPARTA) framework of the Aerospace Corporation [139]. We present the first algorithmic description of NRS for characterizing its strengths and weaknesses, via two real-world space cyber attacks. This algorithmic description has been incorporated into SPARTA, since version 1.6. (**Chapter IV**)
- We propose a set of desired properties that should be satisfied by a competent space cyber risk analysis and mitigation tool. We demonstrate their usefulness by applying them to systematically assess the two current conceptual space cyber risk analysis and mitigation tools. Our case study sheds light on designing competent space cyber risk analysis and mitigation tools that fulfill those properties, as evidenced by the novel space cyber risks analysis and mitigation framework presented in the subsequent chapter. (**Chapter V**)

- We propose a novel and actionable framework to analyze and mitigate space cyber risks, where *actionable* means it provides detailed representations and models for real-world adoption. Two salient features of the framework are: its *explicit* modeling of space cyber attack cascading effects, and its two core algorithms for *mission risk analysis* and *mission hardening*, respectively. We demonstrate the actionability and usefulness of the framework by implementing the algorithms and applying them to manage space cyber risks; the effectiveness are validated via our space testbed. (**Chapter VI**)

## 1.3 Publications

The four contributions of the Dissertation mentioned above respectively correspond to the following publications or manuscripts:

1. **E. Ear**, J. L. C. Remy, C. Chang, A. Feffer, and S. Xu. "Characterizing cyber attacks against space infrastructures with missing data: Framework and case study," full version, which is to be submitted to IEEE Transactions on Dependable and Secure Computing (TDSC) for peer review, of the paper which appeared in *Proceedings of IEEE Conference on Communications and Network Security (CNS)*, 2023 [27]. (**Chapter III**)

2. **E. Ear**, B. Bailey, and S. Xu. "The notional risk scores approach to space cyber risk management," *under conference peer review.* Notably, The Aerospace Corpo-ration has already incorporated our research into their update release of SPARTA 1.6 [24] (**Chapter IV**)

3. **E. Ear**, B. Bailey, and S. Xu. “Space cyber risk management: desired properties,” *under conference peer review.* [25]. (**Chapter V**)

4. **E. Ear**, C. Chang, and S. Xu. “An actionable framework for space cyber risk management,” *manuscript to be submitted for conference peer review.* (**Chapter VI**)

The following journal or book-chapter publications during my PhD study are not incorporated into this Dissertation:

1. J. L. C. Remy, **E. Ear**, and S. Xu. “Quantifying and reducing system non-resilience: methodology, metrics, and case study,” invited book chapter in Springer Book on Cyber Resilience, to appear in 2025 [120].

2. T. Longtchi, R. Montanez Rodriguez, K. Gwartney, **E. Ear**, D. P. Azari, C. P. Kelley, and S. Xu. “Quantifying psychological sophistication of malicious emails,” in *IEEE Access*, 2024 [135].

3. **E. Ear**, C. Chang, J. Butler, J. Mejia, S. Xu, H. Echelmeier, and N. Wells. “Characterizing Russia’s cyber operations in Ukraine through the lenses of cyber attack tactics, techniques, and procedures,” in *USCYBERCOM CyberRECon*, 2024 [26]. Notably, this work won the **USCYBERCOM CyberRECon’2024 Hunter Award**.

4. L. Zeien, C. Chang, **E. Ear**, and D. S. Xu. "Characterizing advanced persistent threats through the lens of cyber attack flows," in *Military Cyber Affairs*, 2024 [171].

The following conference publications during my PhD study are not incorporated into this Dissertation:

1. J. L. C. Remy, **E. Ear**, C. Chang, and S. Xu. "SoK: Space Infrastructures Vulnerabilities, Attacks and Defenses," accepted to *Proceedings of IEEE Conference on Security & Privacy (IEEE S&P)*, 2025.

2. J. L. C. Remy, C. Chang, **E. Ear**, and S. Xu. "Space cybersecurity testbed: fidelity framework, example implementation, and characterization," in *Workshop on the Security of Space and Satellite Systems (SpaceSec)*, 2025.

3. R. Montanez Rodriguez, T. Longtchi, K. Gwartney, **E. Ear**, D. P. Azari, C. P. Kelley, and S. Xu. "Quantifying psychological sophistication of malicious emails," in *Proceedings of International Conference on Science of Cyber Security*, 2023 [93].

4. **E. Ear**, J. L. C. Remy, and S. Xu, "Towards automated cyber range design: Characterizing and matching demands to supplies," in *Proceedings of IEEE Conference on Cyber Security and Resilience (CSR)*, 2023 [28].

The following manuscripts, which are to be submitted for peer review or are under peer review, are not included in this Dissertations.

1. **E.Ear**, J. L. C. Remy, C. Chang, J. Carey, S. Xu, and D. A. Bodnar. "CR2M-JPP: a methodology for hardening joint tactical missions against cyber threats, with a case study," *under review and possible publication by USCYBERCOM*.

2. **E. Ear**, T. Munoz, and S. Xu. "Shifting from fires-centric to maneuver-centric cyberspace operations: a new paradigm for incorporating cyber maneuver into multi-domain operations," *to be submitted for journal peer review*.

3. B. Hancock, H. Nguyen, O. Karpoyan, **E. Ear**, and S. Xu. "Understanding Russia's cyber policies, strategies, and doctrines," *to be submitted to Military Cyber Affairs*.

4. S. Kettner, C. Chang, **E. Ear**, and S. Xu. "Charactering cyber attacks against operational technology infrastructures through the lens of attack flows," *to be submitted to Military Cyber Affairs*.

5. M. Cao, K. Teoh, **E. Ear**, and S. Xu. "Characterizing Caldera's adversary emulation abilities," *to be submitted to Military Cyber Affairs*.

# CHAPTER II

## TERMINOLOGY AND LITERATURE REVIEW

In this chapter, we define terminology used throughout this Dissertation. Terminology and concepts specific to a chapter are presented in the respective chapter. Then, we review the prior studies that are related to the research presented in this Dissertation.

### 2.1 Terminology

We adopt the following terms from NIST [56]:

- *Risk*: it refers to the measure of the degree that an entity (e.g., a space infrastructure) may be threatened, and is represented by a function of the impact or damage that may be sustained from the threat and the likelihood that this impact would occur.

- *Threat*: it refers to a circumstance or event that could potentially negatively impact an entity and may use unauthorized access, destruction, disclosure, modification of information, and/or denial of service to affect operations, assets, individuals and/or other organizations.

- *Vulnerability*: it refers to a weakness in the design, implementation, or configuration of an information system that could lead to exploitation by a threat. Specifically, a vulnerability may be exhibited in hardware, firmware, software, or humans [41]. While we do consider how the vulnerabilities that are inherent to humans may relate to space infrastructures, we focus on the vulnerabilities that arise in hardware-firmware-software interactions and do not analyze humans' susceptibility to cyber social engineering attacks; we refer systematic studies on human vulnerabilities to cyber social engineering attacks to [81, 84].

NIST relates the preceding three terms to each other as follows: risks materialize from threats that take advantage of vulnerabilities. NIST also provides a second characterization of the relationship: a threat source can initiate a threat event that exploits a vulnerability causing adverse impact which produces risk [56].

## 2.2 Related Work

We divide related prior studies into three categories with respect to the three components of the Dissertation, namely identification, analysis, and mitigation of space cyber risks.

### 2.2.1 Related Work on Identifying Space Cyber Risks

We divide prior studies on identifying space cyber risks into three categories: (i) studies that investigate risk taxonomies and research approaches related to space infrastructures; (ii) studies that identify space cyber risks in academic research; and (iii) studies that investigate space cyber risks in real-world incidents.

#### 2.2.1.1 Prior Studies on Space Cyber Risk-Related Taxonomies

Falco and Boschetti [32] present a taxonomy of security risks for commercial space missions that are derived from a dataset of 2,000 space security incidents that are compiled by the authors. The taxonomy considers risk in categories that range from organizational to physical risks, of which cyber is one subcategory of the digital risks that can materialize through operations in cyberspace. More specific to space infrastructures, [152] provides a taxonomy of threats against satellite firmware. The taxonomy organizes cyber threats according to a number of different physical modules in the satellite bus system and payload components. The Dissertation leverages these taxonomies to gain a better understanding of space cyber risks and also using their classification of attack types in our space cyber attack analysis framework. We go beyond classifying space cyber risks to analyzing and mitigating them.

#### 2.2.1.2 Prior Studies on Identifying Space Cyber Risks in Academic Research

There are studies that identify cyber risks by analyzing possible attacks. Falco et al. [35] present a general system architecture of a small cube-shaped satellite (i.e., cubesat) and details three attack trees where attackers attempt to conduct a denial of service, tamper sensitive sensor data, or inject a "kill radio" command to deny communication to the cubesat. Boschetti et al. [9] present four reference architectures of ground station as a service (GSaaS) and conducts a cybersecurity analysis using the attack tree method. Ramdass et al. [118] analyze telerobotic systems for existing non-space-related threats and attack vectors to form theoretical threats against on-orbit servicing, assembly, and manufacturing (OSAM) systems with attack trees that represent disabling payloads, severing deployables, and sending malicious commands through rogue equipment.

The following studies present and analyze system models of specific segments of space infrastructures to identify theoretical cyber risks. Falco [31] discusses the technical feasibility of satellite-to-satellite cyber attacks. It enumerates potential cyber attack consequences, namely failures of actuators, onboard computer systems (OBCS), sensors, communications systems, and power systems. It further asserts the attacking satellites would require special-purpose sensors and actuators, namely situational awareness sensors, electromagnetic pulse actuators, and radio frequency actuators. Falco et al. [33] describe potential attacks against spaceports, namely timing attacks and consumption of resources. It also emphasizes particular cyber risks to spaceports, namely in the spaceports' fuel supply chain, facility management, and launch operation monitoring systems. Lee and Falco [62] enumerate attack vectors for exploitation of end-of-life satellites. This includes exploiting the conditions of reduced staffing, mission fatigue, unpatched software, and satellite cross-links. Our study leverages these studies and goes beyond the theoretical approaches to analyzing real-world space cyber attacks and providing approaches to mitigate them.

There are studies that employ a simulated environment to identify risks, by either using static code analysis or by crafting attack scenarios in dynamic analysis. For example, four studies employ static code analysis. Willbold et al. [152] analyze three firmware images from satellites. It found authentication bypass, arbitrary code execution, data leakage, and parameter tampering vulnerabilities in the three firmware images. It investigates the exploitability of some of the discovered vulnerabilities through static analysis. Curbo and Falco [17] conduct dependency and critical path analyses on the Real-Time Executive for Multi-processor Systems (RTEMS) real-time operating system (RTOS) in the context of its utility with NASA's core Flight Software (cFS), namely cFS' Operating System Abstraction Layer (OSAL). It found OSAL's complexity as a danger to security while it found RTEMS lacks memory protection, task access control, and default file system integrity protection. Falco et al. [34] employ static code analysis to develop an approach for deploying a ransom-type cyber attack against a space vehicle running cFS. The study characterizes cFS using an automated analysis tool to map the dependencies of its modules and application suite. From this, it identified the Software Bus API as the optimal target for a ransomware attack. Hansen et al. [50] is a continuation of [34], where they identify the requirements to deploy a ransomware against a space vehicle and develops a proof-of-concept ransomware against cFS within the NOS-3 simulator. The study further evaluates the attack surface of the flight software, particularly for ransomware-based vulnerabilities and found the SPARTA framework to be accurate concerning ransomware attacks. Our study lever-ages these studies to understand space infrastructures while the experiments in our study are conducted on our fully implemented space testbed.

There are studies that employ dynamic analysis to identify cyber risks, such as the following. Thebarge et al. [143] approach the design of intrusion detection systems for satellites through the lens of penetration testing. It applies the penetration testing process on a notional cubesat to generate attack scenarios for disrupting satellite operations. It found operating system resource metrics, commands, files, and processes require malicious activity monitoring. Schalk et al. [128] present four exploit demonstrations on cFS as implemented in the OpenSatKit (OSK) simulator. The four attack scenarios in its demonstration are replay, denial-of-service, Command Inject deletion, and systematic application deletion. It found a lack of layered permissions on the Software Bus and the existence of single points of failure in the Command Ingest Module. Calabrese et al. [10] conduct a case study that examines the susceptibility of the European Space Agency OPS-SAT cubesat to cyber threats. The study emulates an authenticated malicious user targeting the camera payload of OPS-SAT in an emulated environment. The researchers found OPS-SAT's dynamic linking process vulnerable to malicious library injection. Our study leverages these studies as we employed static and dynamic analysis in our case study experiments.

#### 2.2.1.3 Prior Studies on Identifying Space Cyber Risks in the Real World

There are four studies that leverage real-world environments, including three studies on analyzing cyber threat intelligence (CTI) concerning APT attacks and one study on conducting experiments with a real-world space system. Boschetti et al. [8] describe an attack against ViaSat and presents technical- and organizational-level lessons, where the attack was likely enabled by an unpatched VPN server. Our own study [26] goes into further technical details concerning the ViaSat incident through the lens of attack flows. Thummala and Falco [144] identify the probable vulnerabilities and exploits that enable the Network Battalion 65's (NB65) breach of the ROSCOSMOS satellite imaging payload. Analyzing open-source and cyber threat intelligence, the researchers found the Log4j2 remote code execution exploit was likely used to target the ground station's outdated integration platform server (WSO2 Enterprise Integrator) for API management. Rose et al. [126] apply static and dynamic code analysis techniques that have been employed for terrestrial systems in a case study of a real-world academic experimental cubesat. It found a buffer overflow vulnerability stemming from a *memcpy* call in the cubesat software stack. Our study leverages these studies to implement our high-fidelity case study experiments to validate our space cyber risk analysis and mitigation framework.

### 2.2.2 Related Work on Analyzing Space Cyber Risks

Concerning the analysis of space cyber risks, we divide related prior studies into four categories: analyzing space infrastructure incidents in the real world, analyzing threats against space infrastructures in an abstract model, employment of IT cybersecurity frameworks to space infrastructures, and cybersecurity metrics.

#### 2.2.2.1 Prior Studies on Analyzing Space Cyber Attacks in the Real World

There are studies on analyzing space-related incidents [8, 32, 43, 110, 111, 132]. For example, Pavur and Martinovic [110] consider incidents in terms of the payload, signal, and ground aspects; Fritz [43] provides narrative descriptions concerning NASA, jamming, hijacking, and control attack categories; Falco and Boschetti [32] analyze 1,847 space-related incidents according to their risk taxonomy for space. The study [32] also presents a risk taxonomy of threats against space systems. We leverage information from the presented taxonomy to categorize the incidents compiled within this Dissertation (e.g., hijacking, control, or DoS type of incidents). There are also studies on individual real-world incidents [8, 111, 132]. By contrast, we are the first to provide a unified space cyber kill chain dataset that we constructed through our comprehensive analysis of 108 space cyber attacks.

#### 2.2.2.2 Prior Studies on Analyzing Space Cyber Threats in Abstract Models

The Center for Strategic and International Studies Aerospace Security Project's 2021 assessment [51] goes into significant details about various threats to space, especially concerning physical and electronic capabilities of various nation-states. While it provides categories for cyber threats, it provides no specific cyber capabilities of these nation-states against space infrastructures, which is similar to several other studies that discuss more physical and electronic threat vectors [32, 132, 136, 137]. There are studies on threat models for space applications, such as leveraging LEO constellations to jam GEO targets [119], creating attack trees against cubesats [35], demonstration of com-mand injection via a software-defined radio [78], characterizing the transmission layer's susceptibility to eavesdropping [123], leveraging the ATT&CK framework [108], and nanosatellites as attack platforms [112]. By contrast, we analyze real-world cyber attacks, while aiming to leverage our findings to make future abstract threat models more realistic and holistic. Our Dissertation goes beyond a general threat model or singular applications to define specific unified space cyber kill chains of real-world incidents that provide detailed insights for analyzing space cyber threats.

#### 2.2.2.3 Prior Studies on the Employment of IT Cybersecurity Frameworks to Space Infrastructures

There are studies concerning the application of IT cybersecurity frameworks to the space domain, such as for defined space mission areas with cybersecurity overlaid upon them [16, 149, 170], a more hybrid approach combining both mission areas and explicit threat models [7], and applying pre-existing IT controls to map to space systems [60, 126, 147, 168]. These studies help us to consider space cybersecurity holistically. However, they typically treat cybersecurity requirements in abstract and generalized concepts. Our study applies threat-centric cybersecurity taxonomies to real-world space cyber attacks.

#### 2.2.2.4 Prior Studies on Cybersecurity Metrics

There are systematic studies on IT cybersecurity metrics [15, 113, 166]. However, there are very few studies on cyber attack sophistication metrics for space-related incidents. One study referenced an Aerospace resiliency framework which provides categories where cybersecurity metrics could be developed [137]. Tedeschi et al. [136] provide performance metrics of physical layer defense. Pendleton et al. [114] provide a survey of current and proposed cybersecurity metrics that is most useful for our study, especially in its discussion of measuring attacks, evasion techniques, evasion capability, obfuscation sophistication, and power of targeted attacks. We leverage these studies to define new sophistication metrics and apply them to the space-related cybersecurity incidents. Moreover, our framework is innovative, including metrics that have not appeared in the literature [15, 92, 114, 166].

### 2.2.3 Related Work on Mitigating Space Cyber Risks

Concerning the mitigation of space cyber risks, we divide related prior studies into three categories: the broader risk mitigation context, the space cyber risk context, and concerning space cyber risk analysis and mitigation tools.

#### 2.2.3.1 Prior Studies Related to the Broader Cyber Risk Context

The National Institute of Standards and Technology (NIST) has proposed two frameworks: (i) the Cyber Security Framework (CSF) [103], which offers six functions—identify, protect, detect, respond, recover, and govern; and (ii) the Risk Management Framework (RMF) [55], which implements CSF with security controls [40–42, 56]. However, these frameworks provide high-level guidance without considering space infrastructure specifics.

#### 2.2.3.2 Prior Studies Related to the Space Cyber Risk Context

NIST has also published a sequence of guidance on adapting CSF to the space domain [130], to space-based positioning, navigation, and timing (PNT) services [6], to the satellite/spacecraft command and control systems in the ground segment [77], and to hybrid satellite networks [90]. Moreover, NASA [99] offers security controls to reduce cyber risks to satellite command authority (i.e., the ability to issue commands to satellites) and PNT [2, 58, 100]. Although these efforts consider space specifics, they are still high level and up to practitioners to interpret them when turning them into actionable procedures or software tools. This can be justified by the introduction of SPARTA NRS [139], which is a significant first step at addressing the space cyber risk analysis and mitigation problem while leveraging strengths of the frameworks mentioned above.

#### 2.2.3.3 Prior Studies Related to Space Cyber Risk Analysis and Mitigation Tools

The most closely related prior study is NRS [139], which is built on top of government efforts including the NIST Cybersecurity Framework (CSF) [102], NIST Risk Management Framework (RMF) [55], and NASA STD 1006A entitled "Space System Protection Standard" that establishes high-level requirements to protect space systems against threats that include cyber attacks [99]. Another study [58] proposes the Counterspace Threat Assessment Process (CTAP), applying NASA STD 1006A from a systems engineering perspective to conduct risk assessments, where cyber is among other categories of threats. Our study is different from all these prior studies because it presents the first systematic set of desired properties for competent space cyber risk analysis and mitigation tools. Our study shows that NRS is largely a manual process that substantially relies on subject matter experts (in four out of its six steps), meaning that different experts will likely produce different outcomes. Moreover, NRS and its predecessors focus on individual space systems, rather than the entire space infrastructure. This Dissertation aims to make a significant step beyond NRS and towards an objective and repeatable space cyber risk analysis and mitigation solution, by proposing an actionable framework that can be automated in 12 out of its 13 steps, while assuming the availability of certain inputs to its algorithms and noting that the remaining step is for identifying security controls to mitigate attack techniques (the automation of which is orthogonal to our focus and left for future research).

## 2.3 Chapter Summary

This chapter provided terminology used throughout the Dissertation as well as a literature review of studies related to the identification, analysis and mitigation of space cyber risks.

# CHAPTER III

## ANALYZING REAL-WORLD SPACE CYBER ATTACKS

**Chapter Abstract.** Cybersecurity of space infrastructures is an emerging topic, despite that there are space-related cybersecurity incidents recorded as early as 1977, with the hijacking of a satellite transmission signal. However, there is no single dataset that documents cyber attacks against space infrastructures that have occurred in the past; instead, these incidents are often scattered in media reports while missing many details, which we dub the *missing-data* problem. Nevertheless, even "low-quality" datasets containing such reports would be extremely valuable because of the dearth of space cybersecurity data and the sensitivity of space infrastructures which are often restricted from disclosure by governments. This prompts a research question: How can we characterize real-world cyber attacks against space infrastructures? In this chapter, we address the problem by proposing a framework, including metrics, while also addressing the missing-data problem by leveraging methodologies such as the Space Attack Research and Tactic Analysis (SPARTA) and the Adversarial Tactics, Techniques, and Common Knowledge (ATT&CK) to "extrapolate" the missing data in a principled fash-ion. We show how the extrapolated data can be used to reconstruct "hypothetical but plausible" *space cyber kill chains* and *space cyber attack campaigns* that have occurred in practice. To show the usefulness of the framework, we extract data for 108 cyber attacks against space infrastructures and show how to extrapolate this "low-quality" dataset containing missing information to derive 6,206 attack technique-level space cyber kill chains. Our findings include: cyber attacks against space infrastructures are getting increasingly sophisticated; successful protection of the link segment between the space and user segments could have thwarted nearly half of the 108 attacks. We will make our dataset available.

## 3.1 Chapter Introduction

Space infrastructures have become an underpinning of modern society because they provide services that support many land, air, maritime, and cyber operations, such as Positioning, Navigation, and Timing (PNT) for the global stock market [88], Satellite Communications (SATCOM) for global beyond-line-of-sight (BLOS) terrestrial voice and data requirements, and remote sensing for space domain awareness and planetary defense (e.g., detecting and deflecting large debris from hitting Earth). At a high level, space infrastructures can be understood via their four segments: space (e.g., satellites), ground (e.g., radar facilities), user (e.g., radio receivers), and link (e.g., radio frequency signals between the other segments). Moreover, space infrastructures can be described at multiple levels of abstraction [122].

The space domain is interwoven with and enabled by the cyber domain, with real-time cyber-physical systems comprising the space segment. Consequently, cyber attacks can affect space infrastructures, as evidenced by numerous space-related security incidents [32, 43, 53, 110]. Space incidents have occurred as early as 1977, with the hijacking of a satellite's audio transmission to broadcast the attacker's own message [43]. In 1998, a U.S.-German RoSat (sensing satellite) experienced a malfunction that led to the satellite turning its x-ray sensor towards the sun, causing permanent damage. While it is debatable whether this incident was caused by cyber attacks, the cyber attack at the Goddard Space Flight Centre, where the RoSat is controlled, shows that cyber attacks can cause physical damage to space infrastructures [151]. Although cyber attacks against space infrastructures have become a reality, there is no systematic understanding of real-world cyber attacks against space infrastructures, likely due to the lack of data.

In this chapter, we initiate the study on the problem of characterizing *real-world* cyber attacks against space infrastructures *with missing data*. The problem is important to deepen our understanding of cyber attacks against space infrastructures, which are not yet understood, and to gain insights into making future space infrastructures secure. It would be ideal that we have well documented attacks with significantly detailed descriptions (e.g., through digital forensics) to serve as input to this characterization study. Unfortunately, cyber attacks, especially those against space infrastructures, are rarely well-documented owing to a variety of reasons (e.g., sensitivity), which explains why we must embrace the fact of missing data.

Note that the perspective of our study is different than textbook cyber threat models, which typically *assume* what an attacker attempts to achieve and what capabilities are available to an attacker. Moreover, a competent design is often able to thwart the attacks specified in well-defined threat models; for instance, a cryptographic protocol can be rigorously proven to thwart the attacks specified in a rigorously defined threat model. However, the fact attacks succeeding in the real world means that the threat model considered at the design phase are incomplete and/or the design (of the employed defense) is not competent.

**Chapter contributions**. We make two technical contributions. First, we propose a novel framework to characterize real-world cyber attacks against space infrastructures *with missing data*. The framework has three characteristics: (i) It is *general* because it can accommodate both cyber attacks against space infrastructures that occurred in the past and cyber attacks that may occur in the future. (ii) It is *practical* because it explicitly deals with missing details of attacks, which is often the case with real-world datasets. At a high level, our idea is to leverage the Aerospace Space Attack Research and Tactic Analysis (SPARTA) [1] and the MITRE Adversarial Tactics, Techniques, and Common Knowledge (ATT&CK) [133] frameworks to extrapolate the missing details of attacks. (iii) It offers three *metrics* for measuring *attack consequence*, *attack sophistication*, and *Unified Space Cyber Kill Chain (*USCKC*) likelihood*, where the concept of USCKC will be elaborated later. These metrics might be of independent value because they could be adapted to other kinds of infrastructures and networks despite that they are tailored to space infrastructures in the present study.

Second, we show the usefulness of the framework by applying it to characterize the cyber attacks described in a dataset, which is prepared by this study and is, to our knowledge, the first comprehensive dataset of cyber attacks against space infrastructures. The dataset includes 108 attacks. Among the 108 attacks, 72 are manually extracted from four publicly available datasets documenting space-related incidents, which are not geared toward cyber attacks; the other 36 attacks are obtained via our own manual Internet search and analysis. These 108 attacks naturally lead to 108 *attack tactic*-level USCKCs (i.e., one USCKC per attack), but the extrapolation of the missing data leads to 6,206 probable *attack technique*-level USCKCs, where we use the standard concepts of attack tactic and attack technique described in [134]. This allows us to draw a number of insights, such as: attacks against space infrastructures can be effectively mitigated by hardening the ground segment; average-sophisticated cyber attacks have been be effective against space infrastructures; attacks are getting increasingly sophisticated, perhaps because of the increasing employment of defenses; defending the links between the space segment and the user segment could have thwarted nearly half of the 108 attacks. We will make this dataset publicly available.

**Chapter outline**. Section 3.2 describes terms and concepts used throughout the chapter. Section 3.3 presents the framework. Section 3.4 presents a case study. Section 3.5 concludes the chapter.

## 3.2 Terminologies and Concepts

**Cyber Attack Campaign**. A cyber attack campaign is an attacker's activities against an infrastructure or network (e.g., IT network, space network) for attempting to accomplish certain objectives. These activities are typically manifested as multi-step attacks. The first attack step typically compromises a space infrastructure unit (at an appropriate level of abstraction), dubbed *entry node*, whereby the attacker penetrates into the space infrastructure. The last attack step typically compromises another space infrastructure unit (at the same level of abstraction), dubbed *objective node*, where the attacker accomplishes its overall objectives. Each objective is often divided into multiple sub-objectives.

**ATT&CK**. We adopt the following terms from ATT&CK [134]: an *attack tactic* specifies *what* an attacker attempts to accomplish (i.e., an attacker's sub-objective); an *attack technique* specifies *how* the attacker attempts to accomplish an attack tactic; an *attack procedure* is a concrete implementation of an attack technique (as an attack technique could have many concrete implementations or instantiations). At a high level, ATT&CK describes possible attack tactics, techniques, and procedures against enterprise IT networks [133,134]. We use $\text{ATT}_{\text{TA}}$ to denote the set of ATT&CK tactics, meaning $\text{ATT}_{\text{TA}} = \{$Reconnaissance, Resource Development, Initial Access, Execution, Persistence, Privilege Escalation, Defense Evasion, Credential Access, Discovery, Lateral Movement, Collection, Command & Control, Exfiltration, Impact$\}$. We use $\text{ATT}_{\text{TE}}$ to denote the set of ATT&CK techniques, which are too many to be listed here but can be found at [134]. Real-world use cases of ATT&CK include: Cyber Threat In-telligence (CTI) analysts use it to extract useful information from raw data [109]; digital forensics and incident response analysts use it to characterize cyber incidents [148].

**Cyber Kill Chain and Unified Kill Chain (UKC)**. The concept of *Cyber Kill Chain* [89] aims to help see the "big picture" of cyber attack campaigns. Attack tactics can be chained together to formulate tactic-level Cyber Kill Chain, and attack techniques can be chained together to formulate technique-level Cyber Kill Chain [89]. Both tactic- and technique-level Cyber Kill Chains present an abstract description of cyber attack campaigns. The concept of Unified Kill Chain (UKC) [117] improves upon the concept of Cyber Kill Chain by providing three major phases of a cyber campaign: access into a victim network, denoted by in; pivoting through the victim network, denoted by through; and achieving overall objectives of the attack, denoted by out. This is reasonable because a cyber attack campaign typically begins with an in phase, followed by one or multiple through phases, and concluding with an out phase. However, some attack campaigns may not have all the three phases.

**SPARTA**. Since ATT&CK is geared towards enterprise IT networks, it is not well suited for space infrastructures and systems. This motivates the introduction of SPARTA [1], which describes attack tactics, techniques, and procedures against space infrastructures and systems, especially for those which have no counterpart in enterprise IT networks, such as attacks against satellites and satellite-ground communications. We use $\mathrm{SPA_{TA}}$ to denote the set of SPARTA tactics, meaning $\mathrm{SPA_{TA}} = \{$Reconnaissance, Resource Development, Initial Access, Execution, Persistence, Defense Evasion, Lateral Movement, Exfiltration, Impact$\}$. Note that $\mathrm{SPA_{TA}} \subset \mathrm{ATT_{TA}}$. We use $\mathrm{SPA_{TE}}$ to denote the set of SPARTA techniques, which are too many to be listed here but can be found at [1].

Real-world use cases of SPARTA include: space CTI analysts use it to extract useful information from raw data as shown in the present chapter, and digital forensics and incident response analysts use it to characterize space cyber incidents; both are similar to how ATT&CK is used for similar purposes. In principle, SPARTA-based tactics and techniques can be chained together as in the case of ATT&CK. However, the resulting chains may not be sufficient because SPARTA is geared towards space infrastructures *without* the ground systems. This explains why we propose incorporating both ATT&CK and SPARTA to formulate more comprehensive kill chains.

**Unified Space Cyber Kill Chain** (USCKC). We propose adapting the concept of UKC to the concept of USCKC, which cuts across all the four segments of space infrastructures (including the space segment and the ground segment), justifying our proposal of leveraging SPARTA and ATT&CK. As a result, we can see the "big picture" of space cyber attack campaigns that may be waged from ground systems but target the space segment. Moreover, SPARTA and ATT&CK attack tactics can be chained together to formulate tactic-level USCKCs; SPARTA and ATT&CK attack techniques can be chained together to formulate USCKCs at the technique level. USCKCs present abstract descriptions of space cyber attack campaigns. To unify notations, we denote by $\text{JAT}_{\text{TA}} = \text{SPA}_{\text{TA}} \bigcup \text{ATT}_{\text{TA}} = \text{ATT}_{\text{TA}}$, while recalling $\text{SPA}_{\text{TA}} \subset \text{ATT}_{\text{TA}}$, the set of *joint SPARTA-ATT&CK attack tactics*, and by $\text{JAT}_{\text{TE}} = \text{SPA}_{\text{TE}} \bigcup \text{ATT}_{\text{TE}}$ the set of *joint SPARTA-ATT&CK attack tactics*.

**Reconstructing USCKCs via Activities.** At a high level, a USCKC is described by an in phase, one or multiple through phases, and an out phase. Our goal is to reconstruct USCKCs from CTI reports on space cyber attack campaigns. To accomplish this, we propose adapting the following concepts from the digital forensics community [54].

- *Objective activities*: Two attack tactics in $\text{JAT}_{\text{TA}}$, *Exfiltration* and *Impact*, are objective activities because they describe the overall impact of a space cyber attack campaign.
- *Milestone activities*: Three attack tactics in $\text{JAT}_{\text{TA}}$, *Initial Access*, *Lateral Movement*, and *Credential Access*, are milestone activities because they advance an attack campaign towards an objective activity.
- *Enabling activities*: Six attack tactics in $\text{JAT}_{\text{TA}}$, *Resource Development*, *Execution*, *Privilege Escalation*, *Persistence*, *Command & Control*, and *Defense Evasion*, are enabling activities because they seek to establish or modify a current state of a system environment to facilitate an objective or milestone activity.
- *Information discovery activities*: Three attack tactics in $\text{JAT}_{\text{TA}}$, *Reconnaissance*, *Discovery*, and *Collection*, are information discovery activities because they seek to provide necessary information to support an objective, milestone, or enabling activity.

## 3.3 Framework

### 3.3.1 Framework Requirements and Overview

To analyze cyber attacks against space infrastructures, we need a framework that ideally satisfies the following requirements:

- **Requirement 0:** It is *general* enough to accommodate all of the past cyber attacks (i.e., the ones that have been observed) and possible future attacks against space infrastructures.
- **Requirement 1**: It is *practical* by accommodating real-world datasets with missing attack details.
- **Requirement 2:** It provides metrics for characterizing cyber attacks against space infrastructures.

To address Requirement 0, we propose a system model that can be leveraged to describe past and future attacks. To address Requirement 1, we propose "extrapolating" a given description of a space cyber attack to incorporate *hypothetical but plausible* missing details. To address Requirement 2, we define *attack consequence*, *attack sophistication*, and *attack likelihood* metrics.

Figure 3.1 highlights the framework, which has four major components: (i) designing system model; (ii) preparing dataset; (iii) defining metrics to characterize attacks; and (iv) leveraging metrics to answer research questions.

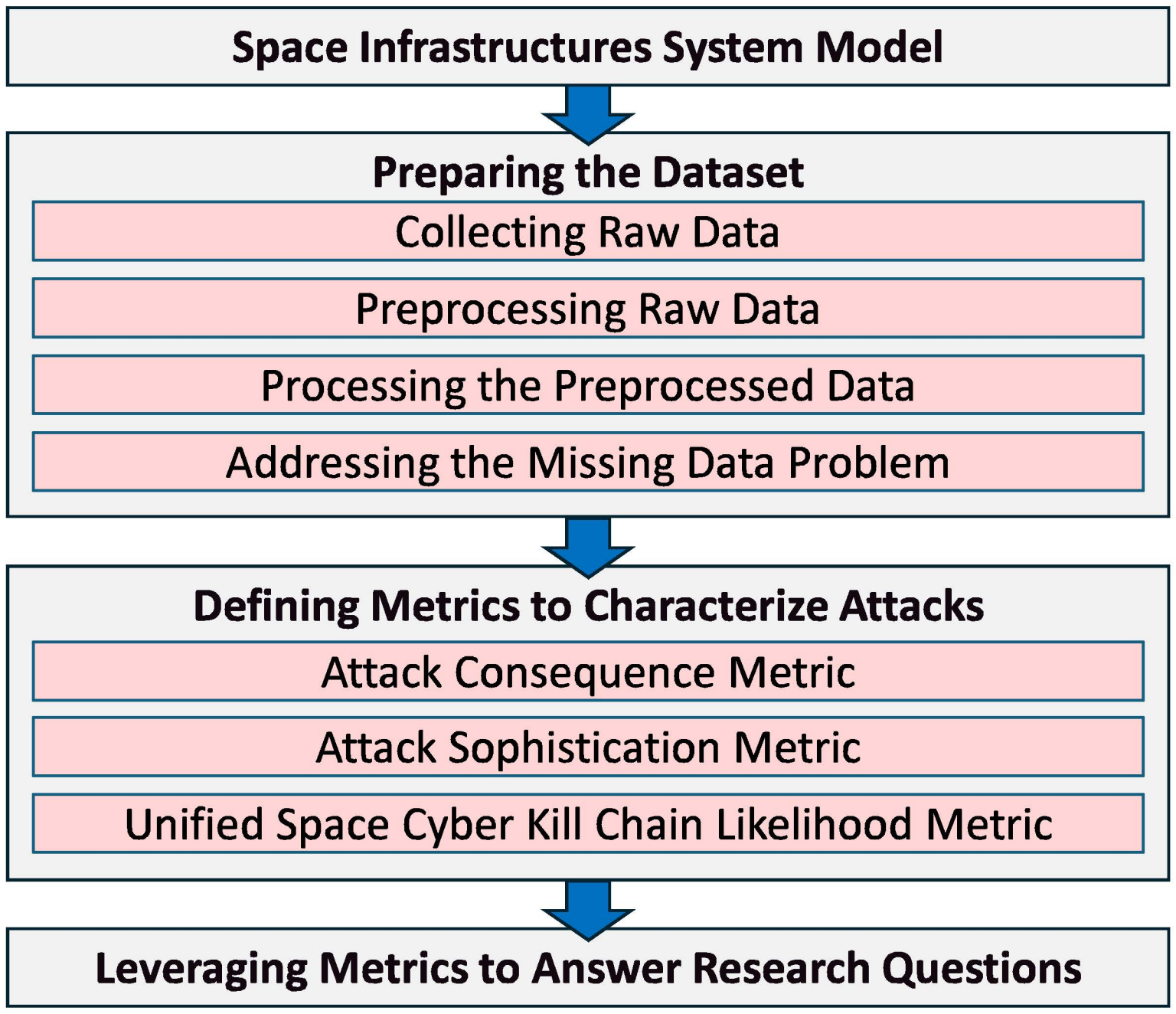

Figure 3.1: The space cyber attack analysis framework.

### 3.3.2 Space Infrastructure System Model

As mentioned above, a space infrastructure often consists of four segments: *space*, *link*, *ground*, and *user* [43, 47, 136]. This prompts us to propose the system model highlighted in Figure 3.2. We advocate separation between the ground segment and the user segment because the participants of each segment are different (e.g., an aerospace engineer versus a research scientist) and users often interact with a payload of space infrastructures whereas the ground segment interacts with both the payloads and the bus. Further, this separation increases the precision of cybersecurity design and analysis. For example, certain defense tools may be best employed at one segment but not the other, and cyber threat actors may desire to attack users without inflicting impact upon the ground and/or space segments.

The space segment includes satellites, spacecrafts, and space stations. A *Bus System* component facilitates tracking, telemetry, and command requirements (TT&C) and typically contains the following modules [14, 106]: *electrical power*, *attitude control*, *communication*, *command and data handling*, *propulsion*, and *thermal control*. These modules may be attacked through the cyber domain. The *Payload* component of space infrastructures include the following modules: *communication* (e.g., antennas and transmitters for relaying voice and data), *navigation* (e.g., Global Navigation Satellite System (GNSS) receivers for position and timing), *scientific experiment* (e.g., telescopes and spectrometers for research), *remote sensing* (e.g., sensors and cameras for terrestrial environmental monitoring), and *defense* (e.g., national security or military equipment for reconnaissance). Note that some space infrastructures may employ legacy technologies because they often cannot be replaced after being launched into the space. Note also that services such as data centers may be established in space in the future especially because of the increasingly low launch costs [132].

In the ground segment, the *Ground Station* component contains the hardware and software to transmit and receive RF communication signals, as well as tracking and ranging objects in space. The *Mission Control* component processes telemetry data to assess the health of modules in the space segment, sends commands to control the modules in the space segment, conducts analyses to plan orbital maneuvers and assess conjunctions, and manages other aspects of space infrastructure operations. The *Data Processing Center* component conducts deeper analyses of space infrastructure missions and processes payload data. The *Remote Terminal* component provides a light-weight software stack and network connectivity to the other elements of the ground segment. The ground segment often reflects the architecture of traditional IT enterprise networks, with the addition of space-specific hardware and software, such as satellite terminals, modems, flight software. The increasing commercialization of space acquisition has caused the supply chain for space-specific systems to trend towards enterprise network requirements. However, the geographical dispersion of ground stations presents unique challenges because stations across the globe have varying degrees of staff occupancy and physical security [132].

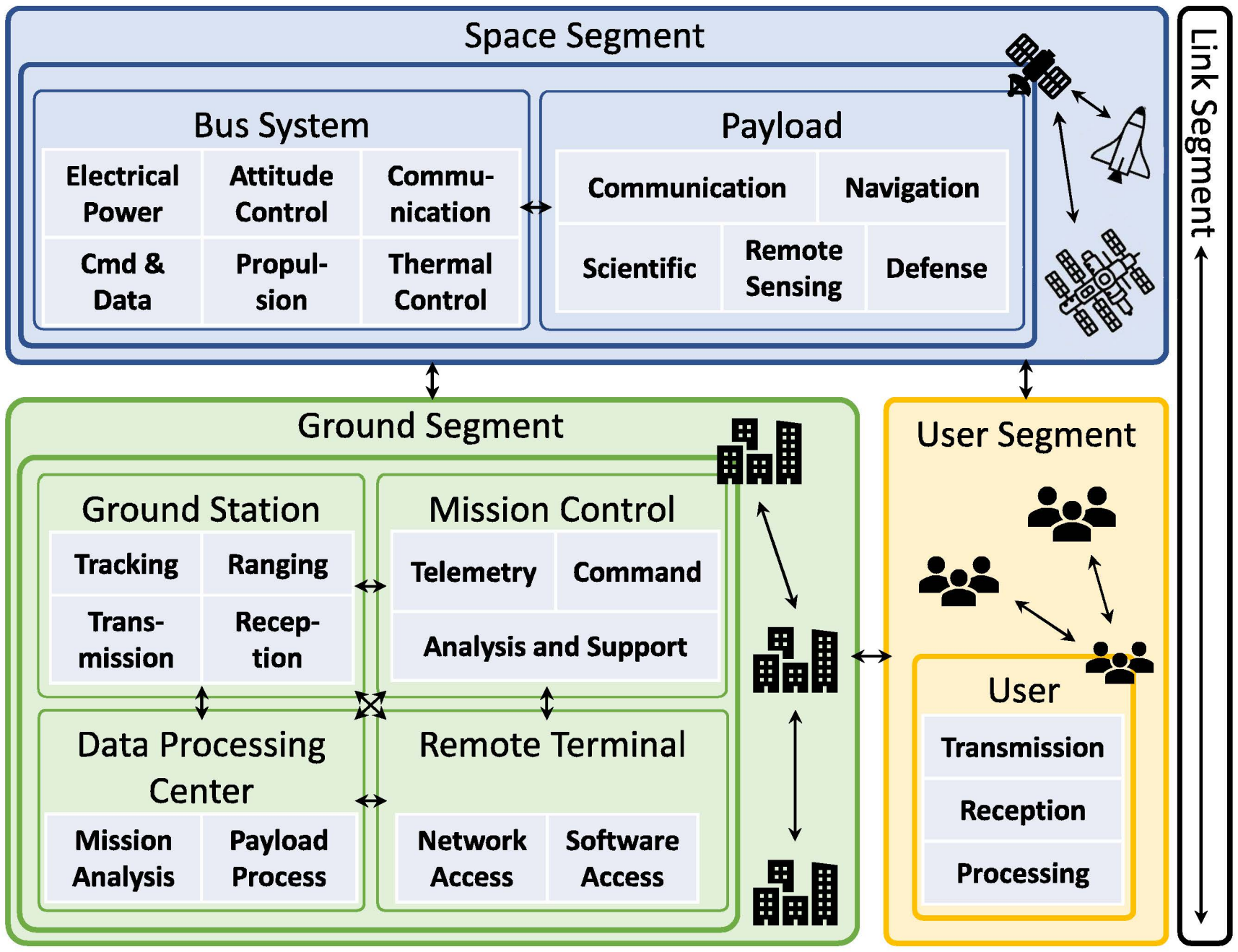

Figure 3.2: System model showing the four segments of space infrastructures.

The user segment is also geographically dispersed across continents and oceans, often requiring space infrastructure services around the clock. GNSS is an example of such service used across the land, air, and maritime domains in automobiles, airplanes, and ships across the world. The user segment typically receives data directly from GNSS satellites while SATCOM (Satellite Communications) users also transmit traffic to satellites. At times, the user segment may process data, such as researchers processing images from scientific satellites hosting telescopes.

The link segment is concerned with inter-segment (e.g., satellite to ground) and intra-segment (i.e., satellite to satellite) data connections. Satellites communicate by passing data through the link segment with other satellites, ground stations, as well as directly to user terminals, potentially via its payload (e.g., for voice and data transfer) or bus (for TT&C). It possesses a variety of physical and electromagnetic properties across the spectrum. For our purpose, it suffices to consider the link segment from a cyber perspective, namely the data transiting the link segment, rather than the physical properties of the link segment. For example, satellites regularly communicate with the ground and user segments, as well as other satellites, for TT&C and mission execution of payloads. Elements of the ground segment may share data with each other as part of the same facility or across wide area networks, as well as forward and retrieve data to the user segment. Members of the user segment may also collaborate with other users, e.g., scientists processing data or internet service providers broadcasting packets.

We represent a space infrastructure at the module level of abstraction, as a directed graph $G_{\mathsf{infra}} = (V_{\mathsf{infra}}, E_{\mathsf{infra}})$, where $V_{\mathsf{infra}}$ is the set of nodes representing space infrastructure modules, and $E_{\mathsf{infra}}$ is the set of arcs representing communication relationships such that $(v_1, v_2) \in E_{\mathsf{infra}}$ means $v_1$ can initiate unidirectional or bidirectional communications with $v_2$.

### 3.3.3 Preparing Dataset

It would be ideal that a dataset of cyber attacks against space infrastructures possesses the following properties:

- *Comprehensiveness*: This property deals with the coverage of a dataset, in terms of the aspects that are important to describe space cyber attacks. Ideally, a dataset should contain every step of every attack, including attack tactics, techniques, and procedures. This detailed description makes it possible to reconstruct the USCKC as a useful abstraction for further studies. This means that the data source should have conducted forensics at the greatest details possible. Clearly, this property is hard to guarantee in the real world because of the lack of deployed sensors, which also hinders the effectiveness of forensics techniques and the fact that data about space infrastructures are typically classified by governments (i.e., not provided to the public).

- *Accuracy*: The details about attack tactics, techniques, and procedures described in a dataset should be accurate, leaving no room for ambiguity or misinterpretation. This property is also hard to guarantee because descriptions in real-world datasets are often ambiguous (e.g., different people may use different terms when documenting space cyber attacks and many sources often come from news reporting rather than rigorous investigations).

- *Zero Missing Data*: This means every aspect of an attack that should be described in a dataset is indeed described in the dataset. While the accuracy property implies that the provided descriptions should be accurate, the present property is different because a dataset can be accurate on the provided description but there can be missing data.

Note that defining what must be covered (i.e., assuring the comprehensiveness property) and how to make the descriptions accurate (i.e., assuring the accuracy property) are two open problems on their own, which are orthogonal to the focus of the present study, which addresses the missing-data problem. Unfortunately, real-world cyber attack datasets often miss a lot of detailed technical descriptions that are required to properly understand the cyber attacks, as evidenced by the specific dataset we will analyze.

#### 3.3.3.1 Collecting Raw Data

Real-world space cyber attack data is often buried in raw CTI (Cyber Threat Intelligence) reports in the Internet, typically in the form of non-technical news reports as well as blog and social media posts. This means such data can be collected via Internet search, using keywords such as "space AND cyber AND incident OR {system, infrastructure, satellite, attack, interference, eavesdropping, spoofing, weapon, Jamming, hijacking, seizure, corruption, interception, denial, deception}." Given that there can be many ways to collect such data, in this chapter we assume that an analyst is given a set of raw CTI reports with claimed pertinence to space cyber attacks, while bearing in mind that these reports may be low quality and could be incorrect in the worst-case scenario. The collected data is often in the form of unstructured narratives, and must go through a preprocessing step to assure relevance.

#### 3.3.3.2 Preprocessing Raw Data

This is to assure the relevance of each raw CTI report because some reports may be included by mistake, possibly incurred by errors of Internet search engines or incompetency of data sources. For this purpose, one approach is to leverage domain experts to filter the collected raw CTI reports according to the following attributes:

- *Incident identification*: Each incident should be assigned a unique incident identifier for reference purposes, while avoiding duplicates.
- *Attack type*: This refers to the attack type associated with an incident according to a chosen cyber risk taxonomy, such as the one presented in [32], which includes the following *attack types*: High-powered Laser, High-powered Microwaves, RF Interferences, Eavesdropping, Spoofing, Ultrawideband Weapon, Electromagnetic Pulse (EMP) Weapon, Jamming, Signal Hijacking, Seizure of Control, Data Corruption/Interception, Denial of Service (DoS), and Space Situational Awareness (SSA) Deception.
- *Date*: This is the date when an incident takes place, ideally including a unified time zone (e.g., the Greenwich Mean Time or GMT) if available.
- *Geographic locations*: This describes where an incident takes place, ideally including facility, city, state, and country. The presence of such location information is an indicator of incident credibility.

- *Attack description*: This is the description of an incident, the more detailed the better. It is often unstructured text in a raw CTI report and thus requires substantial manual verification to ensure the description makes a technical sense.
- *Attacker identity*: This is the name, alias, and/or country of origin of the attacker that incurred an incident. The attacker can be an individual, organization, company, or agency.
- *Victim identity*: This is the name and industry of the victim involved in an incident. This is another indicator of the credibility of the raw CTI reports.
- *Data sources*: These are the sources that report an incident. Multiple independent sources reporting the same incident indicate a higher incident credibility.

The result is a preprocessed dataset of unique space cyber attack incidents, where every incident is represented by a row of the attributes defined above. In the ideal case, the preprocessed dataset is *comprehensive*, *accuracy*, and containing *zero missing data* as described above.

#### 3.3.3.3 Processing Preprocessed Data with Full Information

The task is to construct a USCKC from the description of a space cyber attack, while leveraging the afore-reviewed cyber threat taxonomies (i.e., SPARTA [1], ATT&CK [133], and UKC [117]). We reiterate the choice of these taxonomies as follows: (i) they are widely used by practitioners, making our results easily adoptable by them; and (ii) they collectively allow us to create USCKCs across the four segments because ATT&CK focuses on enterprise IT networks in the ground segment, SPARTA

**Algorithm 1:** Constructing USCKC of a space cyber attack with no missing data

**Input:** attack description AD of the attack; space infrastructure graph $G_{\mathsf{infra}} = (V_{\mathsf{infra}}, E_{\mathsf{infra}})$; the joint SPARTA-ATT&CK attack tactics set $\mathrm{JAT}_{\mathrm{TA}}$; the joint SPARTA-ATT&CK attack techniques set $\mathrm{JAT}_{\mathrm{TE}}$

**Output:** USCKC $= \{\mathrm{PH}, \mathrm{AC}, \mathrm{TA}, \mathrm{TE}\}$, where PH $= \{ph_1, \ldots, ph_n\}$ is the ordered set of attack phase types, AC $= \{ac_1, \ldots, ac_n\}$ is the ordered set of attack activity types, TA $= \{ta_1, \ldots, ta_n\}$ is the ordered set of attack tactics, and TE $= \{te_1, \ldots, te_n\}$ is the ordered set of attack techniques that are employed by the attack

PH $\leftarrow \{\}$, AC $\leftarrow \{\}$, TA $\leftarrow \{\}$, TE $\leftarrow \{\}$ `// initializing empty ordered sets`

partition AD into $n$ attack steps in order

**for** $i = 1$ *to* $n$ **do**

  determine attack phase $ph_i$ to which step $i$ belong and append $ph_i$ to PH

  determine activity type $ac_i$ to which step $i$ belong and append $ac_i$ to AC

  determine attack tactics $ta_i$ to which step $i$ belong and append $ta_i$ to TA

  determine attack technique $te_i$ to which step $i$ belong and append $te_i$ to TE

USCKC $\leftarrow \{\mathrm{PH}, \mathrm{AC}, \mathrm{TA}, \mathrm{TE}\}$

**return** USCKC

focuses on the space segment, and UKC provides information for sequencing cyber attack progressions. Note that our framework can be adapted to accommodate other cyber threat taxonomies (e.g., the ones that may be introduced in the future). However, SPARTA, ATT&CK, and UKC do not provide any explicit method that can be leveraged to compile attack tactics or techniques into a USCKC. This prompts us to propose a *skeleton* algorithm, Algorithm 1, to construct a USCKC for each preprocessed space cyber attack description with full information (i.e., with no missing data), where *skeleton* means that the algorithm is extracted from our manual processing and can guide the design of automated algorithm in the future.

At a high level, Algorithm 1 generates one USCKC, including both tactic-level and technique-level manifestations, per space cyber attack in the preprocessed dataset. (For any entry in the preprocessed dataset with partial information or missing data, we will address it in Section 3.3.3.4.) Its input includes the description AD of a space cyber attack, the context space infrastructure $G_{\mathsf{infra}}$, the set of joint SPARTA-ATT&CK attack tactics $\text{JAT}_{\text{TA}}$ and the set joint SPARTA-ATT&CK attack techniques $\text{JAT}_{\text{TE}}$ that are defined by desired versions of SPARTA and ATT&CK. Its output includes $\mathsf{USCKC} = \{\text{PH}, \text{AC}, \text{TA}, \text{TE}\}$ of $n$ attack steps, where $\text{PH} = \{ph_1, \ldots, ph_n\}$ is an ordered set of *phase types* with $ph_i \in \{\mathsf{in}, \mathsf{through}, \mathsf{out}\}$ for $1 \leq i \leq n$ being the phase type to which the $i$th attack step belong, $\text{AC} = \{ac_1, \ldots, ac_n\}$ is the ordered set of *activity types* with $ac_i \in \{\text{objective}, \text{milestone}, \text{enabling}, \text{information discovery}\}$ being the activity type to which the $i$th attack step belong, $\text{TA} = \{ta_1, \ldots, ta_n\}$ is the ordered set of *attack tactics* with $ta_i \in \text{JAT}_{\text{TA}}$ being the attack tactic to which the $i$th attack step belong, and $\text{TE} = \{te_1, \ldots, te_n\}$ is the ordered set of *attack techniques* with $te_i \in \text{JAT}_{\text{TE}}$ being the attack technique that is used in the $i$th attack step.

More specifically, the algorithm has two stages, *partitioning* (Line 2) and *classifying* (Lines 3-7), as follows. Specifically, Line 1 initializes the ordered empty sets PH, AC, TA, and TE. Line 2 partitions AD into $n$ attack steps. In loop iteration for the $i$th attack step, Line 4-7 determine the corresponding attack phase $ph_i \in \{\mathsf{in}, \mathsf{through}, \mathsf{out}\}$, activity type $ac_i \in \{\text{objective, milestone, enabling, information discovery}\}$, attack tactic $ta_i \in \text{JAT}_{\text{TA}}$, and attack technique $te_1 \in \text{JAT}_{\text{TE}}$. We note that it is non-trivial to automate these steps (i.e., Lines 2 and 4-7), even leveraging Large Language Models. This is because the algorithm requires, for instance, a deep understanding of $G = (V, E)$ and $\text{JAT}_{\text{TE}}$ in order to distinguish an attack phase $ph_i = \mathsf{in}$, meaning the attacker compromises a module $v \in V$ as an *entry node* (typically from outside the space infrastructure), from an attack phase $ph_i = \mathsf{through}$, meaning the attacker cannot directly compromise $v$ from outside the space infrastructure. Correspondingly, the computational complexity of the algorithm depends on how each function is implemented; in the present study, they are realized manually and thus we cannot report the complexity.

#### 3.3.3.4 Addressing the Missing Data Problem

In practice, we often encounter the problem of missing data in the raw and preprocessed datasets, especially the attack description AD missing many technical details. In this case, Algorithm 1, which requires full information, cannot be used. To address this missing data problem, we propose extrapolating the missing data, which is a non-trivial task that demands a deep understanding of space infrastructure $G_{\mathsf{infra}}$, the joint attack tactics set $\text{JAT}_{\text{TA}}$, and the joint attack techniques set $\text{JAT}_{\text{TE}}$. For example, a proper understanding of $G_{\mathsf{infra}}$ is necessary to determine whether a USCKC requires no through phases, which can happen when the attacker's *entry node* and *objective node* are the same, or multiple through phases, which can happen when the attacker's entry node and objective node belong to different modules, components or even segments; a deep understanding of $\text{JAT}_{\text{TA}}$ is required to determine if the identified attack tactics can work together to achieve the attacker's overall objective; and an in-depth understanding of $\text{JAT}_{\text{TE}}$ is required to identify when $te_i$ requires $te_{i-1}$ as a prerequisite or precondition. These explain why we have not been able to automate this algorithm and why we achieve this manually in the present study, while noting that it is non-trivial to apply Large Language Models to tackle the problem.

To address this missing data problem, we propose a *skeleton* Algorithm 2 to construct a set of probable USCKCs for each AD. Its input includes the context space infrastructure $G_{\mathsf{infra}}$, the set of joint attack tactics $\text{JAT}_{\text{TA}}$ and the set joint attack techniques $\text{JAT}_{\text{TE}}$ that are defined by desired versions of SPARTA and ATT&CK. Its output includes $\{\mathsf{USCKC}\}$ where $\mathsf{USCKC} = \{\text{PH}, \text{AC}, \text{TA}, \text{TE}\}$ of $s$ attack steps with $s$ to be determined (unlike Algorithm 1), PH $= \{ph_1, \ldots, ph_s\}$ is an ordered set of *phase types* with $ph_i \in \{\mathsf{in}, \mathsf{through}, \mathsf{out}\}$ for $1 \leq i \leq s$ being the phase type to which the $i$th attack step belong, AC $= \{ac_1, \ldots, ac_s\}$ is the ordered set of *activity types* with $ac_i \in \{\text{objective}, \text{milestone}, \text{enabling}, \text{information discovery}\}$ being the activity type to which the $i$th attack step belong, TA $= \{ta_1, \ldots, ta_s\}$ is the ordered set of *attack tactics* with $ta_i \in \text{JAT}_{\text{TA}}$ being the attack tactic to which the $i$th attack step belong, and TE $= \{te_1, \ldots, te_s\}$ is the ordered set of *attack techniques* with $te_i \in \text{JAT}_{\text{TE}}$ being the attack technique that is used in the $i$th observed or extrapolated attack step.

The algorithm proceeds in three stages, *partitioning* (Line 2), *extrapolating* (Lines 3-15), and *combining* (Line 16), as follows. Line 1 initializes the ordered empty sets PH, AC, TA, and TE. Line 2 partitions AD into $n'$ *observed* attack steps, while noting that there may be *hidden* or *unobserved* steps. For the $i$th observed attack step ($1 \leq i \leq n'$), Lines 4-11 obtain the observed attack step, denoted by subscript $(i, 1)$ and extrapolate the $|J_i|$ missing attack steps prior to the observed attack step. The total number of steps is $s = \sum_{i=1}^{n'}(1 - J_i)$ because an observed attack step $i$ is extrapolated to $(1 - J_i)$ steps (including the observed step), while noting that $J_i \leq 0$. By renaming the sequence $(X_{1,J_1+1}, \ldots, X_{1,1}, X_{2,J_2+1}, \ldots, X_{2,1}, \ldots, X_{n',J_{n'}+1}, \ldots, X_{n',1})$ as $(X_1, \ldots, X_s)$, where $X \in \{ph, ac, ta, te\}$, and by renaming $(k_{1,J_1+1}, \ldots, k_{1,1}, k_{2,J_2+1},$ $\ldots, k_{2,1}, \ldots, k_{n',J_{n'}+1}, \ldots, k_{n',1})$ as $(k_1, \ldots, k_s)$, Line 16 generates up to $K = \prod_{z=1}^{s} k_s$ USCKC's, where each USCKC is similar to what is generated in Algorithm 1 but with length $s$. Note that in our manual approach for Line 16, we may eliminate some extrapolated attack steps to make better space cybersecurity sense for a particular USCKC, reducing the number of attack step from $s$ to $s'$, where $s' < s$. Similar to Algorithm 1, the computational complexity of the algorithm depends on how each function is implemented; in the present study, they are realized manually and thus we cannot report the complexity.

### 3.3.4 Defining Metrics to Characterizing Attacks

We define three metrics to characterize cyber attacks against space infrastructures: *attack consequence*, *attack sophistication*, and USCKC *Likelihood* with each being a multi-dimensional vector which can be aggregated into a single number if desired. These metrics are useful, for example, by employing the former to compare multiple attacks in terms of their consequences or damages and employing the latter to characterize how advanced each of those attacks are.

#### 3.3.4.1 Attack Consequence Metric

Given that space infrastructures have four segments and that attack consequences may be manifested at some or all of the four segments, we define a vector of vectors, denoted by $(\vec{s}_{\mathbf{S}}, \vec{g}_{\mathbf{G}}, \vec{u}_{\mathbf{U}}, \vec{l}_{\mathbf{L}})$, to represent the attack consequences to the space segment (**S**), ground segment (**G**), user segment (**U**), and link segment (**L**), respectively. The four vectors are defined as follows.

**Algorithm 2:** Constructing USCKCs of space cyber attacks with missing data

**Input:** description AD of an attack with missing data; space infrastructure graph $G_{\mathsf{infra}} = (V, E)$; the joint SPARTA-ATT&CK attack tactics set $\text{JAT}_{\text{TA}}$; the joint SPARTA-ATT&CK attack techniques set $\text{JAT}_{\text{TE}}$

**Output:** a set $\{\mathsf{USCKC}\}$ of probable USCKCs extrapolated from AD, where $\mathsf{USCKC} = \{\text{PH}, \text{AC}, \text{TA}, \text{TE}\}$

PH $\leftarrow \{\}$, AC $\leftarrow \{\}$, TA $\leftarrow \{\}$, TE $\leftarrow \{\}$ `// initializing empty ordered sets`

partition AD into $n'$ attack steps of attack techniques, in sequential order $1, 2, \ldots$ `// the` $n'$ `steps are given or observed`

**for** $i = 1$ *to* $n'$ **do**

  $j_i \leftarrow 1$ `// keeping extrapolated steps`

  **repeat**

    determine attack phase $ph_{i,j_i}$ to which the current attack step $(i, j_i)$ belong

    determine activity type $ac_{i,j_i}$ to which the current attack step $(i, j_i)$

    determine attack tactics $ta_{i,j_i}$ to which the current attack step $(i, j_i)$ belong

    determine a set of $k_{i,j_i}$ probable techniques $\text{TE}_{i,j_i} = \{te_{i,j_i,1}, \ldots, te_{i,j_i,k_{i,j_i}}\}$ that can be used at the present attack step $(i, j_i)$ `// where` $|\text{TE}_{i,1}| = 1$

    $j_i \leftarrow j_i - 1$

  **until** *no more prior steps need to be extrapolated*;

  $J_i \leftarrow j_i$ `// bookkeeping,` $J_i \leq 0$

  PH $\leftarrow \{ph_{i,J_i+1}, ph_{i,J_i+2}, \ldots, ph_{i,1}\}$

  AC $\leftarrow \{ac_{i,J_i+1}, ac_{i,J_i+2}, \ldots, ac_{i,1}\}$

  TA $\leftarrow \{ta_{i,J_i+1}, ta_{i,J_i+2}, \ldots, ta_{i,1}\}$

$\{\mathsf{USCKC}\} \leftarrow \{\text{PH}, \text{AC}, \text{TA}, \text{TE}\}$ where $\text{TE} = \text{TE}_{1,J_i+1} \times \ldots \times \text{TE}_{1,1} \times \ldots \times \text{TE}_{n',J_i+1} \times \ldots \times \text{TE}_{n',1}$ and each USCKC makes space cybersecurity sense

**return** $\{\mathsf{USCKC}\}$

**Consequence to space segment ($\vec{s}_{\text{S}}$).** We define the attack consequence to a Space Segment **S** as $\vec{s}_{\text{S}} = (\vec{s}_{\mathsf{BS}}, \vec{s}_{\mathsf{PL}})$, where vector $\vec{s}_{\mathsf{BS}}$ denotes the consequence to the Bus System (BS) and vector $\vec{s}_{\mathsf{PL}}$ denotes the consequence to the Payload (PL).

- We define $\vec{s}_{\mathsf{BS}} = (s_{\mathsf{BS},1}, \ldots, s_{\mathsf{BS},6})$ as a vector of attack consequences to the six components of the Bus System: electrical power ($s_{\mathsf{BS},1}$), attitude control ($s_{\mathsf{BS},2}$), communication ($s_{\mathsf{BS},3}$), command & data ($s_{\mathsf{BS},4}$), propulsion ($s_{\mathsf{BS},5}$), and thermal control ($s_{\mathsf{BS},6}$). We define $s_{\mathsf{BS},j} \in [0,1]$, $1 \leq j \leq 6$, as the degree of the functionality (i.e., availability) of the corresponding component being degraded because of the attack in question, where $s_{\mathsf{BS},j} = 0$ (or 1) means the functionality is 0% (or 100%) degraded.

- We define $\vec{s}_{\mathsf{PL}} = (s_{\mathsf{PL},1}, \ldots, s_{\mathsf{PL},5})$ as a vector of attack consequences to the five payload components: communication ($s_{\mathsf{PL},1}$), navigation ($s_{\mathsf{PL},2}$), scientific application ($s_{\mathsf{PL},3}$), remote sensing ($s_{\mathsf{PL},4}$), and national security ($s_{\mathsf{PL},5}$), with $s_{\mathsf{PL},j} \in [0,1]$ in the same fashion as $s_{\mathsf{BS},j}$.

The preceding definition of $\vec{s}_{\mathsf{S}}$ has several salient features, which also apply to the subsequent metrics corresponding to the other segments. First, we differentiate "defining what to measure" from "how to measure what we need to measure." The present study addresses the former. Note also that our definitions remain valid when considering the inter-dependencies between different segments. Concerning the latter, it would be ideal that measurements of these metrics are provided to analysts for purposes such as our present study; or alternatively, there is a community-wide agreement on the degrees of degradation mentioned above. In the absence of both, one can use their own domain expertise to estimate these metrics. Second, we make the number of components specific to the system model described in Figure 3.2 (e.g., six components in the Bus System) to make the definitions easier to follow. The definitions can be trivially generalized to accommodate an arbitrary number of components. Third, the "fine-granularity" of $\vec{s}_{\mathsf{BS}}$ and $\vec{s}_{\mathsf{PL}}$ makes them suitable to compare the consequences of multiple attacks and to make statements like "Attack 1 is more powerful than Attack 2." Fourth, one can aggregate the vector metrics into a single number. For example, $\vec{s}_{\mathsf{BS}}$ can be aggregated into $\bar{s}_{\mathsf{BS}}$ via some mathematical function $f$, namely $\bar{s}_{\mathsf{BS}} = f(s_{\mathsf{BS},1}, \ldots, s_{\mathsf{BS},6})$, where $f$ can be for instance the (weighted) algebraic average function that makes cybersecurity sense because they deal with the same property (i.e., availability). We can similarly aggregate $\bar{s}_{\mathsf{BS}}$ and $\bar{s}_{\mathsf{PL}}$ into a single number.

**Consequence to ground segment** ($\vec{g}_{\mathsf{G}}$). We define the attack consequence to a Ground Segment **G** as $\vec{g}_{\mathbf{G}} = (\vec{g}_{\mathsf{GS}}, \vec{g}_{\mathsf{MC}}, \vec{g}_{\mathsf{DPC}}, \vec{g}_{\mathsf{RT}})$, where $\vec{g}_{\mathsf{GS}}$ is the consequence to the Ground Station (GS), $\vec{g}_{\mathsf{MC}}$ is the consequence to Mission Control (MC), $\vec{g}_{\mathsf{DPC}}$ is the consequence to the Data Processing Center (DPC), $\vec{g}_{\mathsf{RT}}$ is the consequence to the Remote Terminal (RT).

- We define $\vec{g}_{\mathsf{GS}} = (g_{\mathsf{GS},1}, \ldots, g_{\mathsf{GS},4})$ as a vector of consequences to the four components of the Ground Station: tracking ($g_{\mathsf{GS},1}$), ranging ($g_{\mathsf{GS},2}$), transmission ($g_{\mathsf{GS},3}$), and reception ($g_{\mathsf{GS},4}$), where $g_{\mathsf{GS},j} \in [0,1]$, $1 \le j \le 4$, and 0 (1) means 0% (100%) functionality degradation.

- We define $\vec{g}_{\mathsf{MC}} = (g_{MC,1}, g_{MC,2}, g_{\mathsf{MC},3})$ as a vector of consequences to the three components of the Mission Control: telemetry processing ($g_{\mathsf{MC},1}$), commanding ($g_{\mathsf{MC},2}$), and analysis and support ($g_{\mathsf{MC},3}$), where $g_{\mathsf{MC},j} \in [0,1]$, $1 \le j \le 3$, in the same fashion as $g_{\mathsf{GS},j}$.

- We define $\vec{g}_{\mathsf{DPC}} = (g_{\mathsf{DPC},1}, g_{\mathsf{DPC},2})$ as a vector of consequence to the two components of the Data Processing Center: mission analysis ($g_{\mathsf{DPC},1}$) and payload processing ($g_{\mathsf{DPC},2}$), where $g_{\mathsf{DPC},j} \in [0,1]$ as similiar to $g_{\mathsf{GS},j}$.

- We define $\vec{g}_{\textsf{RT}} = (g_{\textsf{RT},1}, g_{\textsf{RT},2})$ as a vector of consequence to the components of the Remote Terminal: network access ($g_{\textsf{RT},1}$) and software access ($g_{\textsf{RT},2}$), where $g_{\textsf{RT},j} \in [0,1]$ as with $g_{\textsf{GS},j}$,

Note that $\vec{g}_{\mathbf{G}}$ has the same salient features as $\vec{s}_{\mathbf{S}}$.

**Consequence to user segment ($\vec{u}_{\mathrm{U}}$).** We define the attack consequence to a User Segment **U** as $\vec{u}_{\mathbf{U}} = (u_1, u_2, u_3)$, whose elements respectively measure the consequence to components of **U**: transmission ($u_1$), reception ($u_2$), and processing ($u_3$), with $u_j \in [0,1]$, $1 \leq j \leq 3$, as with $s_{\textsf{BS},j}$. Note that $\vec{u}_{\mathrm{U}}$ has the same salient features as $\vec{s}_{\mathbf{S}}$ or specifically $\vec{s}_{\textsf{BS}}$.

**Consequence to link segment ($\vec{l}_{\mathrm{L}}$).** We define the attack consequence to a Link Segment as $\vec{l}_{\mathbf{L}} = (\{\vec{l}_{\mathbf{S}}\}, \{\vec{l}_{\mathbf{G}}\}, \{\vec{l}_{\mathbf{SS}}\}, \{\vec{l}_{\mathbf{GG}}\}, \{\vec{l}_{\mathbf{SG}}\}, \{\vec{l}_{\mathbf{SU}}\}, \{\vec{l}_{\mathbf{GU}}\}, \{\vec{l}_{\mathbf{UU}}\})$, where the elements respectively correspond to a set of links within a space infrastructure affected by an attack, within a ground wide-area network (WAN), between two satellites, between two ground WANs, between a Space Segment and a Ground Segment, between a Space Segment and a User Segment, between a Ground Segment and a User Segment, and between two users. We further define:

- $\vec{l}_{\mathbf{S}} = (l_{\mathbf{S},\mathrm{C}}, l_{\mathbf{S},\mathrm{I}}, l_{\mathbf{S},\mathcal{A}})$ as the consequences to a link between the Bus System and the Payload, where $l_{\mathbf{S},\mathrm{C}}, l_{\mathbf{S},\mathrm{I}}, l_{\mathbf{S},\mathcal{A}} \in [0,1]$ are respectively the consequence to the confidentiality, integrity, and availability assurance of the link, with 0 (1) meaning 0% (100%) degradation.

- $\vec{l}_{\mathbf{G}} = (\vec{l}_{\textsf{GS,MC}}, \vec{l}_{\textsf{GS,DPC}}, \vec{l}_{\textsf{GS,RT}}, \vec{l}_{\textsf{MC,DPC}}, \vec{l}_{\textsf{MC,RT}}, \vec{l}_{\textsf{DPC,RT}})$ as the consequences to links of components in a Ground Segment, where $\vec{l}_{\textsf{GS,MC}} = (l_{\textsf{GS,MC};\mathrm{C}}, l_{\textsf{GS,MC};\mathrm{I}}, l_{\textsf{GS,MC};\mathcal{A}})$ $\in [0,1]^3$ are respectively the consequence to the confidentiality, integrity, and availability of the link between a Ground Station (GS) and a Mission Control (MC), with 0 (1) meaning 0% (100%) degradation. The other elements of $\vec{l}_{\mathbf{G}}$ are defined in the same fashion.

- $\vec{l}_{\mathbf{SS}} = (\vec{l}_{\mathbf{SS},\mathrm{C}}, \vec{l}_{\mathbf{SS},\mathrm{I}}, \vec{l}_{\mathbf{SS},\mathcal{A}})$ as the consequences to a link between two satellites, where $\vec{l}_{\mathbf{SS},\mathrm{C}}, \vec{l}_{\mathbf{SS},\mathrm{I}}, \vec{l}_{\mathbf{SS},\mathcal{A}} \in [0,1]^3$ are respectively the consequence to the confidentiality, integrity, and availability assurance of the link, with 0 (1) meaning 0% (100%) degradation.

- $\vec{l}_{\mathbf{GG}} = (\vec{l}_{\mathbf{GG},\mathrm{C}}, \vec{l}_{\mathbf{GG},\mathrm{I}}, \vec{l}_{\mathbf{GG},\mathcal{A}})$ as the consequences to a link between two ground WANs, where $\vec{l}_{\mathbf{GG},\mathrm{C}}, \vec{l}_{\mathbf{GG},\mathrm{I}}, \vec{l}_{\mathbf{GG},\mathcal{A}} \in [0,1]^3$ are respectively the consequence to the confidentiality, integrity, and availability assurance of the link, with 0 (1) meaning 0% (100%) degradation.

- $\vec{l}_{\mathbf{SG}} = (\vec{l}_{\mathbf{SG},\mathrm{C}}, \vec{l}_{\mathbf{SG},\mathrm{I}}, \vec{l}_{\mathbf{SG},\mathcal{A}})$ as the consequences to a link between a Space Segment and a Ground Segment, where $\vec{l}_{\mathbf{SG},\mathrm{C}}, \vec{l}_{\mathbf{SG},\mathrm{I}}, \vec{l}_{\mathbf{SG},\mathcal{A}} \in [0,1]^3$ are the consequence to the confidentiality, integrity, and availability assurance of the link, with 0 (1) meaning 0% (100%) degradation.

- $\vec{l}_{\mathbf{SU}} = (\vec{l}_{\mathbf{SU},\mathrm{C}}, \vec{l}_{\mathbf{SU},\mathrm{I}}, \vec{l}_{\mathbf{SU},\mathcal{A}})$ as the consequences to a link between a Space Segment and User Segment, where $\vec{l}_{\mathbf{SU},\mathrm{C}}, \vec{l}_{\mathbf{SU},\mathrm{I}}, \vec{l}_{\mathbf{SU},\mathcal{A}} \in [0,1]^3$ are the consequence to the confidentiality, integrity, and availability assurance of the link, with 0 (1) meaning 0% (100%) degradation.

- $\vec{l}_{\mathbf{GU}} = (\vec{l}_{\mathbf{GU},\mathrm{C}}, \vec{l}_{\mathbf{GU},\mathrm{I}}, \vec{l}_{\mathbf{GU},\mathcal{A}})$ as the consequences to a link between a Ground Segment and User Segment, where $\vec{l}_{\mathbf{SS},\mathrm{C}}, \vec{l}_{\mathbf{SS},\mathrm{I}}, \vec{l}_{\mathbf{SS},\mathcal{A}} \in [0,1]^3$ are the consequence to the confidentiality, integrity, and availability assurance of the link, with 0 (1) meaning 0% (100%) degradation.

- $\vec{l}_{\mathbf{UU}} = (\vec{l}_{\mathbf{UU},\mathrm{C}}, \vec{l}_{\mathbf{UU},\mathrm{I}}, \vec{l}_{\mathbf{UU},\mathcal{A}})$ as the consequences to a link between two users, where $\vec{l}_{\mathbf{UU},\mathrm{C}}, \vec{l}_{\mathbf{UU},\mathrm{I}}, \vec{l}_{\mathbf{UU},\mathcal{A}} \in [0,1]^3$ are the consequence to the confidentiality, integrity, and availability assurance of the link, with 0 (1) meaning 0% (100%) degradation.

Note that $l_{\mathbf{L}}$ has the same salient features as $\vec{s}_{\mathbf{S}}$ except how metrics may be aggregated. For example, it would not make good cybersecurity sense to aggregate $(l_{\mathbf{S},\mathrm{C}}, l_{\mathbf{S},\mathrm{I}}, l_{\mathbf{S},\mathcal{A}})$ via a (weighted) algebraic average because confidentiality, integrity, and availability metrics describe different properties. One may suggest to aggregate them via $1 - [(1 - l_{\mathbf{S},\mathrm{C}})(1 - l_{\mathbf{S},\mathrm{I}})(1 - l_{\mathbf{S},\mathcal{A}})]$, which is reminiscent of the aggregation of Common Vulnerability Scoring System (CVSS) scores [39]. However, this does not appear sound as this aggregation function appears rooted in Probability Theory, which however does not apply here because the events, even if $(l_{\mathbf{S},\mathrm{C}}, l_{\mathbf{S},\mathrm{I}}, l_{\mathbf{S},\mathcal{A}})$ can be interpreted as probabilities, are not independent (e.g., the three assurances may be degraded at will by an attacker, rather than degrading independently of each other). Accordingly, we define:

**Definition III.1** (Attack Consequence)**.** *We define the attack consequence of a cyber attack against space infrastructures as*

$$\left(\cup_{\mathbf{S}}\{\vec{s}_{\mathbf{S}}\},\ \cup_{\mathbf{G}}\{\vec{g}_{\mathbf{G}}\},\ \cup_{\mathbf{U}}\{\vec{u}_{\mathbf{U}}\},\ \cup_{\mathbf{L}}\{\vec{l}_{\mathbf{L}}\}\right),$$

*where the union is over all the Space Segments (**S**), Ground Segments (**G**), User Segments (**U**), and Link Segments (**L**) that are affected by the attack.*

#### 3.3.4.2 Attack Sophistication Metric

We define this metric to characterize how sophisticated a cyber attack against a space system is, through the lens of attack tactics and attack techniques. Suppose for each cyber attack we extrapolate it into a set of $n$ USCKCs, denoted by $\mathsf{USCKC} = \{\mathsf{USCKC_1}, \ldots, \mathsf{USCKC_n}\}$, where $\mathsf{USCKC_i}$ $(1 \leq i \leq n)$ denotes the $i$th USCKC that is possibly associated with the cyber attack in question (when there is missing data describing the attack). Suppose $\mathsf{USCKC_i}$ is associated with a set of $u$ attack *tactics*, denoted by $\text{TA}_i = \{\text{TA}_{i,1}, \ldots, \text{TA}_{i,u}\}$, and with a set of $v$ attack *techniques*, denoted by $\text{TE}_i = \{\text{TE}_{i,1}, \ldots, \text{TE}_{i,v}\}$, where $\text{TA}_{i,j}$ and $\text{TE}_{i,j}$ are respectively associated with a given sophistication score $\alpha_{\text{TA}_{i,j}}$ and $\alpha_{\text{TE}_{i,j}}$. For $\mathsf{USCKC_i}$, we define its tactic sophistication as the maximum element among the element of $\text{TA}_i$, namely $\max(\text{TA}_i) = \max(\{\alpha_{\text{TA}_{i,1}}, \ldots, \alpha_{\text{TA}_{i,u}}\})$, and define its attack technique sophistication as $\max(\text{TE}_i) = \max(\{\alpha_{\text{TE}_{i,1}}, \ldots, \alpha_{\text{TE}_{i,v}}\})$, which respectively correspond to the most sophisticated attack tactic and attack technique used by the attacker. Now we are ready to define:

**Definition III.2** (Attack Sophistication). *For an attack described by a set of probable unified space cyber kill chains* $\mathsf{USCKC} = \{\mathsf{USCKC_1}, \ldots, \mathsf{USCKC_n}\}$*, we define its* possible highest sophistication *as vector* $(\alpha_{TA_+}, \alpha_{TE_+})$*, where* $\alpha_{TA_+} = \max(\{\max(TA_1), \ldots, \max(TA_n)\})$ *which corresponds to the most sophisticated attack tactic that is used among the possible kill chains, and* $\alpha_{TE_+} = \max(\{\max(TE_1), \ldots, \max(TE_n)\})$ *which* *corresponds to the most sophisticated attack technique that is used among the possible kill chains.*

*We define its* possible lowest sophistication *as vector* $(\alpha_{TA_-}, \alpha_{TE_-})$*, where* $\alpha_{TA_-} = \min(\{\max(TA_1), \ldots, \max(TA_n)\})$ *which corresponds to the least sophisticated attack tactic that is necessary for the attack to succeed, and* $\alpha_{TE_-} = \min(\{\max(TE_1), \ldots, \max(TE_n)\})$ *which corresponds to the least sophisticated attack technique that is necessary for the attack to succeed.*

Note that Definition III.2 can be easily extended to define the metric of, for example, *possible average sophistication*.

#### 3.3.4.3 USCKC **Likelihood**

We define this metric to characterize the likelihood that a USCKC may have been used to successfully accomplish a given space cyber attack. This is important because we often encounter missing details of space cyber attacks, which incur uncertainty on the USCKC that is actually used.

Suppose a USCKC has $n$ attack steps at the attack technique level of abstraction, namely one attack technique per attack step, denoted by an ordered set $\mathsf{USCKC} = \{\mathsf{te}_1, \ldots, \mathsf{te}_\mathsf{n}\}$ where $te_i \in$ TE and USCKC may be obtained by Algorithm 2. To compute the likelihood that a USCKC may have been used, namely $L(\mathsf{USCKC_j})$, we assume a likelihood for each attack technique is given, either by domain experts or via a principled approach (the latter is an interesting future research problem). We observe that $L(\mathsf{USCKC_j})$ is bounded by the lowest $L(te_i)$ because all attack techniques within the $\mathsf{USCKC}$ must succeed for the entire $\mathsf{USCKC}$ to succeed. In the case of $\{\mathsf{USCKC}\}$, only one $\mathsf{USCKC} \in \{\mathsf{USCKC}\}$ must succeed. Hence, we define:

**Definition III.3** ($\mathsf{USCKC}$ Likelihood)**.** *For a* $\mathsf{USCKC}$ *described at the technique level, namely* $\mathsf{USCKC} = \{\mathsf{te}_1, \dots, \mathsf{te}_\mathsf{n}\}$*, the likelihood that the* $\mathsf{USCKC}$ *succeeds is defined as* $L(\mathsf{USCKC}) = \min(\mathsf{L}(\mathsf{te}_1), \dots, \mathsf{L}(\mathsf{te}_\mathsf{n}))$*, where* $L(te_\ell)$ *for* $1 \leq \ell \leq n$ *is the likelihood that attack technique* $te_\ell$ *is successful. For a set* $\{\mathsf{USCKC}\} = \{\mathsf{USCKC}_1, \dots, \mathsf{USCKC}_\mathsf{m}\}$*, the likelihood that* $\{\mathsf{USCKC}\}$ *succeeds is defined as* $L(\{\mathsf{USCKC}\}) = \max(\mathsf{L}(\mathsf{USCKC}_1), \dots, \mathsf{L}(\mathsf{USCKC}_\mathsf{m}))$.

### 3.3.5 Leveraging the Metrics to Answer Research Questions

Given a dataset of cyber attacks against space infrastructures, the metrics defined above can be leveraged to specify interesting research questions, such as: (i) What is the attack consequence of each space cyber attack that has occurred in practice? Is there any trend exhibited by the attacks in terms of their attack consequence? (ii) What is the attack sophistication of each space cyber attack that has occurred in practice? Is there any trend exhibited by the attacks in terms of their attack sophistication or attack entry points? (iii) What is the uncertainty associated with the extrapolated USCKCs? What is the likelihood of each extrapolated USCKC?

## 3.4 Case Study

We apply the framework to characterize space cyber attacks based on a dataset of real-world space cyber incidents.

### 3.4.1 Preparing Dataset

#### 3.4.1.1 Collecting Raw Data

In search of publicly available datasets, we first query “satellite incidents”, “space attack data”, “space incidents data” in Google Scholar (with no limiting parameters), which return 45 papers. Among these 45 papers, three papers [32, 43, 110] are referred to by the other 42 papers, in terms of real-world space incidents that have affected space infrastructures; this means that these 42 papers do not provide any new incidents. The 42 papers also refer to an online database [53] of space incidents. Specifically, 70 incidents are reported in [53], 112 in [110], 59 in [43], and 1,847 in [32].We gathered the references provided in these four sources for each incident, leading to 272 raw CTI reports in total. Then, another Google search with query “satellite, space, incidents, attack, dataset” returns many items, which include popular news articles and blogs, for which we reviewed the top 200 items. From the 200 items, we identify 65 relevant items that lead to 65 raw CTI reports. This provides us a total of $272 + 65 = 337$ raw CTI reports. We manually review these 337 raw CTI reports, which are very diverse in terms of the number of incidents they contain. For instance, one CTI report found in [32] actually contains 672 “single event upset” incidents where an electronic bit flips typically due to radiation in space, which are not about cyber attacks. Many other reported incidents found in the 337 CTI reports are also not about cyber attacks, such as incidents related to solar flare anomalies; these incidents are eliminated based on relevance. In total, we identify 162 space cyber attacks that make up our *raw dataset*.

Among these 162 space cyber attacks, 72 are reported in the conference version of the present chapter [27] and 90 are newly added in this version.

#### 3.4.1.2 Preprocessing Raw Data

We use domain expertise to manually extract space cyber attack incidents from the 337 raw CTI reports resulting from the preceding step. After eliminating duplicates, we obtain a dataset of 108 unique space cyber attack incidents, among which: (i) 65 are extracted from three sources [43, 53, 110] while noting that these three sources substantially overlap with each other, seven are extracted from [32], while noticing that these $65 + 7 = 72$ incidents are reported in the conference version [27]; and (ii) 36 are extracted from the 90 incidents resulting from our Internet search mentioned above and are added in the present version.

We observe that the 108 attacks have different descriptions. Specifically, three datasets [32, 53, 110] describe attacks via: (i) year of incident occurrence; (ii) category of the attack / incident, such as jamming; (iii) attack target, such as the USA-193 recon satellite; (iv) attacker identification, such as nation-state; (v) narrative of facts concerning the incident, which are all brief; (vi) high-level attack objective, such as state-espionage; and, (vii) source references, such as news articles reporting the incident. On the other hand, the fourth dataset [43] only provides (i), (ii), (v), and (vii). Even so, they all miss some data. Note that for (iv), the datasets provide attribution for 76% (86 out of the 108) space cyber attacks at varying degrees, ranging from country of origin to individual cyber threat organizations. Furthermore, the 36 attacks are generally described in diverse ways in narrative form, where a few are lengthy case studies (e.g., of the ViaSat attack [18]), while the majority are short and lacking significant details.

Consequently, it is difficult to gauge the veracity of the 108 attacks. Moreover, different incidents may lead to different interpretations. For example, the RoSat incident is attributed to a causation between the cyber intrusion of the Goddard Space Flight Centre and the satellite's subsequent destructive maneuver in [30,43,129,151], but this causation is refuted by a scientist on the RoSat project in an interview [91], which is resonated by another source [132]. This means that we must be able to conduct the proposed research in the *absence* of ground-truth attack details, and that we must embrace the uncertainty corresponding to the missing data while striving to leverage the available data to infer as much as we are able.

Still, we manage to derive that the 108 space cyber attacks comprise eight categories: 26 Data Corruption/Interception attacks; nine Denial of Service (DoS) attacks; three Eavesdropping attacks; two High-powered Laser attacks; 41 Jamming attacks; three Seizure of Control attacks; 15 Signal Hijacking attacks; and nine Spoofing attacks. We also observe that 73% of the attacks fall into three related categories: political, state espionage, and criminal; this indicates that space cyber attacks are likely state-sponsored. We refer to the resulting dataset of 108 space cyber attacks as the *preprocessed* space cyber attack dataset.

#### 3.4.1.3 Processing Preprocessed Data with Full Information

Given that all the 108 attacks have missing data, Algorithm 1 is not applicable to our case study. However, this does not mean the algorithm is useless. Instead, the algorithm is useful to researchers and practitioners who know the details of space cyber attacks (e.g., operators of space infrastructures).

#### 3.4.1.4 Addressing the Missing Data Problem

As mentioned above, the description of all 108 attacks have missing data, which prompts us to extrapolate attack phases, activities, tactics and techniques as described in Algorithm 2, as follows. We manually partition the description of each of the 108 attacks into $n'$ observed or reported individual attack steps (Line 2), where $n'$ ranges from one to nine steps, with average $\bar{n'} = 2$. For example, the description of the RoSat attack contains the statement that the attacker caused the satellite to "*miscalculate its alignment and turned toward the sun . . . NASA was able to correct*" [151]. For this attack, our manual analysis extracts $n' = 9$ observed attack steps at the technique-level.

For each of the $n'$ observed steps, we manually determine whether and how to extrapolate (Lines 3-15). Among all the observed steps and the 108 attacks, we extrapolate one to 10 attack steps, meaning that we always extrapolate at least one step preceding a given attack step. Consider again the RoSat attack and the 5th (i.e., $i = 5$) attack step of the observed $n' = 9$ attack steps. This 5th attack step requires network data to be successfully executed at the network access module, meaning the data must be provided by the previous step. This prompts us to extrapolate an attack step prior to the one in question, leading to attack phase $ph_{5,0} = \mathsf{through}$, activity type $ac_{5,0} = \text{Information Discovery}$, attack tactic $ta_{5,0} = \text{Reconnaissance}$; in terms of attack techniques, we hypothesize that the attacker requires network data to successfully bypass defense (i.e., to accomplish an *enabling* activity) leads to $k_{5,0} = 2$ probable attack techniques, namely $te_{5,0,1}$ = T1595 (Active Scanning ATT&CK technique) and $te_{5,0,2}$ = T1590 (Gather Victim Network Information ATT&CK technique). We manually determine whether all the combinations obtained via the Cartesian product (Line 16) make space cybersecurity sense.

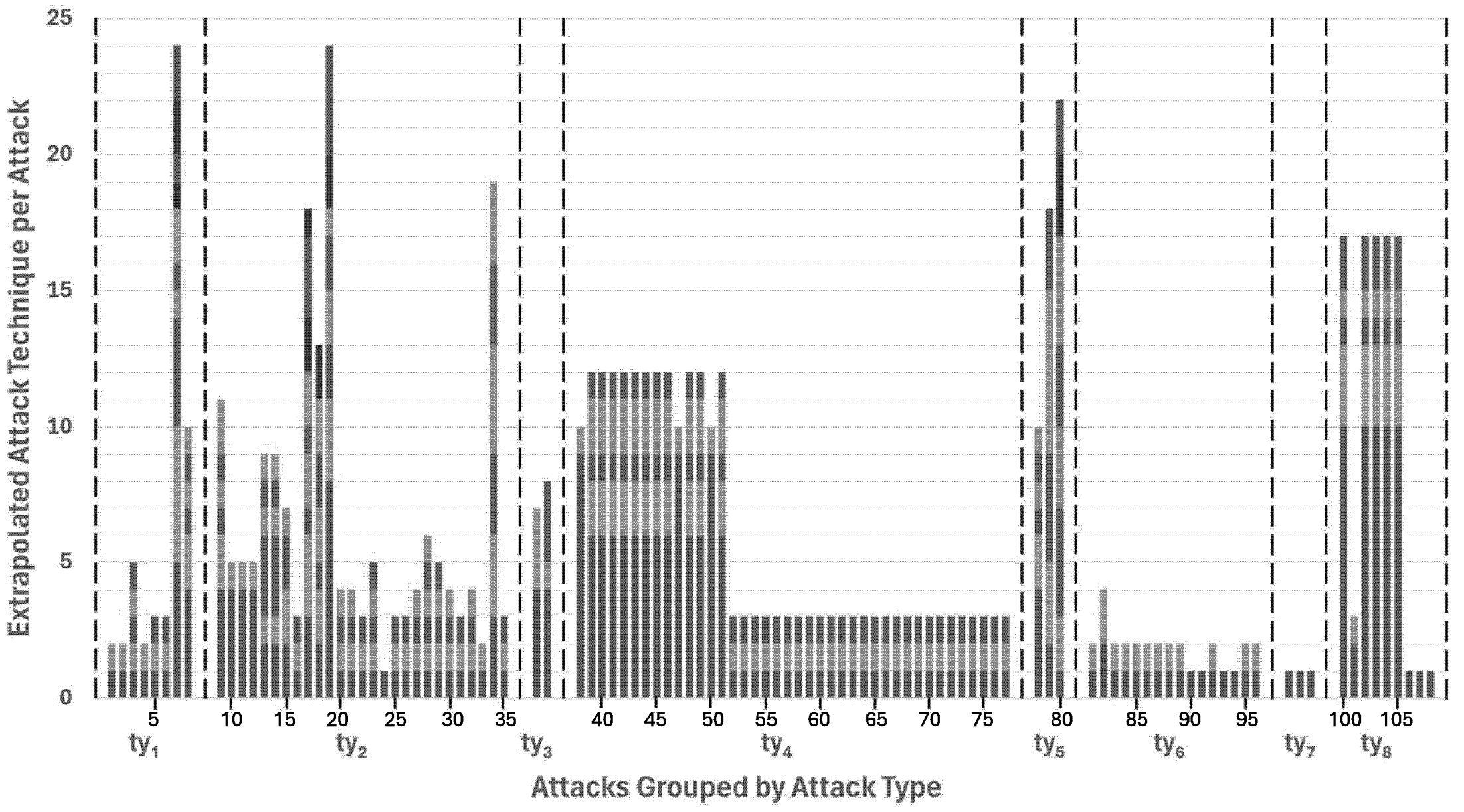

Figure 3.3: The number of extrapolated attack techniques ($y$-axis) for each of the 108 attacks ($x$-axis). For each bar, the number of colors corresponds to the number of extrapolated attack steps, meaning that the length of a color corresponds to some $k_{i,j_i}$ in Algorithm 2; the product of the lengths of sub-bars in different colors is up to $|\{\mathsf{USCKC}\}|$ in Algorithm 2, namely the total number of probable USCKCs extrapolated from an attack, because some USCKCs may not make space cybersecurity sense. For instance, we extrapolated the RoSat attack (i.e., attack #79) for five attack steps, leading to $2 \times 3 \times 4 \times 6 \times 3 = 432$ probable USCKCs in total (i.e., they all make space cybersecurity sense). The 108 attacks are grouped according to their *attack type*, showing that eight (out the aforementioned 13) attack types are observed, dubbed $\text{ty}_1$ (Denial of Service), $\text{ty}_2$ (Data Corruption/Interception), $\text{ty}_3$ (High-powered Laser), $\text{ty}_4$ (Jamming), $\text{ty}_5$ (Seizure of Control), $\text{ty}_6$ (Signal Hijacking), $\text{ty}_7$ (Eavesdropping), and $\text{ty}_8$ (Spoofing).

Figure 3.3 plots the extrapolation results for the 108 attacks, where each bar corresponds to one attack and the attacks are organized according to the *attack type* to which they belong. For instance, the RoSat attack (#79) leads to $n' = 9$ and $s = 14$, meaning five attack steps are extrapolated. For these five steps, two, three, four, six, and three attack steps are probable, leading to $2 \times 3 \times 4 \times 6 \times 3 = 432$ probable $\mathsf{USCKC}$s in total. This large size of $\{\mathsf{USCKC}\}$ is indicative of the complexity of this attack and the large amount of missing data. We observe that attack types 1, 2, 5, and 8 have a larger number of probable $\mathsf{USCKC}$s, perhaps because they target the ground or user segment, where humans operate and interact with space infrastructures and offer attackers opportunities to wage cyber social engineering attacks, as opposed to the single entry node and deterministic behavior of the space segment or the physics-bound link segment. We further observe that attack types 1 and 3 target the ground segment, where attack capabilities effective against enterprise IT systems and networks are reusable against the ground segment (i.e., attackers can reuse their attack capabilities); attack type 5 is complex because these attackers often seeks to control the ground segment before seeking to control the space segment, as evidenced by the three attacks of this type in our dataset; and attack type 8 employs payloads that are often specific to their targets in the user segment. In summary, we observe that the ground and user segments provide a larger attack surface than the space and link segments because they offer attackers with opportunities to use more mature and more kinds of attack capabilities.

**Insight 1.** *There is a larger number of probable* $\mathsf{USCKC}$*s that target the ground and user segments than the ones that target the space and link segments perhaps because sophisticated attacks that directly target the space and link segments (without pivoting from the ground and user segments) are yet to be uncovered.*

We caveat our results with the consideration that ATT&CK, which is geared towards the ground and user segments, is more mature than SPARTA, which is geared towards the space and link segments, contributing to the larger number of USCKCs for attack types 1, 2, 5, and 8.

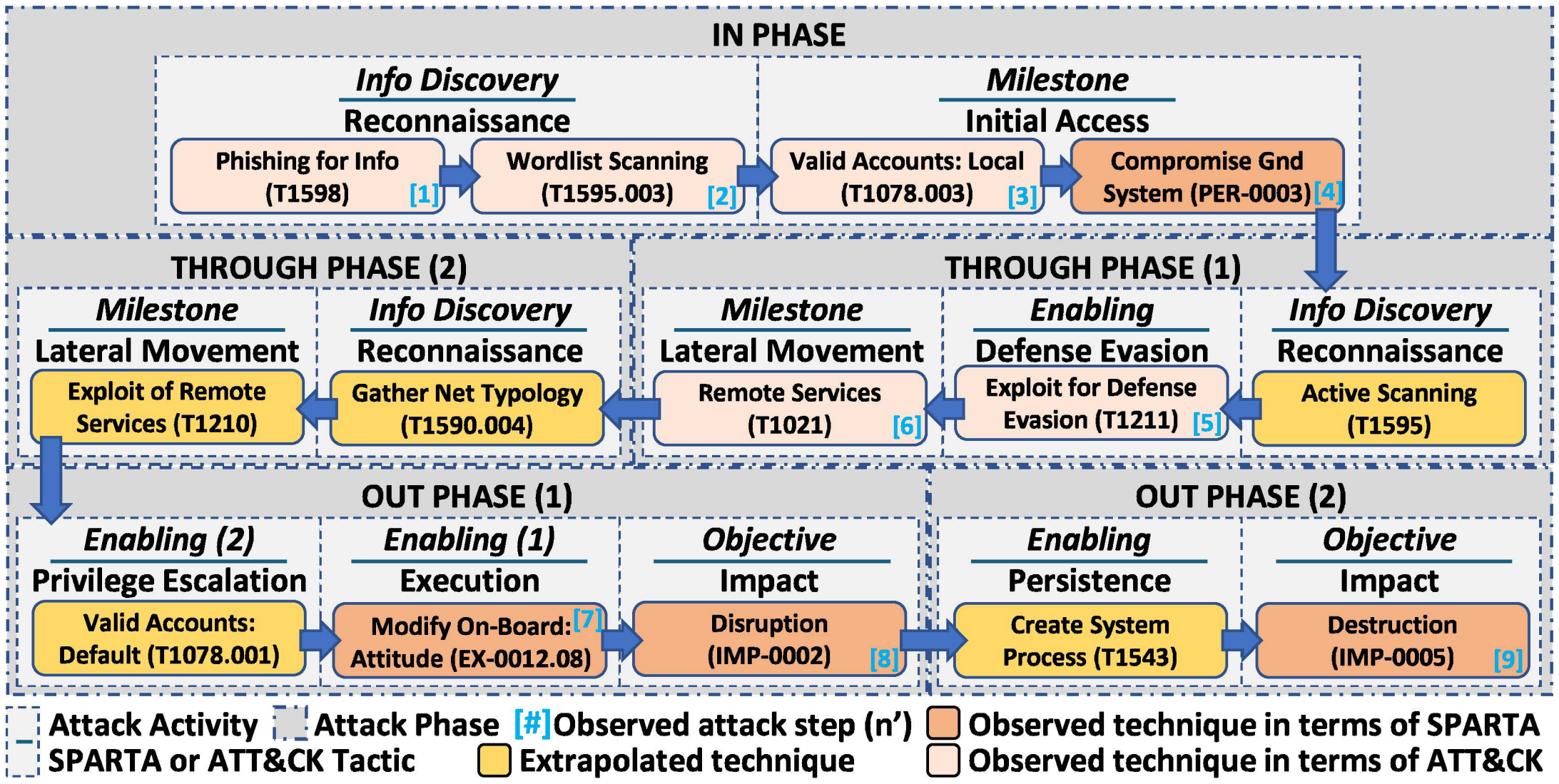

Figure 3.4: One example USCKC (out of the 432 probable USCKCs) we extrapolated for the RoSat 1998 attack, with $n' = 9$ and $s = 14$.

Figure 3.4 highlights one USCKC (out of 432) extrapolated from the RoSat attack. It can be understood as follows. During the in phase, the attacker uses techniques T1598 (Phishing for Information) and T1595.003 (Wordlist Scanning) to conduct a thorough reconnaissance, and use techniques T1078.003 (Valid Accounts: Local) and PER-0003 (Compromise Ground System) to gain initial access into the space infrastructure. In the first through phase, the attacker uses an extrapolated technique T1595 (Active Scanning) to conduct further reconnaissance, uses technique T1211 (Exploit for Defense Evasion) to evades the defense, and uses technique T1021 (Remote Services) to support its lateral movement to get deeper into the space infrastructure. During the second through phase, the attacker uses an extrapolated technique T1590.004 (Gather Network Typology) to accomplish further reconnaissance, and then uses an extrapolated technique T1210 (Exploit of Remote Services) to laterally moves to a module that can affect the space segment. In the first out phase, the attacker uses an extrapolated technique T1078.001 (Valid Accounts: Default) to escalate its privileges, uses technique EX-0012.08 (Modify On-Board: Attitude) to execute its malware, and uses technique IMP-0002 (Disruption) to temporarily impact the RoSat satellite. In the second out phase, the attacker uses an extrapolated technique T1543 (Create System Process) to establish persistence in the space infrastructure, and then uses technique IMP-0005 (Destruction) permanently to impact the RoSat's x-ray sensing device.

Figure 3.5 presents another example of the 432 probable USCKCs extracted from the RoSat attack. It can be understood as follows. During the in phase, the attacker uses techniques T1598 (Phishing for Information) and T1595.003 (Wordlist Scanning) to conduct reconnaissance, and uses techniques T1078.003 (Valid Accounts: Local) and PER-0003 (Compromise Ground System) to gain initial access to the space infrastructure. In the first through phase, the attacker uses an extrapolated technique T1590 (Gather Victim Network Information) to conduct reconnaissance, uses technique T1211 (Exploit for Defense Evasion) to evade the defense, and uses technique T1021 (Remote Services) to move laterally. In the second through phase, the attacker uses an extrapolated technique T1590.004 (Gather Network Typology) to conduct further reconnaissance, ans uses an extrapolated technique T1210 (Exploit of Remote Services) to support its lateral movement. In the first out phase, the attacker uses an extrapolated technique T1078.001 (Valid Accounts: Default) to escalate its privileges, and uses technique EX-0012.08 (Modify On-Board: Attitude) to execute its malicious payload, and further uses technique IMP-0002 (Disruption) to temporarily impact the satellite. In the second out phase, the attacker uses an extrapolated technique T1098 (Account Manipulation) to achieve persistence, and uses technique IMP-0005 (Destruction) to enable the attacker future access to impact the RoSat's x-ray sensing device.

The preceding two example USCKCs are extrapolated from the same $n' = 9$ observed attack steps. While differing in details, they both are probable because: in the case of Figure 3.4, the extrapolated T1595 ($te_{5,0,1}$) provides the needed network data to inform T1211 ($te_{5,1}$), while the extrapolated technique T1543 ($te_{9,0,1}$) provides the needed persistent access to enable IMP-0005 ($te_{9,1}$); in the case of Figure 3.5, T1590 ($te_{5,0,2}$) provides the needed network data to inform T1211 ($te_{5,1}$), while T1098 ($te_{9,0,2}$) provides the needed persistent access to enable IMP-0005 ($te_{9,1}$).

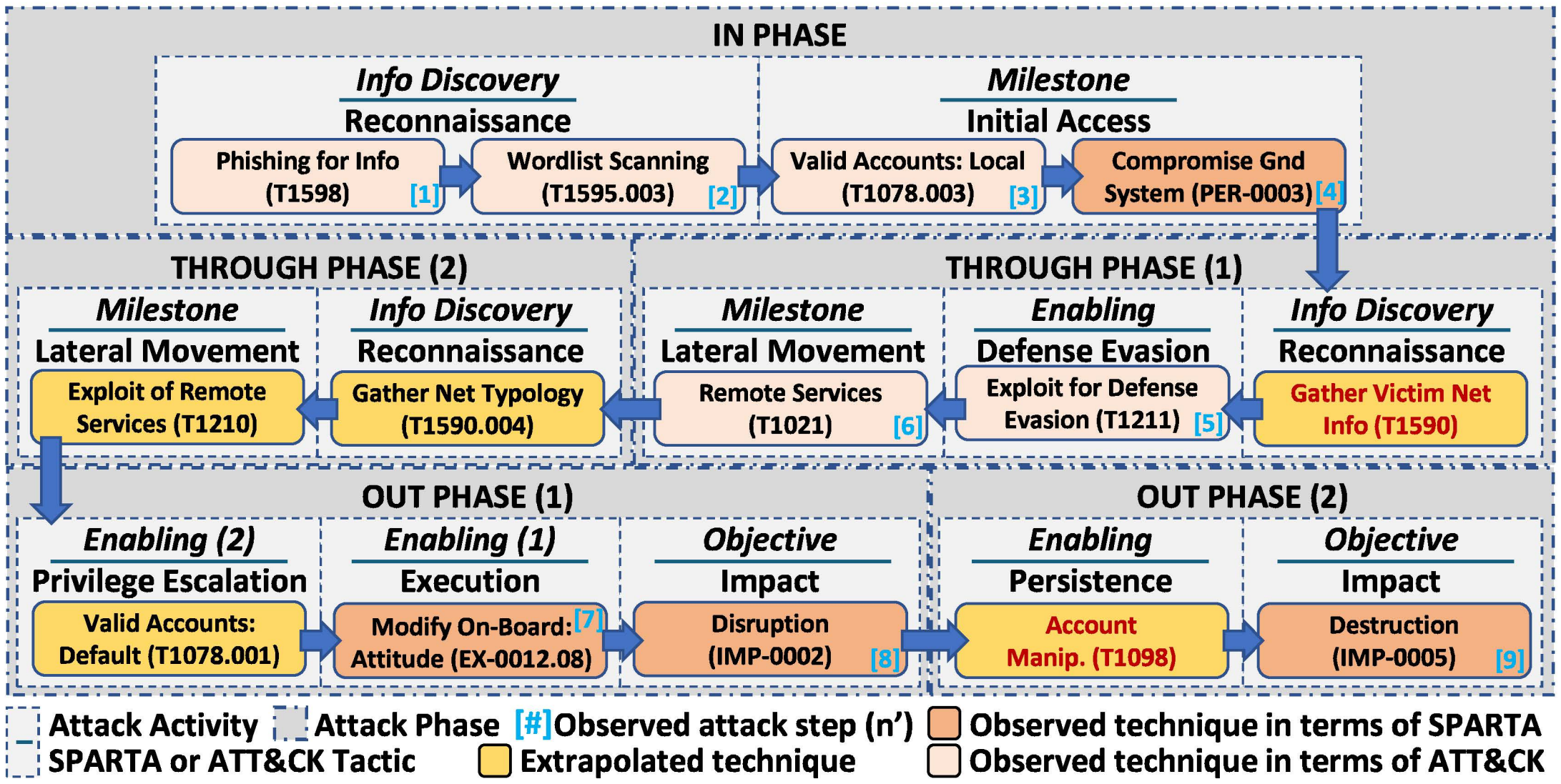

Figure 3.5: Another example USCKC (out of the 432 hypothetical but plausible USCKCs) we extrapolated for the RoSat 1998 attack.

In total, we extrapolate 6,206 USCKCs from the 108 attacks. We call this the USCKC *dataset*, where one row per USCKC. We will make this dataset publicly available.

### 3.4.2 Leveraging Metrics to Answer Research Questions

#### 3.4.2.1 Attack Consequence Analysis

We leverage Definition III.1, our space infrastructure system model, our raw cyber attack dataset, and our extrapolated USCKC dataset to holistically consider the attack consequence via its manifestation in the space, ground, user, and link segments. For each attack, we leverage our domain expertise to assess the attack consequence in each segment, by assigning a score between 0 and 1 from least to most consequential. More specifically, we assign a score smaller than or equal to 0.3 when the attack impact is superficial (e.g., it is instantly recoverable); we assign a score between 0.3 and 0.8 when the attack impact is temporary (i.e., recoverable); we assign a score equal to or greater than 0.8 when the attack impact is non-recoverable.

Figure 3.6 shows that attack consequence for the space segment is generally low, where less than 18% of the attacks (i.e., 19 out of the 108) have a consequence score of 0.7 or higher: one Signal Hijacking, 12 Jamming, three DoS, and three Seizure of Control. We observe Seizure of Control attacks achieve the highest consequence against the space segment, where two attacks score 0.7 and one attack scores 1.0 (i.e., the 1998 RoSat incident scored the highest as it demonstrated the ability to physically destroy an asset in the space segment via cyber means). We observe Jamming is the most common attack type to produce a consequence in the space segment.

Figure 3.6 also shows that attack consequence for the link segment has been consistent over the years. We observe that 61% of the attacks (i.e., 66 out of the 108) have a consequence score of 0.3 or higher, including one DoS attack, 41 Jamming attacks, 15 Signal Hijacking attacks, and nine Spoofing attacks. The 41 Jamming attacks have a high attack consequence score (0.7) because they degrade the conditions (e.g., the noise level) of the channel employed by the legitimate signal, preventing the legitimate user and ground reception modules from receiving a proper signal. We also observe that the Signal Hijacking and Spoofing attacks achieve lower consequence scores (i.e., 0.3 for Signal Hijacking and 0.4 for Spoofing) because these attacks typically leverage the signal to produce a consequence on the other three segments than the link segment. The attacker in a Signal Hijacking or Spoofing attack cannot tolerate the downgrade of channel conditions, thus the consequence on the link segment is lower than what is incurred by Jamming attacks.

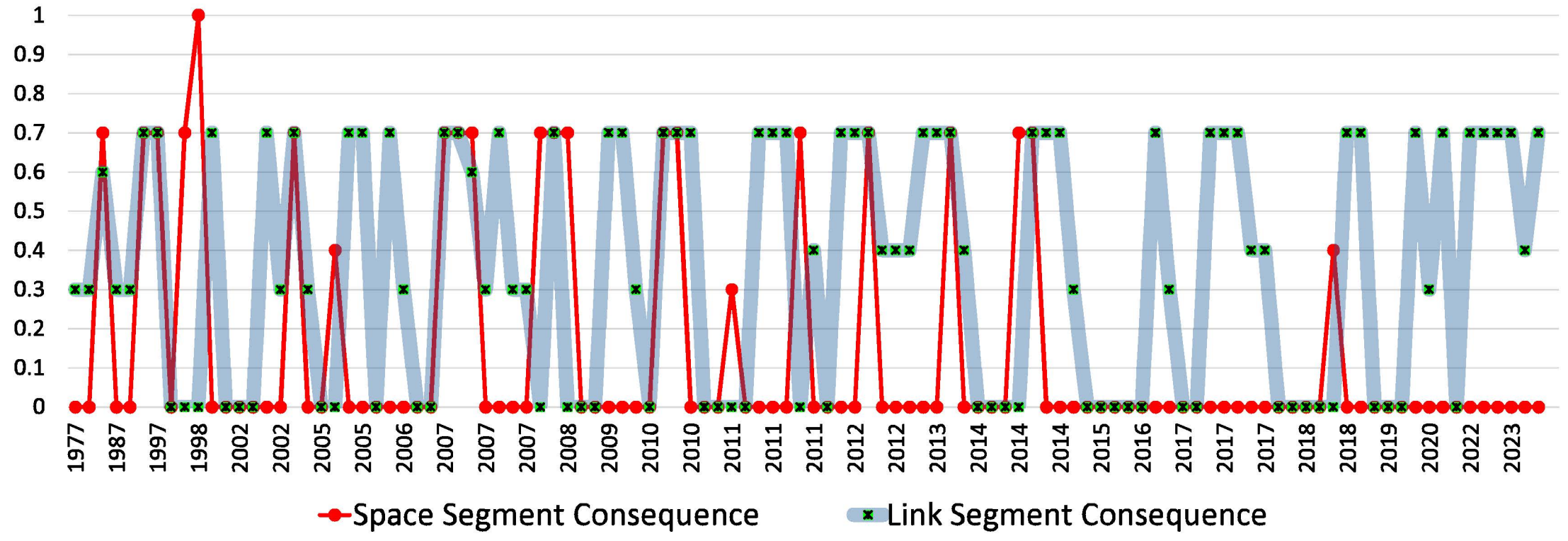

Figure 3.6: Attack consequence in the space and link segments incurred by the 108 attacks over time.

Figure 3.7 shows that 58% of the attacks (i.e. 63 out of the 108) achieve a consequence of 0.4 or higher against the ground segment, including 26 Data Corruption/Interception attacks, nine DoS attacks, two Eavesdropping attacks, 11 Jamming attack, three Seizure of Control attacks, 11 Signal Hijacking attacks, and one Spoofing attack, while noting that the remaining 45 (out of the 108) attacks have no consequences in the ground segment. We also observe that 11 (out of the 26) Data Corruption/Interception attacks achieve the highest consequence against the ground segment, with a consequence score of 0.8. Figure 3.7 also shows that 49% of the attacks (i.e. 53 out of the 108) have a consequence of 0.4 or higher against the user segment, including four DoS attacks, one Eavesdropping attack, 36 Jamming attack, four Signal Hijacking attacks, and eight Spoofing attacks, while the remaining 55 (out of the 108) attacks have no consequences against the user segment. We also observe that four (out of the eight) Spoofing attacks achieve the hightest consequence against the ground segment, including one Spoofing attack achieving a consequence of 1.0 (i.e., the University of Texas Radionavigation Laboratory demonstrated in 2012 the ability to make a rotorcraft dive to the ground by spoofing GPS signals).

**How can we prioritize hardening to mitigate space cyber attacks?** We observe that 54% (34 out of 63) of the attacks that have consequence in the ground segment also leverage entry nodes from the ground segment into the space infrastructure; the 34 attacks include 24 Data Corruption/Interception attacks, seven DoS attacks, and three Seizure of Control attacks. Further, the extrapolated USCKC dataset reveals victims exhibiting potentially weak security hygiene (e.g., default credentials) and unpatched public facing application vulnerabilities. This leads to the following insight:

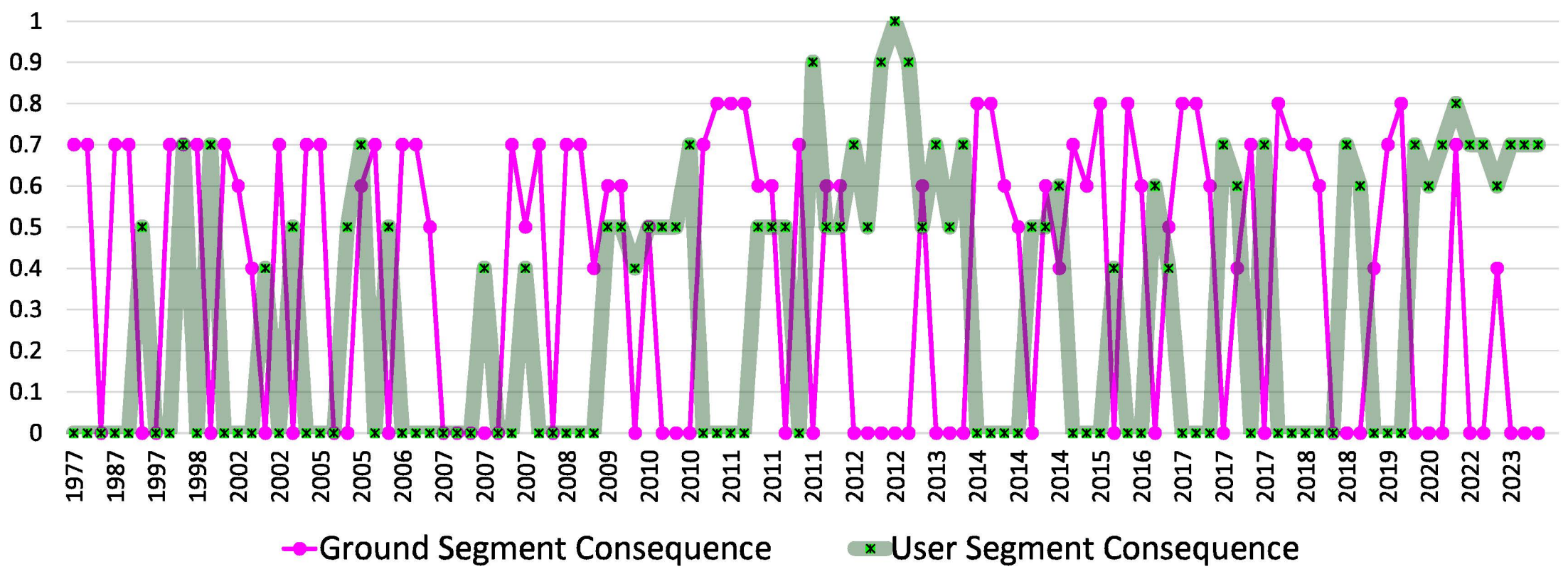

Figure 3.7: Attack consequence in the ground and user segments incurred by the 108 attacks over time.

**Insight 2.** *Attack consequences in the ground segment can be mitigated by prioritizing hardening measures in the ground segment (to prevent attackers from pivoting from the ground segment to the space segment).*

#### 3.4.2.2 Attack Sophistication Analysis

We use the sophistication metric (Definition III.2), namely $(\alpha_{\mathrm{TA}_+}, \alpha_{\mathrm{TE}_+})$, to quantify the *possible highest sophistication* for each attack according to its extrapolated attack tactic chain TA and its associated attack technique chains TE's, while taking as input the measurement of the sophistication score of each attack tactic and each attack technique, which are currently assigned based on our domain expertise and depicted in Figure 3.8. Our results are as follows.

From the 108 attacks, we identify 14 attack tactics and 107 attack techniques in total. For the 14 attack tactics, we manually score each from 0 to 1, while considering 0.5 as the average sophistication required to accomplish an attack tactic. For example, we score the *Initial Access* tactic as 0.5 because majority of the 108 attacks should gain initial access to its target. We assign a score of 0.8 or higher for the tactics requiring a high sophistication. For example, the *Defense Evasion* tactic is scored at 0.9 as it requires additional effort and more advanced capabilities. For the 107 attack techniques, we assign sophistication scores while bearing in mind whether the score should be less than, the same as, or higher than, that of the associated attack tactic. Phishing is commonplace and may require little technical capabilities when compared to the other Initial Access techniques, and hence receives a score of 0.3. We then compute attack sophistication according to Definition III.2.

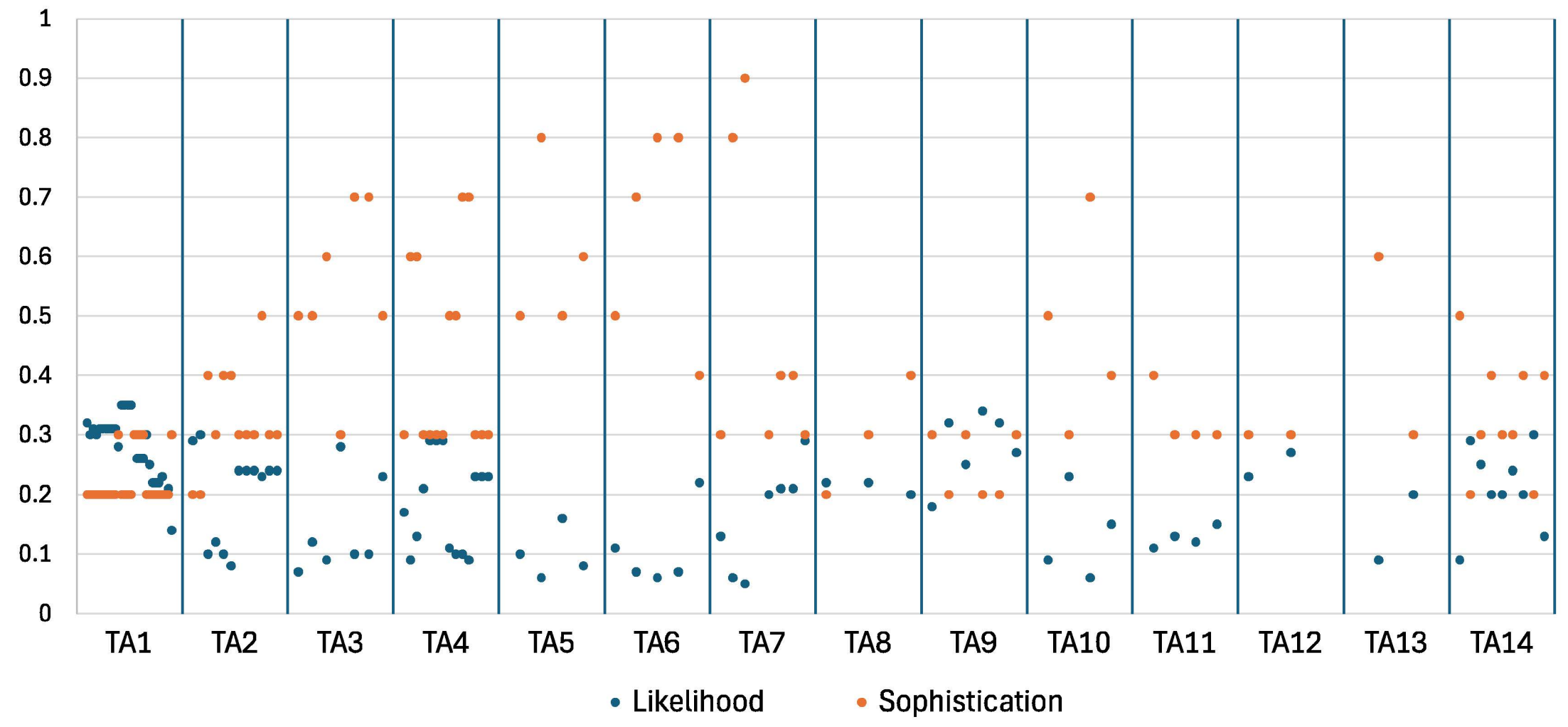

Figure 3.8: Sophistication and likelihood measurements for the 107 attack techniques. TA1 = Reconnaissance; TA2 = Resource Development; TA3 = Initial Access; TA4 = Execution; TA5 = Persistence; TA6 = Privilege Escalation; TA7 = Defense Evasion; TA8 = Credential Access; TA9 = Discovery; TA10 = Lateral Movement; TA11 = Collection; TA12 = Command and Control; TA13 = Exfiltration; TA14 = Impact.

Figure 3.9 depicts the possible highest sophistication of the 108 attacks via $\alpha_{\mathrm{TA}_+}$, while noting that several plots overlap and hide other plots from view. We observe that Signal Hijacking attacks consistently score 0.5, with one exception. In 2007, the Russian Turla Hacking Group's employment of the C2 tactic (ATT&CK ID TA0011) by leveraging SATCOM connections scores 0.8, which we consider as a high sophistication. Approximately 70% of the Jamming attacks score 0.5, and the remainder of Jamming attacks score 0.9 due to the inclusion of the Defense Evasion Tactic (ATT&CK ID TA0005). Each Seizure of Control attack attains the high sophistication score of 0.9. Each of these attacks successfully employs the Defense Evasion tactic (ATT&CK ID TA0005) to overcome ground control station defenses. Majority (19 out of the 26) of Data Corruption/Interception attacks score above 0.5, with one incident in 2011 scores 0.9 because of its employment of both Defense Evasion and Persistence tactics to enable a series of 46 subsequent attacks against the ground segment. DoS attacks consistently score 0.5 except for three attacks that employ the Lateral Movement (ATT&CK ID TA0008), Persistence (ATT&CK ID TA0003), and Defense Evasion (ATT&CK ID TA0005) tactics. All High-powered Laser attacks in our dataset score 0.7 and all Eavesdropping and Spoofing attacks in our dataset score 0.5. Overall, space cyber attacks are getting more sophisticated over time.

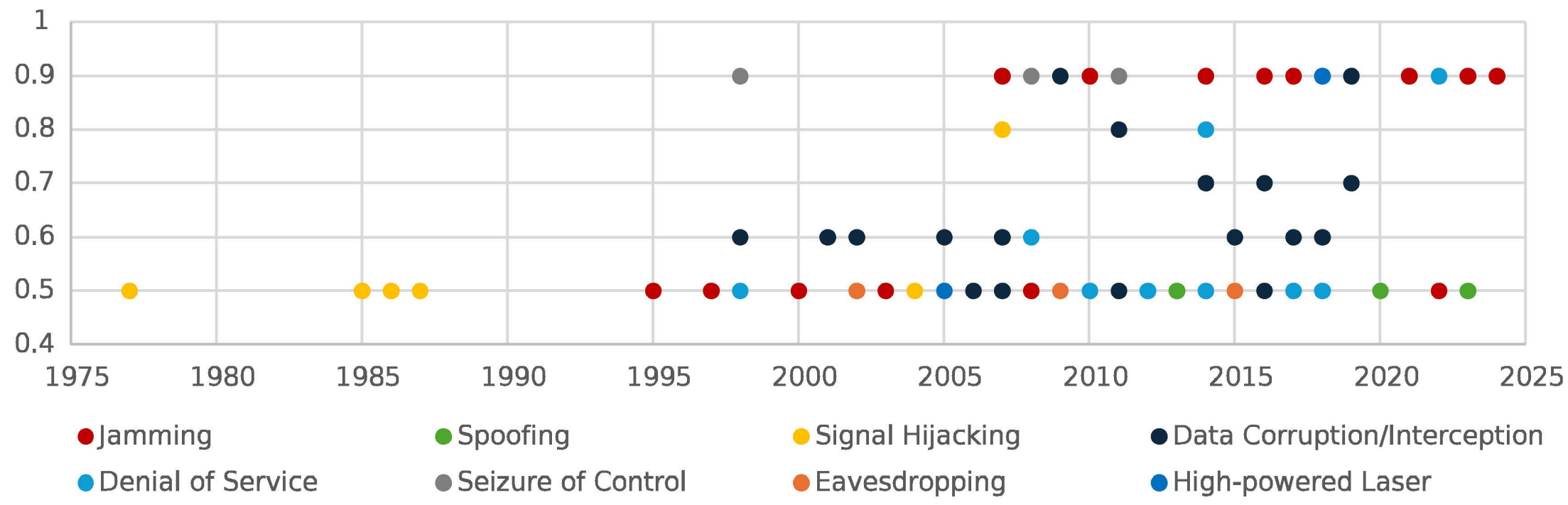

Figure 3.9: Attack tactic sophistication of the 108 attacks over time.

Figure 3.10 depicts the possible highest sophistication of each attack via $\alpha_{\mathrm{TE}_+}$. We observe that $\alpha_{\mathrm{TE}_+}$ and $\alpha_{\mathrm{TA}_+}$ identify the same set of highly sophisticated space cyber attacks. The notable attack techniques employed by these attacks to support the Defense Evasion (T1211) tactic are: Indicator Removal (T1070) and Exploit (i.e., of a vulnerability). The attacks that employ the Persistence tactic also employ the following attack techniques: Event Triggered Execution (T1546), Create or Modify System Process (T1543), and Exploit Hardware/Firmware Corruption (EX-0005).

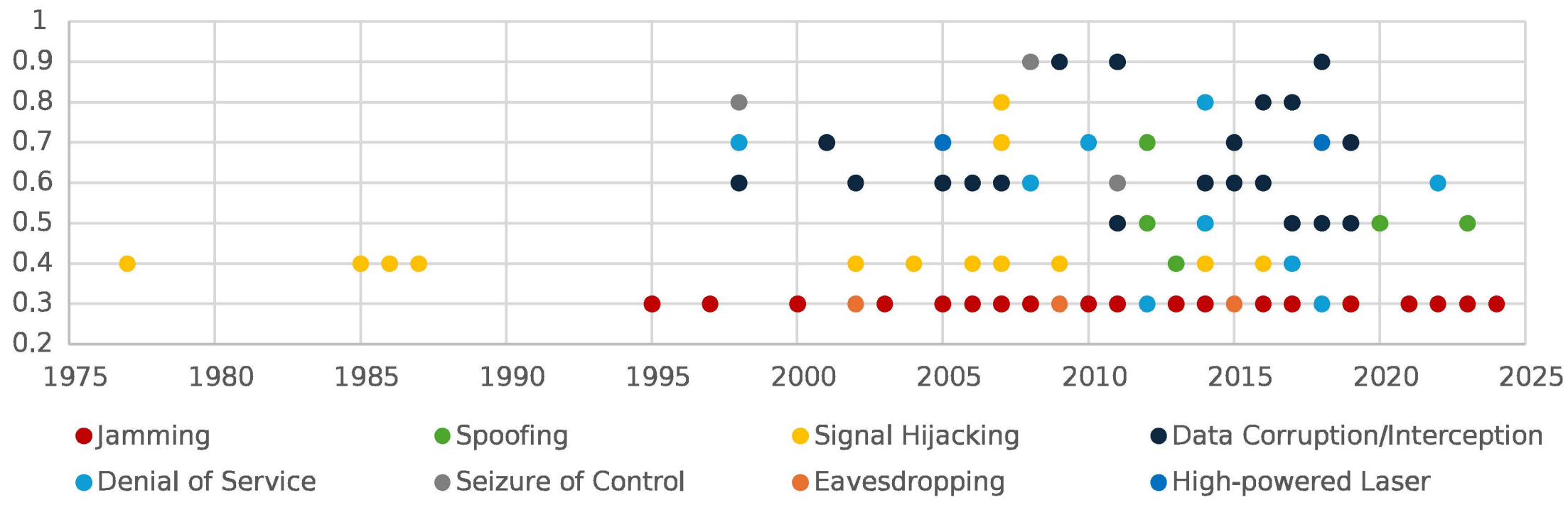

Figure 3.10: Attack technique sophistication of the 108 attacks over time.

By comparing $\alpha_{\text{TA}_+}$ and $\alpha_{\text{TE}_+}$, we observe that these two metrics generally follow the same trend per incident, but $\alpha_{\text{TE}_+}$ scores are slightly more dispersed than $\alpha_{\text{TA}_+}$. This is reasonable because attack tactics are at one level of abstraction higher than attack techniques. Hence, $\alpha_{\text{TE}_+}$ will exhibit greater sensitivity in measurement. Nevertheless, 55% (60 out of the 108) attacks have both $\alpha_{\text{TA}_+}$ and $\alpha_{\text{TE}_+}$ scores of 0.5 or less. This leads to the following insight:

**Insight 3.** *Space cyber attacks of average sophistication can be successful (i.e., space cyber defenses have yet to eliminate the "low hanging fruits" that benefit attackers).*

**How many attacks would have been stopped by using simple countermeasures?** We observe that 44% (48 out of 108) of attacks employ techniques to establish an adversary-in-the-middle attack position, and that the $\alpha_{\text{TE}_+}$ scores for these attacks are 0.4 or less. Although these attacks typically target the user segment, they leverage the weakness of the link segment. This leads to the insight:

**Insight 4.** *Proper security of the link segment between the space and user segments (e.g., using cryptography) could have thwarted nearly half of the 108 space cyber attacks.*

However, this insight does not imply the ease of employing cryptosystems for space infrastructures due to various barriers that need to be overcome (e.g., low compute capabilities onboard satellites, incorporation of legacy systems). Yet, it does point to the urgency to overcome such barriers.

We observe that 21% (23 out of 108) of attacks employ cyber social engineering techniques; among the 23 attacks, 20 have consequences in the ground segment where $\vec{g}_{\mathrm{G}}$ scores range from 0.4 to 0.7, while their $\alpha_{\mathrm{TE}_+}$ scores range from 0.4 to 0.9. The average sophistication score of all the social engineering attacks is 0.3, meaning that mitigations against unsophisticated cyber social engineering attacks can thwart even sophisticated attacks, as well as attacks with significant consequence. However, this observation is made at the attack technique level of abstraction, while noting that sophistication of cyber social engineering attack techniques is highly nuanced at the procedure level [85]. This leads to the following insight:

**Insight 5.** *Traditional IT security controls against cyber social engineering attacks could have thwarted 32% (20/63) of the cyber attacks that compromise the ground segment.*

#### 3.4.2.3 Attack Likelihood Analysis

As mentioned above, the 108 space cyber attacks collectively employ 14 attack tactics and 107 attack techniques. We compute the likelihood that a $\mathsf{USCKC}$ corresponds to the ground-truth of the corresponding incident, denoted by $L(\mathsf{USCKC}) \in (0, 1]$, as follows. First, we apply our domain knowledge to attain measurements for the likelihood of each attack technique involved in the $\{\mathsf{USCKC}\}$ resulting from Algorithm 2. For this purpose, and as depicted in Figure 3.8, we assign each of the 107 attacks techniques a likelihood score from 0 to 1, while considering 0.2 the average likelihood because there are many obstacles attackers must overcome, such as fulfilling the attack technique's data and access requirements. We apply the measurement to each $te \in$ TE corresponding to a specific $\mathsf{USCKC}$ (i.e., one entry in the dataset), and use Definition III.3 to compute the likelihood each $\mathsf{USCKC} \in \{\mathsf{USCKC}\}$, namely $L(\mathsf{USCKC})$. Then, we compute the overall likelihood of all $\mathsf{USCKC}$s for each attack, leading to $L(\{\mathsf{USCKC}\})$. For example, a $\mathsf{USCKC}$ for a Data Corruption/Interception attack against a NASA network in the ground segment in 2019 contains the following attack techniques (with accompanying likelihood scores): Valid Accounts (0.22), Exploitation of Remote Services (0.09), Indicator Removal (0.05), and Resource Hijacking (0.25). The likelihood of the $\mathsf{USCKC}$ is bounded by the least likely technique, leading to $L(\mathsf{USCKC}) = 0.05$. Given that $L(\mathsf{USCKC}) = 0.05$ is the highest likelihood score among all the probable and plausible $\mathsf{USCKC}$s of this attack, we have $L(\{\mathsf{USCKC}\}) = 0.05$.

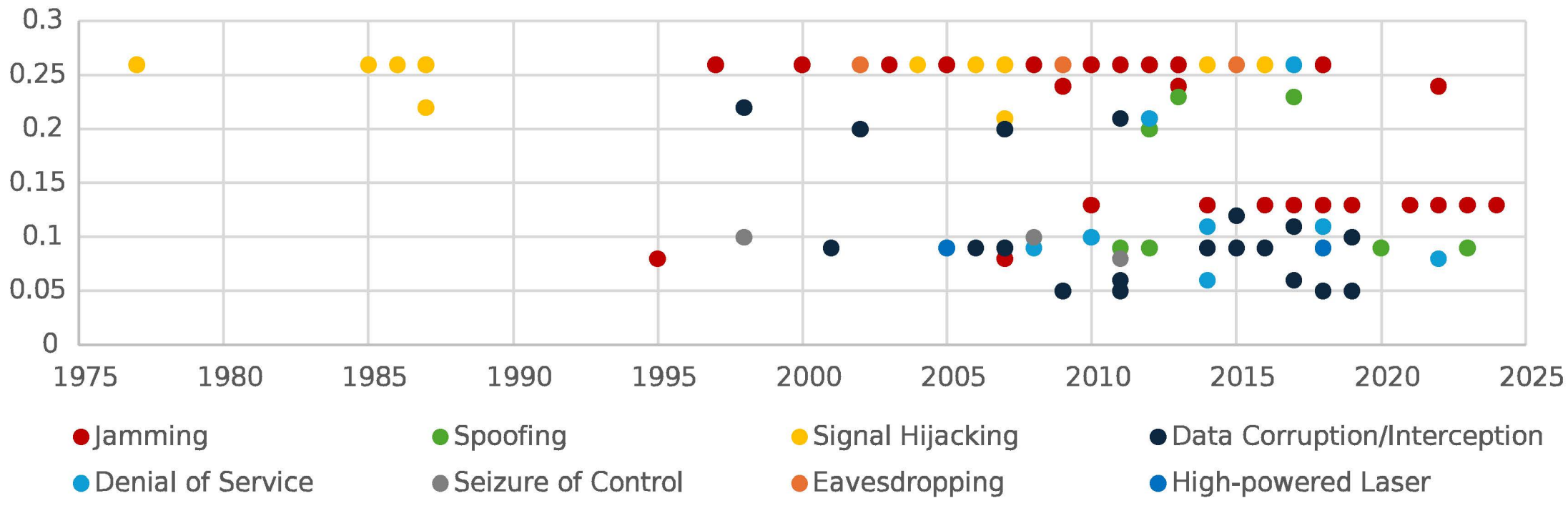

Figure 3.11: USCKC likelihood of the 108 attacks over time.

Figure 3.11 depicts $L(\{\text{USCKC}\})$ of the 108 space cyber attacks. When considering $L(\{\text{USCKC}\})$ in relation to attack sophistication $\alpha_{\text{TE}_+}$, we observe attacks that have a higher $L(\{\text{USCKC}\})$ often have a lower $\alpha_{\text{TE}_+}$. For example, out of the 57 attacks where $L(\{\text{USCKC}\}) \geq 0.2$, 90% (51 out of 57) of the attacks have $\alpha_{\text{TE}_+} \leq 0.5$. This leads to the following insight:

**Insight 6.** *Less sophisticated unified space cyber kill chains are more widely used by real-world attackers.*

Insight 6 is intuitive in the sense that attackers often seek the "low hanging fruit." For example, in the case of the 2014 Data Corruption/Interception attack against JPL where malware was used to compromise a data processing server, the attacker accomplishes the Privilege Escalation tactic, for which we extrapolate three ATT&CK techniques: T1611 (Escape to Host), T1631 (Process Injection), and T1078 (Valid Accounts). Given that the attacker has the capability to accomplish any of the three techniques, the most readily executable technique is T1078 (the "low hanging fruit"), which has the highest $L(te) = .22$ among the 3, while also being among the least sophisticated from the 107 $te$'s that are used by the 108 attacks. Our analysis is corroborated by [172], which shows that Advanced Persistent Threats (APT) in real-world cyber attacks tend to employ less sophisticated attack techniques (e.g., APT28 employs T1078 all six times in the 6-step attack against the Democratic Congressional Campaign Committee).

## 3.5 Chapter Summary

We presented an initial study on characterizing cyber attacks against space infrastructures, which is extracted from four public datasets and other sources across the Internet documenting broader space-related incidents. We proposed an innovative framework with precisely defined metrics, while addressing the missing-data problem with conceptual algorithms because these real-world attacks are poorly documented. We prepared the first dataset of space cyber attacks that include hypothetical but plausible attack details. By applying the framework to the dataset, we drew a number of insights.

# CHAPTER IV

## ANALYZING A REAL-WORLD SPACE CYBER RISK ANALYSIS AND MITIGATION SYSTEM

**Chapter Abstract.** To help space cybersecurity practitioners better manage cyber risks, The Aerospace Corporation proposed a space cyber risk analysis and mitigation approach dubbed Notional Risk Scores (NRS) within their Space Attack Research and Tactic Analysis (SPARTA) framework, with the intent to help quantify the cyber risks associated with space infrastructures and systems. While intended for adoption by practitioners, NRS has not been analyzed with real-world scenarios, putting its effectiveness into question. In this chapter we present an algorithmic description of NRS and characterize its use and effectiveness in a case study of space-related cyber incidents and its strengths, weaknesses, and applicability via 2 real-world cyber attack scenarios against space infrastructures and systems.

## 4.1 Chapter Introduction

Real-world cyber attacks against space infrastructures and systems have been reported for over four decades [27, 32, 110]. However, space cybersecurity practitioners still lack tools to effectively understand and manage cyber risks associated with space infrastructures and systems, or *space cyber risks* in short. In a major effort to support space cybersecurity practitioners, The Aerospace Corporation developed and incorporated space cyber *Notional Risk Scores* (NRS) [139] into their Space Attack Research and Tactic Analysis (SPARTA) framework [141], by associating a notional evaluation of cyber risks to attack techniques. NRS and SPARTA are founded on a wealth of industry research and precedent, including traditional enterprise Information Technology (IT) cybersecurity (e.g., security controls [40–42]).

The intention of NRS is to provide practitioners with a starting point for space cyber risk analysis and mitigation, whereby they can tailor NRS to meet their specific space cyber risk analysis and mitigation needs. However, the precise methodology for practitioners to apply NRS is scant. Further, NRS has not been analyzed with real-world scenarios, putting its effectiveness into question. This motivates us to characterize its strengths, weaknesses, and applicability.

**Chapter Contributions**. In this chapter we make three contributions. First, we present an algorithmic description of applying NRS as a space cyber risk analysis and mitigation tool. This is both important and useful because the current description of NRS [139] does not thoroughly explain the required details for practitioners to correctly understand, adopt, and apply NRS. Second, we present a case study of two real-world cyber attack scenarios that demonstrate the use and validate the effectiveness of NRS, where we employ our algorithmic description. This serves as examples for practitioners to follow when adopting or adapting NRS to meet their real-world space cyber risk analysis and mitigation requirements. Third, we objectively characterize the strengths, weaknesses, and applicability of NRS by applying it to 2 real-world cyber attacks against satellites and their associated systems. This provides further insights into the use of NRS in practice and for future improvements to NRS.

**Chapter Organization**. Section 4.2 describes NRS. Section 4.3 leverages a case study to characterize NRS. Section 4.4 concludes the chapter.

## 4.2 A Description of NRS

**Background**. The MITRE ATT&CK framework [138] is geared towards terrestrial networks. To establish a similar framework for space infrastructures and systems, The Aerospace Corporation proposed the SPARTA framework [141]. The two frameworks are complementary to each other in terms of their application settings. In August 2023, The Aerospace Corporation further proposed NRS [139] in an update to its SPARTA framework. NRS is based on [2, 3], which leverages NIST SP 800-53 [57] to create a threat-based approach to space cyber risk analysis and mitigation. SPARTA and NRS have been adopted by some space cyber practitioners.

### 4.2.1 Space Infrastructure System Model and Terminology

While already discussed in Chapter III, the following serves as a summary of terms and concepts used in this chapter.

**Space infrastructure system model**. To avoid ambiguity, we need a clearly defined space infrastructure system model, which is currently not offered by SPARTA nor by NRS. This prompts us to adopt the space system model recently presented by researchers in [27] because it is the most comprehensive space infrastructures and systems model we find from the literature. The model is depicted in Figure 4.1. At a high level, space infrastructures contain four segments, namely *space*, *ground*, *user*, and *link*; each segment contains one or multiple components (e.g., the space segment consists of two components: *bus system* and *payload*); each component contains multiple modules (e.g., the bus system component contains six modules). We refer to Chapter III for detailed descriptions of segments, components, and modules.

**Terminology**. For describing cyber attacks against, and defenses for, space infrastructures and systems with respect to the system model described above, we adopt the following terminology used by SPARTA and NRS [4, 5, 138]. *Tactics* are cyber threat actors' goals or objectives. *Techniques* are the actions cyber threat actors take to accomplish their tactics. *Sub-techniques* are more specific instances of a parent technique. *Procedures* are the step-by-step implementations of (sub-)techniques. *Attack chains* are ordered sets of SPARTA techniques employed by attackers to accomplish their attack objectives. *Countermeasures* are protective measures, which are instantiated as *secu-*

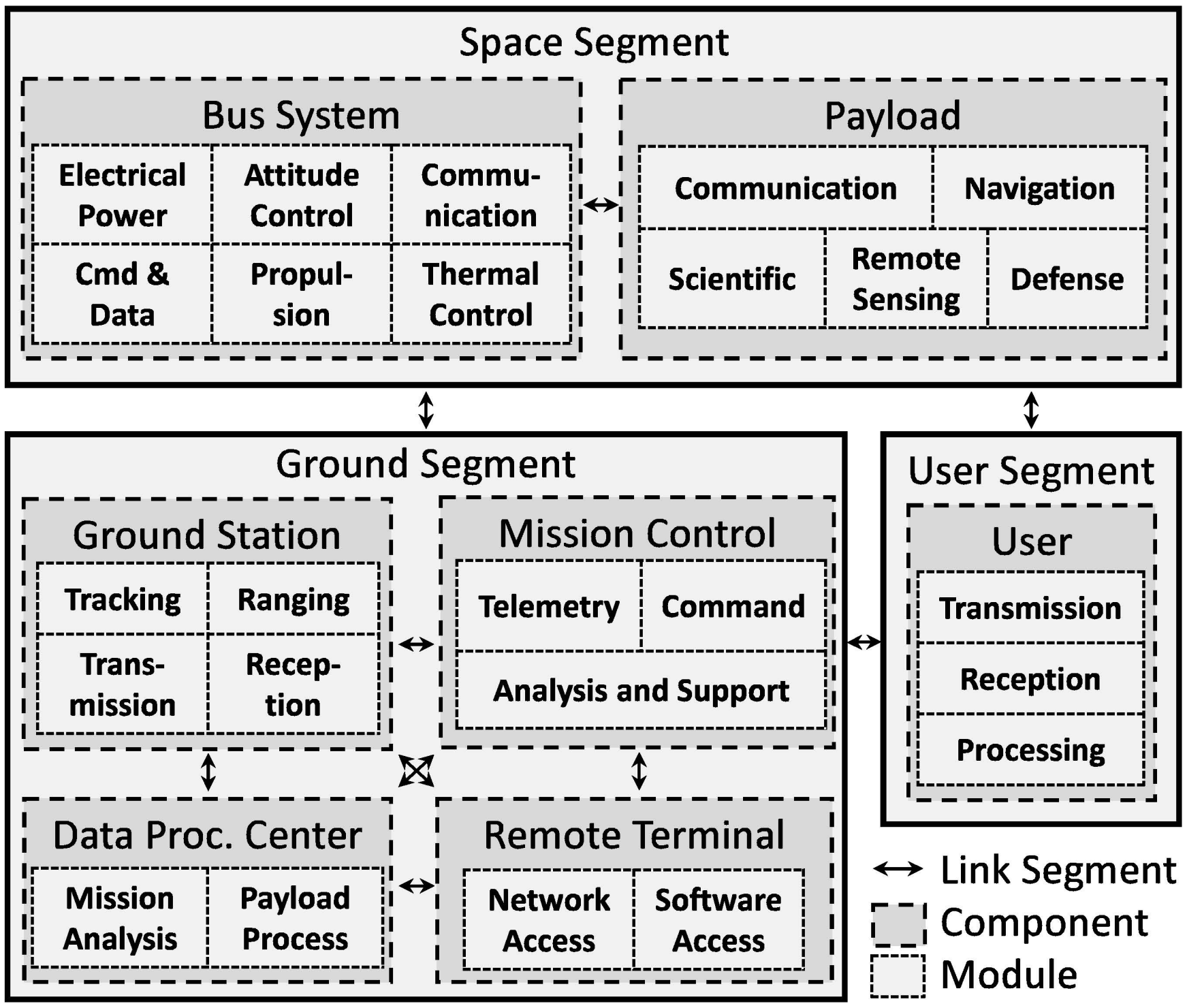

Figure 4.1: Space infrastructure system model.

*rity controls*, namely safeguards for protecting systems and data. NIST [41] provides 20 families of security controls.

To characterize NRS, we introduce the following terms. *Attack flows* are ordered sets of space system components/modules that adversaries must compromise in order to achieve their objectives. Note that *attack flows* are related to, but different from, *attack chains*. *Mission control flows* are ordered sets of space infrastructure units required to send commands that accomplish a mission. Note that the word "control" in *mission control flow* has nothing to do with the same word in "security control" because the former deals with functions and the later deals with security. *Mission data flows* are ordered sets of space infrastructure units required to send data items according to a mission.

### 4.2.2 Overview of the Space Cyber NRS

The SPARTA framework provides technical information about how cyber attackers may compromise spacecraft (e.g., satellites) and helps space cyber defenders understand space cybersecurity through the lenses of tactics, techniques, and procedures (TTPs). It provides nine tactics; each tactic contains between five to 18 techniques; some techniques contain between one to nine sub-techniques; and, each (sub-)technique is mapped to countermeasures and NIST SP 800-53 security controls [41].

At a high level, NRS provides, for each system criticality, base risk scores for each SPARTA technique, determined by (i) technique likelihood of successful attack and (ii) technique impact incurred, via a risk matrix. In addition, NRS specifies countermeasures and security controls for mitigating specific SPARTA techniques.

#### 4.2.2.1 System Criticality

There are three categories in system criticality: *high*, associated with systems related to critical functions, military purposes, and intelligence activities; *medium*, associated with civil, science, weather, and commercial systems; and, *low*, associated with academic and research systems [140]. Aerospace Corporation experts assign a base risk score for each SPARTA technique at each system criticality level. Practitioners subjectively determine the criticality level of each space system unit impacted by a SPARTA technique to attain the SPARTA technique's base risk score.

#### 4.2.2.2 SPARTA Technique Likelihood

The evaluation of technique likelihood includes three aspects: (i) adversary motivation, influenced by the system criticality with the assumption that adversaries are more motivated to attack *high* criticality rather than *low* criticality systems; (ii) exploitation difficulty, based on technique complexity; and, (iii) adversary capabilities, according to the following seven tiers, in increasing order: script kiddies, hackers for hire, small hacker teams, insider threats, large well-organized teams, highly capable state actors, and most capable state actors [2]. Subjective analysis on these three aspects provides the overall likelihood score which results in a range $\{1, \ldots, 5\}$. Note that NRS, like many practitioners, uses the term likelihood in an intuitive sense that is broader than probability, perhaps because it is not clear on how to precisely define a rigorous probabilistic structure for a complex real-world situation. Nevertheless, the calculations based on likelihood should not violate the laws of probability calculations (e.g., a likelihood, like a probability, should always belong to the range $[0, 1]$).

#### 4.2.2.3 SPARTA Technique Impact

The impact of a technique against a space system unit refers to the consequences, effects, or outcomes resulting from the successful execution of the technique. Subjective analysis considers wide-ranging impact that may include mission disruption, data integrity, loss of control or availability, financial consequences, safety, or even national security implications [140]. Impact is also defined in a range $\{1, \ldots, 5\}$.

#### 4.2.2.4 Risk Matrix Representation (Risk Scores)

This is a 5×5 risk matrix representation of the notional risk scores for the SPARTA techniques evaluated [38]. The matrix provides a risk score with respect to an assessed *impact* score from 1 to 5 (the $x$-axis) and a *likelihood* score from 1 to 5 (the $y$-axis); the risk scores are shown in the respective cells of the matrix and reflect the joint effect of impact and likelihood, according to the 5×5 matrix defined in [38]. Risk scores range from 1 to 25, but are *not* the product of likelihood and impact. Risk scores ranging from 1 to 10 are considered *low*, 11 to 19 considered *medium*, and 20 to 25 considered *high*.

#### 4.2.2.5 Countermeasures and Security Controls

These are integrated into SPARTA and can be employed to thwart SPARTA techniques. Each countermeasure contains space-specific protection measures, which can be instantiated by NIST SP 800-53 security controls to achieve the goal of the countermeasure. For example, SPARTA technique *Memory Compromise (PER-0001)* is mapped to eight countermeasures, meaning that an attack exploiting the technique can be thwarted by one or a combination of multiple countermeasures. For instance, the countermeasure with identifier *CM0021* specifies that the digital signature of flight software is verified prior to installation, while noting that *CM0021* is mapped to 19 security controls (e.g., security control *CM-11* specifies that organizational policies governing the installation of software must be established, enforced, and monitored).

### 4.2.3 Algorithm for Using NRS

Algorithm 3 shows how to use NRS to quantify space cyber risk and identify mitigation, which was implied in [2, 139, 140] but the algorithm is given here for the first time. Line 1 *subjectively* determines the SPARTA techniques that can incur risk to the space infrastructure/system in question. Lines 2-9 assess each applicable technique, where: Lines 3-5 *subjectively* generate the tailored risk score associated with each applicable SPARTA technique, by determining the impact and likelihood of each technique according to the specific environment/conditions of the space infrastructure/system and mapping it to the $5 \times 5$ risk matrix; and, Lines 6-9 determine if a SPARTA technique is tolerable, where if not, *subjectively* select countermeasures and security controls to mitigate the *intolerable* SPARTA techniques.

## 4.3 Characterizing NRS

### 4.3.1 Case Study Based on Real-World Attacks

Data for the following real-world cyber attack scenarios are obtained from our previous study [27]. In the first scenario, we show that if NRS were used as a space cyber risk analysis and mitigation tool when designing the Terra satellite and its mission, the real-world cyber attack against the Terra satellite could have been thwarted. In the second scenario, we show how NRS could have been used as an effective space cyber risk analysis and mitigation tool to assess and respond to threats that include the one imposed by the Turla Hacking Group against satellite communications.

**Algorithm 3:** Using NRS for risk analysis and mitigation

**Input:** SPARTA matrix of techniques; NRS base risk scores $R$, which is a set indexed by technique with element $R_A$ being the basic risk score of technique $A$; a set indexed by SPARTA technique with element $C_A$ being a set of countermeasures to technique $A$; a set indexed by countermeasure, with element $S_{c_A}$ being a set of secure controls fulfilling countermeasure $c_A \in C_A$; specific environment/conditions of the space system; tolerable risk threshold $\tau \in \{`low\text{'}, `medium\text{'}, `high\text{'}\}$

**Output:** a set of security controls that must be employed to mitigate intolerable risks

```
subjectively determine the set 𝒜 of SPARTA techniques applicable to the space system in question
for each technique A ∈ 𝒜 do
    identify its base risk score R_A according to R
    subjectively tailor the base risk score R_A to r_A = (impact, likelihood) to reflect the specific environment/conditions of the space system
    populate r_A on the 5×5 risk matrix
    if the risk incurred by A > τ then
        subjectively select a set of countermeasures to mitigate A, denoted by {c_A} ⊆ C_A
        subjectively select a set of security controls to fulfill the countermeasure, denoted by S_{c_A}
return ⋃_{A∈𝒜} S_{c_A}
```

#### 4.3.1.1 Case Study Scenario 1: Cyber Attack against the Terra Satellite

In 2008, a cyber attack successfully compromised a NASA-managed remote sensing space system, where the attack objective was in the space segment, namely the Terra satellite [43]—part of an earth observation program for terrestrial climate research in low earth orbit [101]. The attack exploited an undisclosed vulnerability in the commercially operated ground station, allowing the attacker to command and control the Terra satellite. A report [76] indicated that the entry point of the attack was at the Svalbard Satellite Station in Spitsbergen, Norway through an Internet connection regularly used to transfer data. However, the operator of the ground station, Kongsberg

Satellite Services, found no indicators of compromise at the ground station [153]. The effects of the cyber attack were observed twice: satellite operators experienced interference for more than two minutes in June, and then more than nine minutes in October [29]. The investigators concluded the attacker gained control over the satellite, but chose not to conduct further exploitation [61]. The attack can be described by two attack flows: one is to gain control of the satellite bus system via the network access module of the remote terminal component of the ground segment, and the other is to conduct a denial-of-service attack against the reception module of the ground station.

Figure 4.2 illustrates the mission control and data flows of Terra, the two attack flows derived from the attack, one hypothesized attack flow, and the attack techniques used in the attacks, overlaid on top of the space system model. Specifically, (i) one set of mission control/data flows, which are highlighted with light blue and blue colors, depict the modules and connections required for Terra to accomplish its mission. The flight mission control flow is required to fly and position the satellite appropriately; and, the mission data flow is required to acquire and transmit sensor data to the users of the satellite. (ii) Three sets of attack flows, which are highlighted with dark red, orange, and red colors, depict the modules and connections required to successfully execute three possible attacks against Terra. The attack flow against the payload is not derived from the attack, but considered for illustrating that other threats can be relevant (i.e., attack modifying sensor data at the payload processing module of the data processing center component). The case study proceeds according to Algorithm 3 as follows.

Corresponding to Line 1 of Algorithm 3, we suppose the analyst subjectively applies cyber threat intelligence (CTI), to identify three possible attack objectives: (i) gaining control of the *bus system* component via the attack chain of SPARTA techniques *IA-0007* and *EX-0012.10* and ATT&CK technique *T1133*; (ii) conducting a downlink disruption against the ground station via the attack chain of a single SPARTA technique *EX-0013*; and, (iii) compromising the integrity of the payload sensor data at the *user* component via the attack chain of a single ATT&CK technique *T1586*.

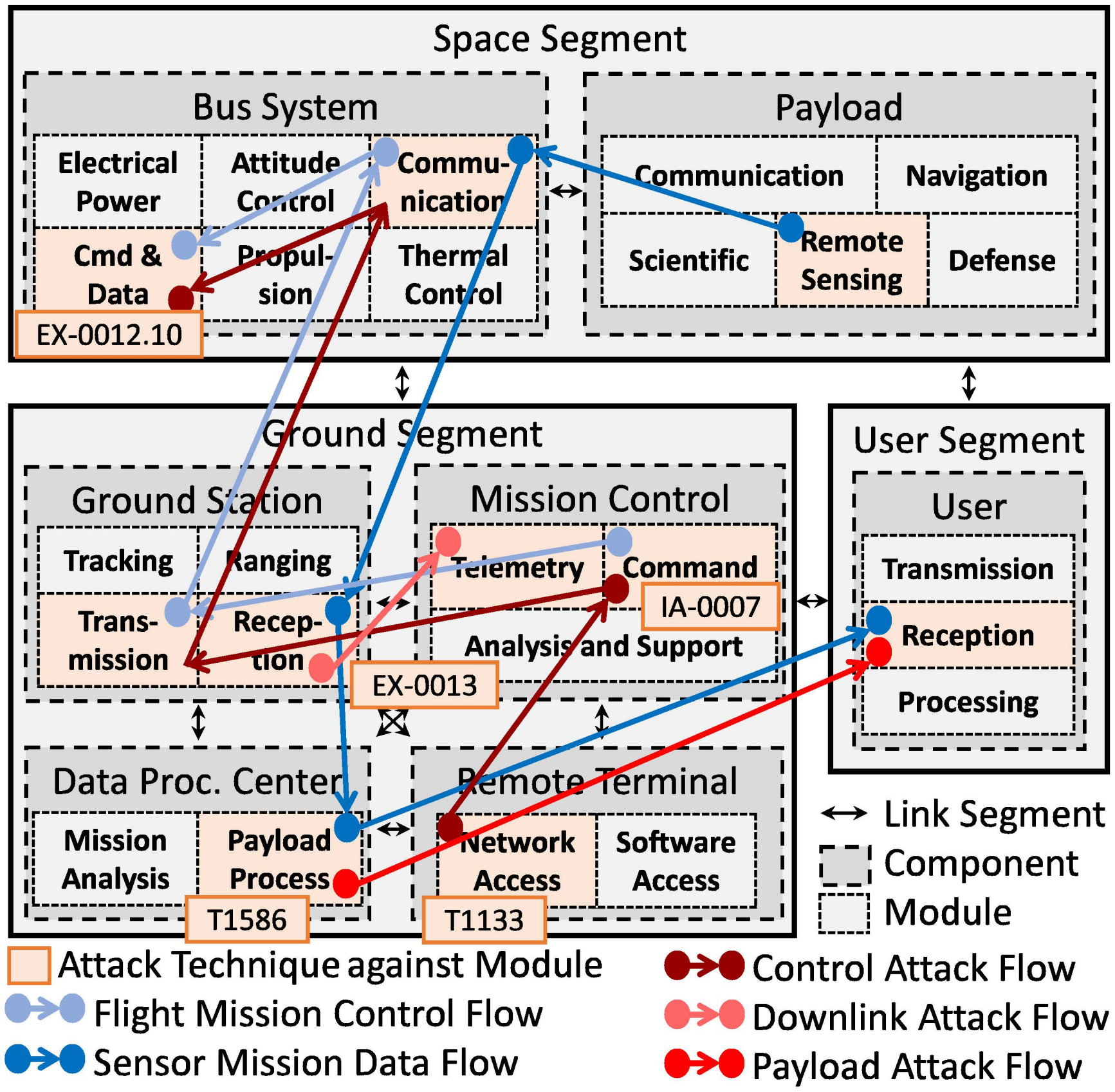

Figure 4.2: System model representing Terra overlaid with mission control/data flows, attack flows, and attack techniques.

Corresponding to Lines 2-5, the analyst needs to determine the risk scores of attacks mentioned above. Figure 4.2 shows how the attack techniques of the attack chains are mapped to the modules that they directly impact, i.e., *EX-0012.10*, *EX-0013*, *IA-0007*, *T1586*, and *T1133* impacting the *command and data*, *reception* (at the ground station), *command*, *payload processing*, and *network access* modules, respectively. Then, the analyst needs to determine the criticality of these affected modules. Suppose the analyst subjectively assesses the criticality of all modules required to accomplish the Terra mission as *high*, meaning that the modules affected by the attacks are all marked as *high* to the mission. This allows the analyst to attain base scores from NRS for each technique (Line 3). Suppose the analyst determines the attacker as Tier VI (i.e., highly capable nation state actor) as the adversary capability (e.g., based on the nature of the mission and/or CTI). To tailor the risk scores (Line 4) correspondingly, the analyst increases (i) the *likelihood*, say increasing *IA-0007* from a likelihood of 4 to 5; and, (ii) the *impact* by assigning *T1133* an *impact* of 5, because T1133 is not assigned a base score in NRS and the network access module permits attackers to easily pivot across the network. The 5×5 matrix is used to determine risk scores from the likelihood and impact values for each technique (Line 5). Figure 4.3 depicts the 5×5 matrix where four SPARTA and ATT&CK techniques (*EX-0013*, *IA-0007*, *EX-0012.10*, and *T1133*) are in the *high* risk region, and one ATT&CK technique (*T1586*) is in the *medium* risk region.

Corresponding to Lines 6-9, suppose the analyst decides that a *medium* level of risk is tolerable but a *high* level of risk is not. Figure 4.3 shows that the analyst must employ countermeasures to mitigate SPARTA techniques *EX-0013*, *IA-0007*, and *EX-0012.10*, and ATT&CK technique *T1133*, but not *T1586* (corresponding to Line 6). For example, *EX-0012.10* is mapped to six countermeasures, of which countermeasure *CM0039* is selected to assure the least privilege principle is enforced in the command and data module (Line 7). SPARTA enumerates 27 NIST security controls that can be used to fulfill this countermeasure. Suppose the analyst selects the following security control (Line 8) because it would be sufficient to fulfill the countermeasure: *CM-7*, which specifies the limitation of functions, ports, protocols, software, and services to only mission essential capabilities. Consequently, the command and data module of Terra (i.e., the bus system component of the space segment) should be able to resist attacks exploiting technique *EX-0012.10*.

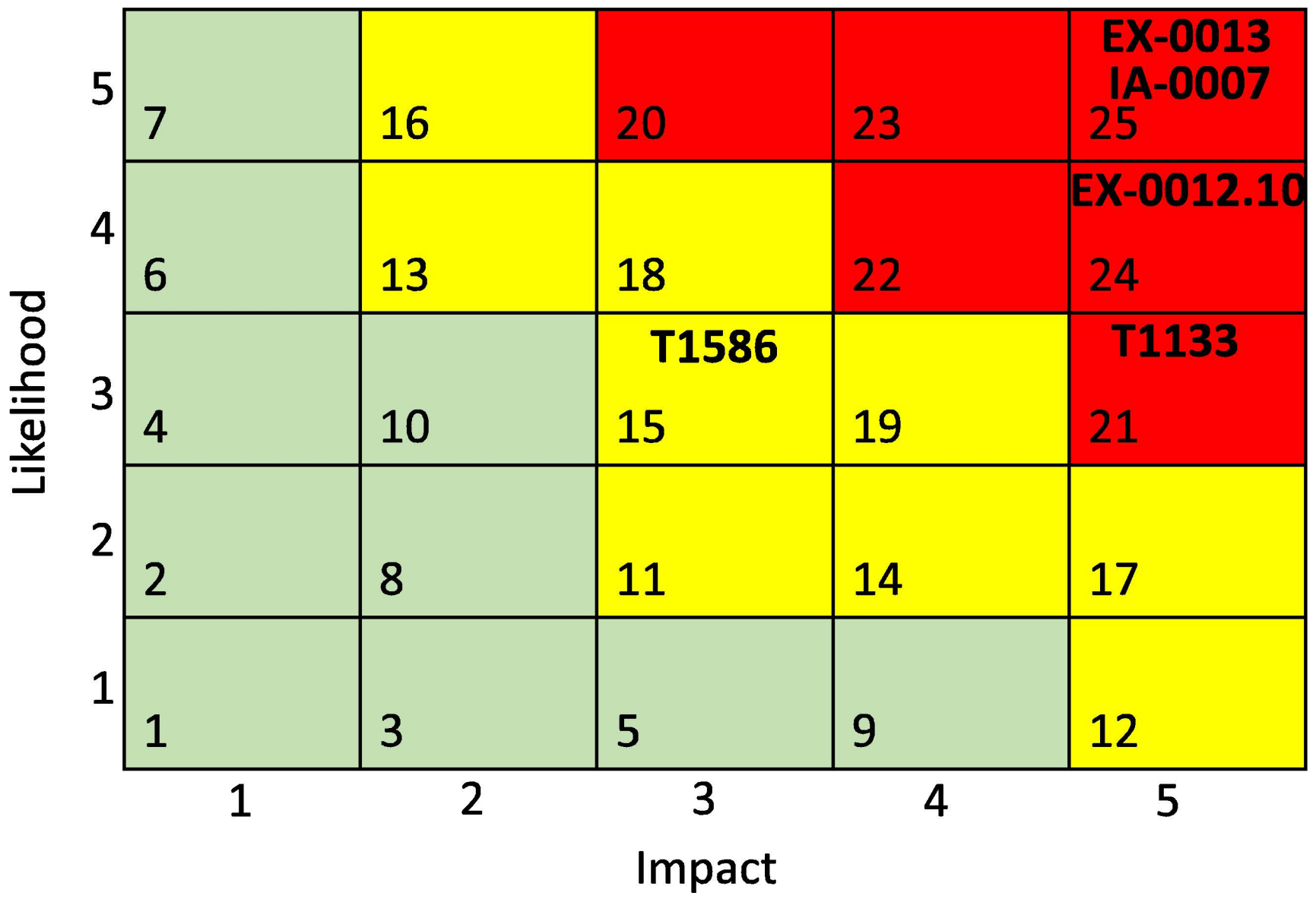

Figure 4.3: 5×5 risk matrix for the Terra space system.

In summary, the NRS-based cyber risk analysis and mitigation tool would have allowed an analyst to identify four techniques (i.e., *EX-0013*, *IA-0007*, *EX-0012.10*, and *T1133*) that must be mitigated with appropriate countermeasures and security controls. Given that these four attack techniques correspond to the control/downlink attack flows that are drawn from the real-world attack, the attack could have been prevented. Note that the countermeasures and security controls do not thwart the third attack mentioned above (i.e., the attack against the payload) because it was deemed tolerable risk.

#### 4.3.1.2 Case Study Scenario 2: Turla Attack against Satellite Communications

Since 2007, Russia's Turla Hacking Group has been reported to exploit the asynchronous nature of satellite internet connections to establish a resilient command and control infrastructure [46]. Turla eavesdrops on the downlink of various commercial satellite communications to identify technical details of the signal while identifying IP addresses and subnets of interest. Once the target satellite communication service-provider companies are identified and selected, Turla crafts malware to execute command and control through these victim companies. Suppose a satellite communications company aims to reduce the risk of similar attacks against its constellation of satellites. The company can employ NRS using Algorithm 3 to identify attack techniques that Turla (or similar attackers) employs and the corresponding countermeasures to harden the company's satellite communications.

Corresponding to Line 1 of Algorithm 3, the analyst can identify from CTI concerning Turla (or its like) that it needs to address: (i) *REC-0005.02* because Turla will intercept its downlink transmissions; (ii) *T1590.005* because Turla will identify its IP subnets; and (iii) *EXF-0010* because Turla will exploit the satellite payload module to broadcast its malicious communications. Corresponding to Lines 2-5, the analyst applies a *medium* system criticality because the satellite constellation has a commercial mission. From the base risk scores, the analyst could increase the likelihood score of *EXF-0010* from 2 to 4 because Turla is definitely targeting similar companies. The analyst establishes *T1590.005* with a likelihood score of 4 and an impact score of 1 because the IP subnet information disclosure does not directly lead to any exploitation.

By leveraging the risk matrix shown in Figure 4.3, the analyst attains the tailored risk scores 22, 24 and 6 for *REC-0005.02*, *EXF-0010* and *T1590.005*, respectively. Corresponding to Lines 6-9, supposing the analyst has determined that the company's risk tolerance is *medium*, the analyst can leverage SPARTA's matrix to identify and employ countermeasures and security controls to mitigate *REC-0005.02* and *EXF-0010*. For example, the analyst may select two of the seven specified countermeasures to mitigate *EXF-0010*: *CM0031* to improve authentication and *CM0036* to ensure proper session termination. Of the five security controls specified in *CM0036*, the analyst may select AC-12 to ensure the user terminal automatically terminates the session once a transmission task is completed, preventing attackers from leveraging active sessions.

**Insight 7.** *Both the attack against the Terra satellite and the Turla attack could have been prevented by applying NRS at satellite design time.*

Note that Insight 7 is derived from an analysis with a specific system model (including the modules and links in the Terra space infrastructure and Turla victim space infrastructure) and threat models (including the attack techniques and attack flows used by attackers). This means that the insight, like any model-driven study, is valid under the premise that the models correctly describe the attackers as well as their attacks. It is an outstanding future research direction to accommodate and quantify the uncertainties associated with the system and threat models, as well as their consequences (e.g., whether the insight still holds when the threat model is not accurate).

### 4.3.2 Characteristics of NRS

From the preceding case study, we observe the following strengths, weaknesses, and applicability of NRS as a space cyber risk analysis and mitigation tool.

#### 4.3.2.1 Strengths of NRS

First, it provides a much-needed starting point for practitioners to incorporate principled cyber risk analysis and mitigation into space system design, development, operation, and maintenance. This is important because space system engineers tend to superficially assess cyber risks as low [60]. As shown above, NRS can be used to identify countermeasures and security controls to mitigate risks that cannot be tolerated.

Second, it provides a baseline for space cyber risk assessment. The large number of SPARTA techniques means a large workforce of DCO experts is required to understand and assess them all. The baseline provided by NRS decreases this workforce burden by identifying the SPARTA techniques most critical to begin cyber risk assessment (i.e., SPARTA techniques with *high* score values). For example, we were able to immediately identify SPARTA techniques *EX-0012.10* and *EX-0013* as priority because of their base scores given by NRS.

#### 4.3.2.2 Weaknesses of NRS

First, it does not give justification on how The Aerospace Cooperation experts assessed the likelihood that a SPARTA technique would succeed against a space infrastructure or system, the impact incurred by the successful use of a SPARTA technique, nor the criticality of components or modules. Consequently, it provides no precedent for practitioners to tailor the likelihood, impact, and criticality to their space system in question. Practitioners are left to subjectively apply their own domain expertise to tailor the base scores.

Second, significant domain expertise from practitioners is required to employ NRS to assess and mitigate cyber risk because it does not provide specific guidance for threat modeling, risk threshold establishment, nor countermeasures and security controls selection. For example, in the preceding discussion SPARTA technique *EX-0013* was subjectively selected, as well as the specific countermeasures and security controls.

Third, risk is considered in terms of individual SPARTA techniques, rather than attack flows (i.e., successful progress from an attack entry point to a destination / impact point). As a result, practitioners must subjectively aggregate the risk scores of the individual SPARTA techniques to derive the risk of an attack. Hence, they must subjectively relate individual SPARTA techniques to components and modules in order to consider a bigger picture of attacks.

#### 4.3.2.3 Applicability of NRS

NRS is applicable to space infrastructures/systems that are subject to cyber attacks, throughout their systems' lifecycle. However, this broad applicability is based on the premise that analysts possess the related *subjective* capabilities indicated in Algorithm 3. First, the analyst knows how to identify the attack techniques that are applicable to the space system in question. This may be difficult because it requires in-depth knowledge of adversary capabilities (across hundreds of SPARTA and ATT&CK techniques) and how they relate to the modules and components of the space system.

Second, the analyst knows how to adjust the base risk score of each technique and how to select appropriate countermeasures. These may be difficult because aerospace experts are required to understand the technical operation of the various modules of the space system while cybersecurity experts are required to understand the impact of cyber attacks against, and countermeasures to protect, those systems.

Third, the analyst knows how to select security controls to instantiate countermeasures. This may be difficult because the analyst must understand how security controls originally developed for IT networks can apply to a space system.

## 4.4 Chapter Summary

Space cyber developers and operators need guidance to accomplish space cyber risk analysis and mitigation. Along this line, NRS represents a good first step. We presented an algorithmic description of NRS and a first characterization on its strengths, weaknesses, and applicability using 2 real-world space-related cyber attack scenarios.

# CHAPTER V

## IDENTIFYING PROPERTIES OF DESIRED SPACE CYBER RISK ANALYSIS AND MITIGATION SYSTEMS

**Chapter Abstract.** Space cyber risk analysis and mitigation is an important problem that space practitioners need help with through principled solutions. In Chapter IV, we analyze The Aerospace Corporation's Notional Risk Scores (NRS) approach which is part of its Space Attack Research and Tactic Analysis (SPARTA) framework and represents the state of the art in space cyber risk analysis and mitigation. However, NRS is just a starting point toward a systematic solution. In this chapter we make a significant step by proposing a set of properties that should be satisfied by competent space cyber risk analysis and mitigation tools. These properties should guide the development of such tools. We apply these properties to assess NRS and the Counterspace Threat Assessment Process that has been proposed in the literature. We also illustrate challenges designers of competent space cyber risk analysis and mitigation tools will encounter when fulfilling these desired properties, namely zero-day vulnerabilities and interdependence.

## 5.1 Chapter Introduction

Space infrastructures and systems are as equally susceptible to cyber attacks as Information Technology (IT) networks [27, 32, 110]. The state of the art is that space cybersecurity is not yet systematically understood, meaning that space cyber risk analysis and mitigation is largely an open problem. This is true despite endeavors such as high level guidance by the National Institute of Standards and Technology (NIST) [55, 103]. In order to harden space infrastructures and systems against cyber threats, we need to identify, quantify, and reduce the cyber risks against them. So far, the most significant effort for space cyber risk analysis and mitigation is The Aerospace Corporation's *Notional Risk Scores* (NRS) [139], which is a part of its Space Attack Research and Tactic Analysis (SPARTA) framework [141]. Both NRS and SPARTA leverage NIST SP 800-53 security controls [40–42] and other industrial endeavors such as [99]. Although NRS offers practitioners with relatively specific guidance on conducting space cyber risk analysis and mitigation, Chapter IV shows that it is just a starting point towards ultimately tackling the space cyber risk analysis and mitigation problem. That is, much research remains before we can provide principled solutions to practitioners. This motivates the present study, which makes a significant step beyond NRS and sheds light on future development of space cyber risk analysis and mitigation tools.

**Chapter Contributions**. In this chapter we make three contributions. First, we propose a principled set of desired properties that should be satisfied by any competent space cyber risk analysis and mitigation tool. To our knowledge, this is the first pro-posal at systematically describing the desired properties, which therefore could guide future studies in designing competent space cyber risk analysis and mitigation tools. Second, we demonstrate the usefulness of the proposed properties, by applying them to assess the two current conceptual space cyber risk analysis and mitigation tools, namely NRS [139] and the Counterspace Threat Assessment Process (CTAP) [58]. We show that NRS is a starting point towards a principled space cyber risk analysis and mitigation tool. Third, to shed light on designing competent space cyber risk analysis and mitigation tools that fulfill those properties, we conduct a case study that addresses two challenges: (i) the accommodation of zero-day vulnerabilities in space systems and (ii) the interdependence between them.

**Chapter Organization**. Section 5.2 discusses terminology. Section 5.3 presents desired properties of space cyber risk analysis and mitigation tools. Section 5.4 assesses NRS and CTAP. Section 5.5 conducts a case study. Section 5.6 concludes the chapter.

## 5.2 Terminology

While already discussed in Chapters III and IV, the following serves as a summary of terms used in this chapter. Space infrastructure is defined in four segments: *space*, *ground*, *user*, and *link*. Each segment contains one or multiple components that accomplish the functions of its segment. Each component contains multiple modules where each accomplishes a specific functionality. *Mission control flows* depict space infrastructure units that work together to send mission-related commands. *Mission* *data flows* depict space infrastructure units that work together to send mission-related data items.

A space cyber risk analysis and mitigation tool is a software and/or hardware system that conducts space cyber risk analysis and mitigation tasks. Such a tool may be a toolbox containing a set of sub-tools applied under different circumstances to accomplish a range of tasks. Moreover, it may contain multiple sub-tools that accomplish the same task with the same input but use different approaches (e.g., mathematical vs. statistical vs. AI/ML), and may aggregate results in a principled fashion.

## 5.3 Desired Properties of Space Cyber Risk Analysis and Mitigation Tools

The characterization in Chapter IV, especially the weaknesses, of the current space cyber risk analysis and mitigation tool, namely NRS, prompts us to propose the desired properties that an ideal space cyber risk analysis and mitigation tool should possess. We propose three classes of desired properties that are orthogonal (and complementary) to each other:

First, Usefulness, which deals with the competency of a space cyber risk analysis and mitigation tool. We define three properties in this case: MISSION-CENTRIC, BACKWARD-COMPATIBILITY, and BROAD APPLICABILITY.

Second, Robustness, which deals with whether a space cyber risk analysis and mitigation tool can cope with errors and adversarial interference, such as manipulated input data. We defined five properties in this class: OBJECTIVITY, RIGOROUSNESS, VALIDITY, UNCERTAINTY QUANTIFICATION, and DYNAMICS.

Third, Usability, which deals with the ease of adopting a space cyber risk analysis and mitigation tool. We define three properties in this class: AUDIENCE, SIMPLICITY, and AUTOMATION.

The three classes of properties are orthogonal in the following sense: A space cyber risk analysis and mitigation tool can be highly useful, but not usable (i.e., hard to use) and/or robust. Moreover, these properties accommodate both the *design* and the *operation* phases of space tools; the distinction between these two phases is explicitly highlighted in [166] because the assumptions made in the design phase may be violated in the operation phase. The desired properties are elaborated below.

### 5.3.1 Properties Related to Usefulness

**Property 1** (MISSION-CENTRICITY)**.** *A space cyber risk analysis and mitigation tool should allow for analyzing risks to missions.*

Property 1 is important because space infrastructures are designed, built, and operated for very specific missions. They are also complex, with some components compromised at any point in time. However, such point-in-time compromises may affect some missions and not others. Therefore, knowing which missions cyber attacks affect will direct defenders' focus during particular periods of time, which is more practical than aiming to secure the entire space infrastructure at every point in time. That is, we advocate *mission-centric* over *infrastructure-oriented* because the latter is too broad and would only be of interest to infrastructure operators. To achieve this, the following should be accommodated:

First, mission specification: This refers to the mission of interest. The specification should include what the mission is, the degree to which cyber attacks can be tolerated by the mission, the conditions under which the mission is disrupted, and whether a disrupted mission can be restored within a desired period of time. The specification should also include the criticality of a mission, as a highly critical mission may be more attractive to attackers [139], and the description of the impact of a successful attack on a mission.

Second, mission control flow: This is one approach to specify a mission in terms of describing how commands are executed in sequence to accomplish a mission. The description of a mission control flow may also include the impact to mission when a control flow is disrupted. This includes the impact of a successful attack. Mission control flows can be specified at multiple levels of abstraction, including the component and the module levels.

Third, mission data flow: This is one approach to specify a mission in terms of describing how data items flow through a space infrastructure. Similarly, the description of a mission data flow may also include the impact to mission when a data flow is disrupted. Mission data flows can also be specified at multiple levels of abstraction.

**Property 2** (BACKWARD-COMPATIBILITY)**.** *A space cyber risk analysis and mitigation tool should accommodate, or co-operate, with existing standards and best practices.*

Property 2 is important because government standards, such as NIST CSF [103], and industry standards, such as ATT&CK [134], SPARTA [141], and NASA's cyberse-curity best practices [100], have been employed by the space sector for years. Therefore, we need to reuse their investment as much as possible, including practitioners' efforts to understand and employ them. Nevertheless, we should note that the existing government/industry standards may contain weaknesses or even flaws from which a space cyber risk analysis and mitigation tool should avoid being misled. It is interesting to note that all the properties proposed in this chapter are partly inspired by NRS and our study analyzing it in Chapter IV, resonating the property of backward-compatibility.

**Property 3** (BROAD-APPLICABILITY). *A space cyber risk analysis and mitigation tool should be applicable to a range of scenarios.*

Property 3 is important because of the high variance and customized nature of space systems. We propose considering the following kinds of applicability:

First, it should accommodate the lifecycle of space infrastructure, which is often *constructed* in a bottom-up fashion, and the lifecycle of space systems, which are often *designed* in a top-down fashion. This is so because the entire space infrastructure consists of space systems owned and operated by many entities (e.g., enterprises if not nations) which do not necessarily coordinate with each other when conceiving their own infrastructures (e.g., SpaceX and Blue Origin do not collaborate in designing and deploying their own constellations).

Second, it should accommodate multiple levels of abstractions of interest. Moreover, the finest level of abstraction for space cyber risk analysis and mitigation purposes should be explored, i.e., whether the module level is appropriate.

Third, it should deal with a range of threat actors with different capabilities. Together with the different degrees of exploitation difficulty, this leads to different *likelihoods* of exploitation, as shown in [139]. For example, it should account for vulnerabilities which are unknown to defenders, rather than known vulnerabilities only. One study [60] shows that expanding threat models to include latent vulnerabilities led to a more effective space cyber risk analysis. More importantly, considering potential zero-day vulnerabilities allows for "what if" analyses that handle potential threats proactively. Such vulnerabilities can be associated with hardware, firmware, and software, and can also propagate through supply chains.

### 5.3.2 Properties Related to Robustness

**Property 4** (OBJECTIVITY). *An ideal space cyber risk analysis and mitigation tool should be as objective as possible.*

Property 4 is important because it reduces unnecessary and error prone subjectivity among the many disparate stakeholders of space infrastructures. As shown in Chapter IV, the current space cyber risk analysis and mitigation tool requires its user to *subjectively* identify: (i) the cyber threat against a space system; (ii) the countermeasures that could be employed to thwart the threat; and (iii) the security controls that can be applied to materialize the countermeasures. Subjectivity is problematic because it largely depends on the risk analyst's technical competency (e.g., one study [60] shows that space engineers lack cybersecurity expertise and often underestimate cyber risk against space systems). Moreover, different experts may have different opinions in assessing cyber threats, highlighting the importance of principled methods to minimize subjectivity (e.g., survey multiple experts and aggregate their responses in some prin-cipled fashion). Principled objective methods also enable repeatability in space cyber risk analysis and mitigation.

**Property 5** (RIGOROUSNESS)**.** *Space cyber risk analysis and mitigation tool design, analysis and validation should be rigorous.*

Property 5 is important because rigorousness allows for clear descriptions of assumptions made by a specific space cyber risk analysis and mitigation tool. This leads to a clear understanding whether a tool is applicable when choosing one or multiple tools for a specific space cyber risk analysis and mitigation task. Rigorousness should be reflected in the following perspectives:

First, input data specification: This describes what data should be input to a space cyber risk analysis and mitigation tool, including the data structure (or representation), the data sources (e.g., raw sensor data or refined cyber threat intelligence). This is relevant when AI/ML models are used to conduct space cyber risk analysis and mitigation tasks because poisoned training data [44] can incorporate back doors for the attacker.

Second, risk analysis method: This can include assumptions made by a space cyber risk analysis method, validation that these assumptions are satisfied by the input data, mathematical soundness of the analysis method, and trustworthiness of the AI/ML methods that are used for risk analysis.

**Property 6** (VALIDITY)**.** *An ideal space cyber risk analysis and mitigation tool should be valid in terms of its models, methods (including algorithms), and parameters.*

Property 6 is important because an unvalidated model or method and arbitrarily assigned parameter values could incur misleading results. Ideally, all assumptions, parameters, and models should be validated through real-world datasets or real-world attack-defense experiments as follows:

First, assumption validation: Any space cyber risk analysis and mitigation tool necessarily makes some assumptions, which should be adequately validated. However, there is no certainty an assumption is always valid because the validation process may not cover all possible scenarios. This highlights the importance of clearly articulating the assumption and the scenarios in which the assumption has been validated. This importance appears to have been recognized only fairly recently [166].

Second, parameter validation: Parameters used in models for space cyber risk analysis and mitigation tools need to be validated. For example, if *tolerable threat* is used as a parameter to indicate what attacks can be tolerated by a space system, then this needs to be validated via experiments to adequately show that these attacks are indeed tolerable (e.g., without comprising the mission in question). Such experiments could use cyber ranges as cybersecurity instruments, as proposed in [28].

Third, model validation: Any space cyber risk analysis and mitigation model used needs to be validated with real-world data and/or experiments. This is particularly relevant to AI/ML models because they are susceptible to adversarial example attacks [45, 154] and poisoned training data attacks [44].

One caveat is that the data reflects past attacks and not current or future attacks, which is an inherent limitation of data-driven characterizations. One mitigating approach is to leverage predictions of the evolution of attack capabilities and consider a distribution of predicted attack capabilities. This necessarily leads to the property of *uncertainty quantification*.

**Property 7** (UNCERTAINTY QUANTIFICATION). *An NRS should provide quantitative measurement of uncertainty associated with its output (e.g., recommendations).*

Property 7 is important because the output produced by a space cyber risk analysis and mitigation tool may have inherent uncertainties. In principle, there are two kinds of uncertainty: *aleatory* vs. *epistemic* [52]. Aleatory uncertainty refers to the inherent uncertainty incurred by irreducible randomness associated with a process or outcome (e.g., the randomness associated with the dice rolling process). This uncertainty, albeit inherent, can be adequately addressed by statistical techniques (e.g., Monte-Carlo methods). On the other hand, epistemic uncertainty deals with the uncertainty incurred by the lack of knowledge. This kind of uncertainty might not be random, but instead inherently biased. As a consequence, the statistical techniques that can adequately cope with aleatory uncertainty (e.g., averaging over many estimates) cannot cope with epistemic uncertainty because all estimates are biased in the same fashion. Thus, it is generally harder to cope with epistemic uncertainty. The associated uncertainties include:

First, assumption uncertainty: This refers to the uncertainty associated with the assumptions that are made by a space cyber risk analysis and mitigation tool. As mentioned above, there is always a chance that an assumption is violated because the validation process cannot cover all possible scenarios (otherwise, it would no longer be an assumption). Thus, principled methods are needed to cope with the scenarios that have not validated the assumption in question. This is a challenging task because there could be exponentially, if not infinitely, many scenarios that are not validated and even infeasible to validate.

Second, parameter uncertainty: If a space cyber risk analysis and mitigation tool uses parameters in its models, then there are uncertainties associated with their measurements.

Third, model certainty: This corresponds to epistemic uncertainty as a model may not reflect the "ground-truth" of how space cyber risk may propagate or evolve within a space infrastructure. For example, it is an open problem to tame the uncertainty against AI/ML models in dealing with adversarial example attacks that aim to evade such models [45, 154].

An ideal space cyber risk analysis and mitigation tool should adequately quantify the uncertainty associated with any prediction or recommendation it makes. For example, when a human decision-maker uses the space cyber risk analysis and mitigation tool as an assistant, it should quantify the uncertainty of its output.

**Property 8** (DYNAMICS)**.** *An ideal space cyber risk analysis and mitigation tool should be able to deal with the evolution of risks.*

Property 8 is important because the space cybersecurity landscape, which includes both attack capabilities and defense capabilities, evolves with time, meaning that the risk resulting from the interactions between these dynamic attack capabilities and dynamic defense capabilities should evolve with time. The importance of explicitly considering dynamic risks has been discussed in a broader context than space cybersecurity (e.g., [160]). Two examples of dynamics are:

First, model assumption dynamics: This refers to the fact that the assumptions that are made and validated at the time of designing a space system may or may not be valid years later when the space system is in orbit.

Second, threat model dynamics: When designing a space system, the designer should not only consider the threats that are currently known, but also consider the threats that may emerge during the lifetime of the space system.

### 5.3.3 Properties Related to Usability

**Property 9** (AUDIENCE). *A space cyber risk analysis and mitigation tool should have clearly defined customers or audiences.*

Property 9 emphasizes that a space cyber risk analysis and mitigation tool should clearly define its purposes because different purposes correspond to different customers or audiences who may have different perspectives when considering cyber risks to space infrastructures and systems. For specifying audiences, we propose considering two dimensions.

This first dimension is the *level of abstraction*. As shown by the space infrastructure system model described in Chapter IV, there are at least four levels of abstraction (from top to bottom): space *infrastructure*, space *system* (e.g., satellite), *component* (e.g., payload in a space system satellite), and *module* (e.g., navigation is a module of a payload component).

The second dimension is *temporal*, meaning that we make explicit distinction between the *design*, *development*, and *operation* (including maintenance) phases, which can be applied to each level of abstraction. For example, the *designer*, *developer*, and *operator*, even at the same level of abstraction, would have different perspectives as evidenced by the fact that the assumptions made in the design phase may be violated in the development and/or operation phase [166].

Thus, a tool should clearly state its audience, ideally specified by the two dimensions mentioned above, with due examples of use cases, to help practitioners decide whether it is appropriate for their own use case.

**Property 10** (SIMPLICITY). *The simpler a space cyber risk analysis and mitigation tool, the better.*

Property 10 highlights the importance of making a space risk analysis and mitigation system easily understandable by its audience of practitioners because this will help them to adopt it. To achieve this, a space cyber risk analysis and mitigation tool should be:

First, *intuitive*, meaning that the tool should be easy for its users to prepare the required input, understand the need for certain kinds of input, interpret the tool's output, and leverage the output (e.g., the recommendations made by the tool).

Second, *transparent*, meaning that an ideal space cyber risk analysis and mitigation tool should make any involved technical details transparent to the practitioners who may adopt them, while providing evidence to show that the transparent process for turning the input into the output is technically sound (e.g., based on a technical document peer reviewed by experts).

**Property 11** (AUTOMATION). *A space cyber risk analysis and mitigation tool should be automated to the fullest extent possible.*

Property 11 is important because manual processes are tedious and allow a large degree of variance, meaning that different analysts would lead to inconsistent, if not contradictory results. Automated analyses can reduce or completely avoid such variance. Moreover, automation is necessary for significantly large data, as demonstrated in prior studies, such as [27,60]. In particular, real-world datasets often lack important details, forcing researchers to extrapolate and causing an explosion in the number of possible attack flows [27]. Automation can be achieved in the following dimensions:

First, data feeding: This means that the input to the space cyber risk analysis and mitigation tool should be based on, for example, the output of other systems, such as some database that may contain sensor data and domain expert's opinions if inevitable.

Second, data processing: This means that the space cyber risk analysis and mitigation tool processes the input data without requiring the user to manually participate in the process.

Third, result interpretation: This means that the space cyber risk analysis and mitigation tool should provide explanation on what the actionable output is and why it is recommended (e.g., why the system recommends a certain security control).

## 5.4 Applying the Properties to Assess Prior Space Cyber Risks Analysis and Mitigation Tools

We are aware of two specific space cyber risk analysis and mitigation tools: NRS [139] which is characterized in Chapter IV from a different perspective than this present chapter, and CTAP proposed in [58], which has not been adopted by practitioners but considered here as a comparison. Both are assessed below for congruence to the desired properties specified in Section 5.3 according to the following ordinal scale: does not satisfy the desired property at all (scored as 1); mostly does not satisfy the property (2);

somewhat satisfies the property (3); mostly satisfies the property (4); and fully satisfies the property (5). Figure 5.1 summarizes the results; details follow.

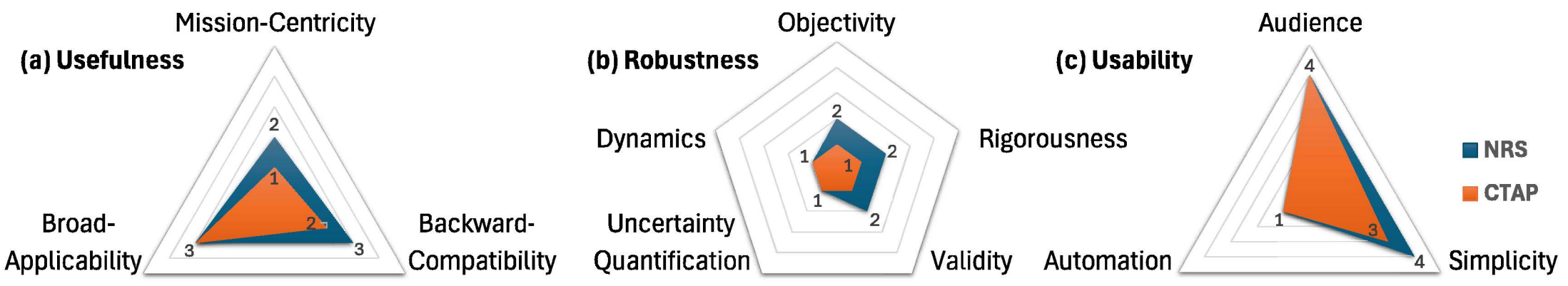

Figure 5.1: Scores of NRS and CTAP with respect to the desired properties in terms of (a) usefulness, (b) robustness, and (c) usability.

### 5.4.1 Assessing NRS

As described in Chapter IV, NRS contains five main components: system criticality, attack technique likelihood, attack technique impact, a 5×5 risk matrix, and countermeasures and security controls. (i) In terms of the usefulness properties, NRS somewhat satisfies them as follows. It mostly does not satisfy MISSION-CENTRICITY because at best its system criticality component addresses missions in a broad sense. NRS somewhat satisfies BACKWARD-COMPATIBILITY and BROAD-APPLICABILITY because its general nature allows for this, but requires significant subjectivity because its baseline risk scores only support SPARTA attack techniques which lack specificity for ground and user segments. (ii) NRS mostly does not satisfy the robustness properties. While the base risk scores and the 5×5 risk matrix provide some OBJECTIVITY, the rest of the algorithm is overly subjective. Its scientific RIGOROUSNESS is lacking even though NRS is founded on industry expertise and analysis. While it leverages over 200 sources of real-world space-related cyber incidents, NRS lacks formal VALIDITY.

It does not provide any UNCERTAINTY QUANTIFICATION nor contain any mechanisms to account for DYNAMICS. (iii) In terms of the usability properties, NRS somewhat satisfies them. It explicitly identifies its AUDIENCE as space system developers, defensive cyberspace operations analysts, and threat intelligence analysts. Its SIMPLICITY is evidenced by the algorithmic description presented in Chapter IV. It currently cannot be AUTOMATED until the subjective components are resolved.

### 5.4.2 Assessing CTAP

CTAP [58] is a three-step process: (i) capability and intent assessment; (ii) counterspace risk scenario generation; and (iii) impact and likelihood assessment. It leverages NASA STD-1006A [99] and proposes incorporation into NASA's Continuous Risk Management process. (i) CTAP mostly does not satisfy the usefulness properties. It is too general to support mission control and data flows for MISSION-CENTRICITY. CTAP contains no technical specifications or capabilities for BACKWARD-COMPABITILITY. It somewhat satisfies BOARD-APPLICABILITY by considering cyber and non-cyber threats. (ii) CTAP does not satisfy the robustness properties at all because: all its steps are subjective (i.e., no OBJECTIVITY); there is a significant lack of RIGOROUSNESS in its cyber risk details and analysis; its base (i.e., NASA STD-1006A) also lacks technical details or VALIDITY; it does not provide any UNCERTAINTY QUANTIFICATION; and CTAP contains no built-in mechanism to deal with the DYNAMICS of risks. (iii) CTAP somewhat satisfies the usability properties. It defines its AUDIENCE as systems engineers who are concerned with all categories of threats. Its SIMPLIC-ITY is evidenced by having only three main steps, but lacks enough details to have transparency. CTAP cannot achieve AUTOMATION because it lacks enough details.

**Insight 8.** *The current space cyber risk analysis and mitigation tools, namely NRS and CTAP, are far from competent or ideal.*

## 5.5 Towards Fulfilling the Desired Properties

We anticipate that fulfilling the desired properties described in Section 5.3 will encounter many challenges. Hence, we illustrate a way to approach two challenges that may be encountered when fulfilling some properties. The two challenges are: the presence of zero-day vulnerabilities, which is related to the BROAD-APPLICABILITY, UNCERTAINTY QUANTIFICATION, and SIMPLICITY properties; and the *interdependence* between space components, which is related to the MISSION-CENTRICITY, VALIDITY, DYNAMICS, and AUTOMATION properties.

### 5.5.1 Tackling Challenges Imposed by Zero-Day Vulnerabilities

Zero-day vulnerabilities, which are security flaws or weaknesses that are unknown to space infrastructure developers, are inevitable. To illustrate this, we leverage the cyber attack against the Terra satellite where the attacker was able to gain full control of the satellite by exploiting its ground station, namely through an "undisclosed" vulnerability in the ground segment, as discussed in Chapter IV. Although this does not necessarily mean a zero-day vulnerability, we notionally assess it as such to discuss how zero-day vulnerabilities can increase both the likelihood and impact of cyber at-tacks in ways that are difficult to predict. Zero-day vulnerabilities potentially increase both the likelihood of attacks by providing the attacker new entryways into gaining or escalating access, and the impact by enabling other attack techniques, such as delivery of more sinister malware. They may also be hard to detect, causing investigators in the Terra incident to reportedly find no indicators of compromise in the ground station whereby the attacker gained control over the satellite.

Designers of future space cyber risk analysis and mitigation tools may address zero-day vulnerabilities by fulfilling the following properties. BROAD-APPLICABILITY is vital because zero-day vulnerabilities may target any component or module of a space infrastructure, across its lifecycle. We propose leveraging "what if" analyses and *predictive analytics* (i.e., forecasting the potential presence of such vulnerabilities) to cope with zero-day vulnerabilities. This naturally leads to the fulfillment of UNCERTAINTY QUANTIFICATION, which requires establishing model uncertainty to cope with the inherent uncertainty about the existence of zero-day vulnerabilities. SIMPLICITY demands that transparency be maintained to not overburden practitioners with the complexity of necessarily large varieties of zero-day vulnerabilities.

### 5.5.2 Tackling Challenges Imposed by Interdependence

To illustrate this issue, we use Turla, a group of cyber attackers, as an example. It has been exploiting satellite communications (SATCOM) space infrastructures for years, as discussed in Chapter IV. Going beyond Chapter IV's discussion, we observe the following *interdependence* challenges that designers of future space cyber risk analysis and mitigation tools must tackle. On the one hand, the user component in the user segment depends on the satellite payload component in the space segment to operate asynchronously to deliver its data to end users. This is because SATCOM users typically do not have other reliable means (e.g., fiber connectivity) to reach their destination, such as rural users. On the other hand, the satellite payload component in the space segment also depends on the user component in the user segment to provide destination details to properly route data. Turla took advantage of both sides of this interdependence because it simultaneously exploited (i) the user's need for asynchronous communication to bypass validity checks on the data and (ii) the satellite's trust of the user's routing information to send its malicious payload to "decoy" user sites.

Designers of future space cyber risk analysis and mitigation tools may address interdependence by fulfilling the following properties. MISSION-CENTRICITY would directly lead to the understanding of how mission requirements relate to users and satellites, such as the asynchronous transmission and routing information requirements. VALIDITY demands enough technical details for parameters and models to adequately capture the dependencies present in pertinent functions performed in the space infrastructure, such as the protocols for data transmission and method for routing verification. DYNAMICS would consider how interdependence can shift over time, such as how growing LEO constellations could shift SATCOM users' dependence from asynchronous protocols to the proliferation of LEO satellites. AUTOMATION could mitigate interdependence, for example, by automating identification of relevant dependencies and corresponding treatments in complex space infrastructures.

## 5.6 Chapter Summary

We have presented a set of desired properties that should be satisfied by any competent space cyber risk analysis and mitigation tool. These properties, or their refinements, will guide development of competent space cyber risk analysis and mitigation tools. To shed light on such future developments, we assessed current tools and discussed how to address the challenges imposed by the presence of *zero-day vulnerabilities* and *interdependence*.

# CHAPTER VI

## AN ACTIONABLE FRAMEWORK: A FIRST STEP TOWARDS AN IDEAL SPACE CYBER RISK ANALYSIS AND MITIGATION SYSTEM

**Chapter Abstract**. Space infrastructures play critical roles in modern society, as demonstrated by their services that include satellite communications (SATCOM) for audio, video and data transmissions. Like the Internet, space infrastructures are vulnerable to cyber attacks, dubbed *space cyber attacks*, highlighting the importance of adequately managing cyber risks to space infrastructures and missions, dubbed *space cyber risks*. However, adequately managing space cyber risks is an outstanding open problem. In this chapter, we propose the first systematic space cyber risk analysis and mitigation framework to describe, quantify, and mitigate space cyber risks. The framework models space infrastructures, space missions, and space cyber attacks, while offering algorithms for space mission risk analysis and hardening. We demonstrate the usefulness of the framework by conducting a case study on three real-world space cyber attacks against the SATCOM infrastructure implemented in our testbed. Our results show, among other things, that: (i) dealing with space cyber attack cascading effects is essential to space cyber risk analysis and mitigation; (ii) the framework can effectively harden space missions; (iii) NIST security controls can effectively mitigate space cyber risks.

## 6.1 Chapter Introduction

Cyber threats against space infrastructures and systems are real as evidenced by many incidents (see, e.g., [27, 32, 110]). This highlights the importance of seeking principled solutions to the space cyber risk analysis and mitigation problem. Towards this goal, the first effective effort is the Space Attack Research and Tactic Analysis (SPARTA) framework [141] developed by The Aerospace Corporation, or more specifically its space cyber Notional Risk Scores (NRS) system [139], which aims to associate notional (i.e., hypothetical but plausible) cyber risks to attack techniques against space systems. However, our study in Chapter IV, which was conducted with industry researchers including the inventor of NRS shows that NRS only represents a first step towards a principled solution to the space cyber risk analysis and mitigation problem. This motivates the present chapter, which aims to propose a new space cyber risk analysis and mitigation framework that goes beyond the NRS and has potential for adoption by industry.

**Chapter Contributions** This chapter makes two contributions. First, we propose a novel and *actionable* space cyber risk analysis and mitigation framework, which is actionable in the sense that it provides detailed representations and models for real-world adoption. The framework has the following features. (i) It is geared towards space infrastructures and the missions they support. (ii) It accommodates two of our recently introduced concepts [122]: *mission control flow*, which specifies how commands are transmitted between units of a space infrastructure, and *mission data flow*, which specifies how data items are processed by, and transmitted between, units of a space infrastructure, at a desired level of abstraction. (iii) It is the first study on *explicitly* modeling cyber attack cascading effects on space infrastructures. (iv) It has two core algorithms, one for *mission risk analysis* and the other for *mission hardening*. Both algorithms can use our sub-routines or incorporate sub-routines desired by analysts in a plug-and-play fashion. (v) It can be applied in the design and operation phase of a space infrastructure, assuming hardening-after-launch is feasible. (vi) It goes much beyond NRS and can be adopted by the industry.

Second, we demonstrate the actionability and usefulness of the framework by implementing its algorithms and applying it to manage space cyber risks, while leveraging our space testbed for validation purposes. Our results show that: (i) the framework can accommodate both ATT&CK and SPARTA *attack techniques* in modeling real-world space cyber attacks; (ii) space cyber attack cascading effects, or *cascading effects* for short, impose as a major challenge to adequate space cyber risk analysis and mitigation; (iii) NIST security controls can be leveraged to harden space infrastructures and their missions; (iv) the framework can effectively reduce space cyber risks to a tolerable level. We will open-source our code to benefit the community and ease industrial adoption.

**Chapter Organization** Section 6.2 models space infrastructures. Section 6.3 models space infrastructure missions. Section 6.4 describes our space cyber risk analysis and mitigation framework. To demonstrate the actionability and usefulness of the framework, Section 6.5 presents a case study via three real-world space cyber attacks and our space testbed. Section 6.6 analyzes our framework with respect to the ideal properties proposed in Chapter V. Section 6.7 concludes the chapter.

## 6.2 Modeling Space Infrastructures

We start with a formal model of space infrastructures because it not only serves as a foundation of space cyber risk analysis and mitigation, but also can guide the development of space cyber risk analysis and mitigation tools with a low chance of arbitrary interpretations / implementations by tool developers—a lesson learned from Heartbleed [21].

### 6.2.1 Terminology and Notational Conventions

A space *infrastructure* typically consists of four segments: space, link, ground, and user. The applications supported by space infrastructures are known as *payloads*, or *application-layer missions*. Payloads are supported by *infrastructure-layer missions*, namely infrastructure-layer services. Each application-layer mission can be specified by one or multiple *control flows*, which may cut across the application and infrastructure layers, and possibly one or multiple *data flows*, which would stay at the application layer [122]. Each infrastructure-layer mission can also have one or multiple *control* *flows* and *data flows*; the latter would stay at the infrastructure layer [122]. In this chapter, *missions* refer to both application-layer and infrastructure-layer missions, unless explicitly stated otherwise.

Our model is based on *graph theory* (see, e.g., [105]), and *directed multigraph graphs* in particular. The notion of directed graph leads to the standard terms of *nodes* (i.e., vertices) and *arcs* (i.e., directed edges). The notion of multigraph means there can be multiple arcs between a same pair of nodes, corresponding to different control and/or data flows. This notational convention also allows us to use the standard set operations, such as the union of node or arc sets.

### 6.2.2 Modeling Space Infrastructures

To represent space infrastructures, we adopt the widely used hierarchical segment-component-module levels of abstraction discussed in [43, 47, 121, 136]. As highlighted in Figure 6.1, a space infrastructure is partitioned into the four *segments* mentioned above (i.e., ground, user, space, and link), which work together to accomplish missions. Within each segment, there are multiple *components*, with each accomplishing some sub-objectives. Each component has one or multiple *modules*, with each accomplishing some of the component's sub-objectives. This leads to the following definition.

**Definition VI.1** (infrastructure representation). *A given space infrastructure is modeled as a directed graph $G_{\mathsf{infra}} = (V_{\mathsf{infra}}, E_{\mathsf{infra}})$, where $V_{\mathsf{infra}}$ is the set of nodes representing space infrastructure modules and $E_{\mathsf{infra}}$ is the set of arcs representing communication relationships such that $(u, v) \in E_{\mathsf{infra}}$ means that $u \in V_{\mathsf{infra}}$ can* directly *communicate with $v_2 \in V_{\mathsf{infra}}$ (possibly via a communication path).*

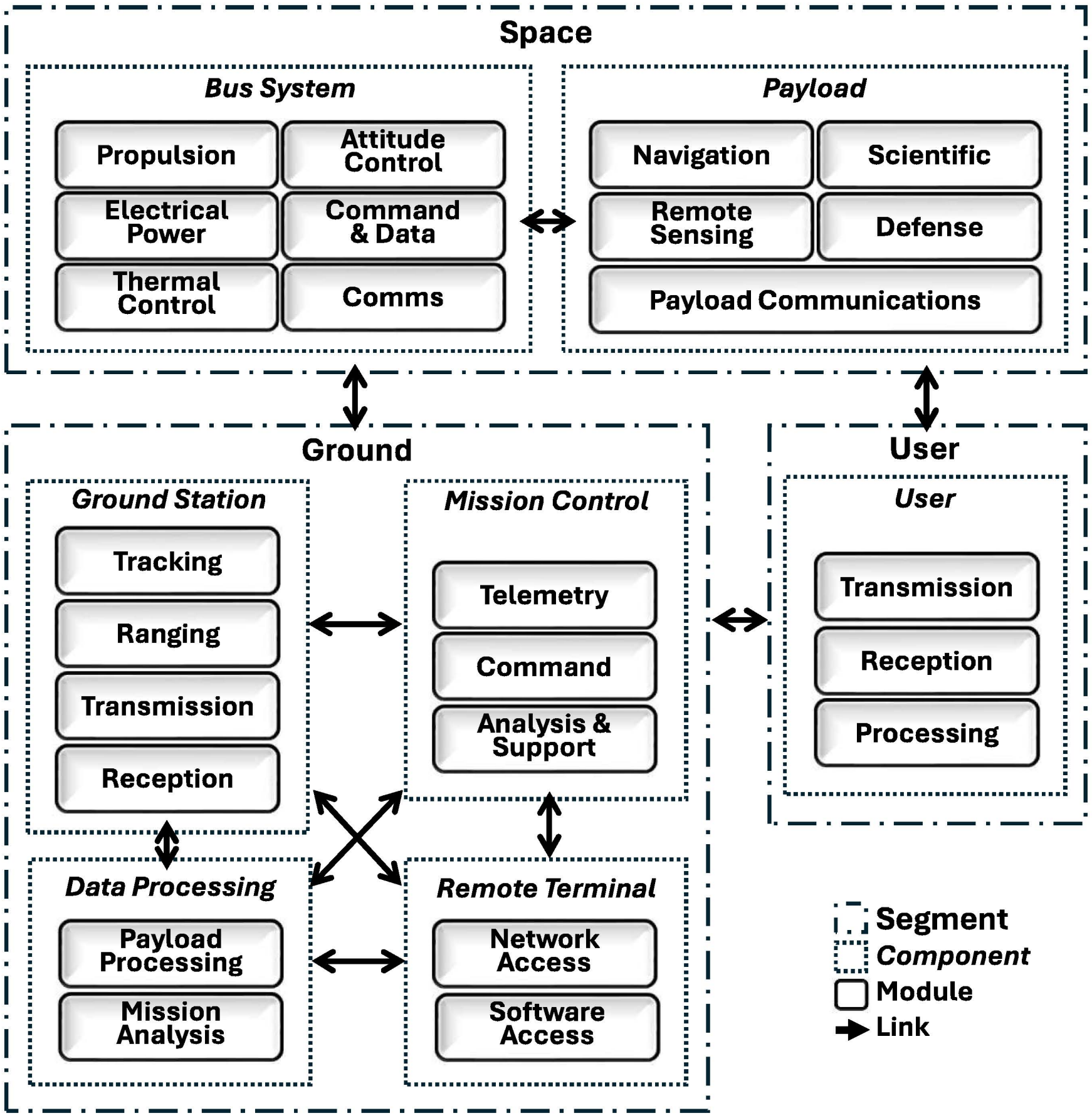

Figure 6.1: Space infrastructure system model

Definition VI.1 focuses on the module level because this is the level of abstraction at which mission control and data flows are described [122], while aligning with the NRS [23, 139]. However, the definition can be easily adapted to describe space infrastructures at other levels of abstraction as needed. Note that $(u, v) \notin E_{\mathsf{af}}$ implies that compromising node $u$ does not enable the attacker to wage a *direct* attack against node $v$ from node $u$, because $u$ cannot communicate with $v$ or the communication from $u$ to $v$ is filtered by a firewall-like mechanism. $G_{\mathsf{infra}}$ can be a multigraph when the communication $(u, v) \in E_{\mathsf{infra}}$ can be implemented with different means or routing paths (e.g., Radio Frequency (RF) vs. Optical communications).

## 6.3 Modeling Missions

Given $G_{\mathsf{infra}} = (V_{\mathsf{infra}}, E_{\mathsf{infra}})$, we formalize the intuitive notions of mission *control flows* and *data flows* introduced in [122] as follows.

**Definition VI.2** (mission control flow). *A mission control flow describes how commands (or instructions) are issued between the nodes in $V_{\mathsf{infra}}$. Formally, a mission control flow is represented as a* directed graph $G_{\mathsf{mcf}} = (V_{\mathsf{mcf}}, E_{\mathsf{mcf}})$*, where $V_{\mathsf{mcf}} \subseteq V_{\mathsf{infra}}$, $E_{\mathsf{mcf}} \subseteq E_{\mathsf{infra}}$, node $v \in V_{\mathsf{mcf}}$ represents a space infrastructure module, and arc $(u, v) \in E_{\mathsf{mcf}}$ indicates commands are issued from node $u$ to node $v$.*

Definition VI.2 indicates that a control flow is *not* necessarily a *line graph*, differing from the intuitive meaning of *flow* that one may think. When the need arises, we use superscript $(i, j)$ to denote the $i$th control flow of the $j$th mission, $\left\{G_{\mathsf{mcf}}^{(i,j)} = \left(V_{\mathsf{mcf}}^{(i,j)}, E_{\mathsf{mcf}}^{(i,j)}\right)\right\}$, where $1 \leq i \leq m_j$ (i.e., mission $j$ has $m_j$ control flows) and $1 \leq j \leq n$.

**Definition VI.3** (mission data flow). *A mission data flow describes how data items are transmitted between the nodes in $V_{\mathsf{infra}}$. Formally, a mission data flow is represented as a* directed graph $G_{\mathsf{mdf}} = (V_{\mathsf{mdf}}, E_{\mathsf{mdf}})$*, where $V_{\mathsf{mdf}} \subseteq V_{\mathsf{infra}}$, $E_{\mathsf{mdf}} \subseteq E_{\mathsf{infra}}$, $v \in V_{\mathsf{mdf}}$ represents a module, and arc $(u, v) \in E_{\mathsf{mdf}}$ indicates data items are sent from $u$ to $v$.*

Definition VI.3 indicates a data flow is not necessarily a line graph, while it is possible $V_{\mathsf{mcf}} \cap V_{\mathsf{mdf}} \neq \emptyset$. When the need arises, we use superscript $(i', j)$ to denote the $i'$th data flow of the $j$th mission, $\left\{G_{\mathsf{mdf}}^{(i',j)} = \left(V_{\mathsf{mdf}}^{(i',j)}, E_{\mathsf{mdf}}^{(i',j)}\right)\right\}$, where $1 \leq i' \leq m'_j$ and $1 \leq j \leq n$.

## 6.4 Managing Space Cyber Risks

### 6.4.1 Modeling Attack Capabilities

We adopt practitioners' way of thinking in modeling attack capabilities via *attack techniques* as described in the SPARTA framework [141] and the MITRE ATT&CK framework [133] (i.e., methods or approaches for achieving some sub-objectives), while noting that the former primarily describes cyber threats against the space and link segments and the latter primarily describes cyber threats against the ground and user segments. We choose to focus on these industrial standards because we hope this will ease the adoption of our framework in practice. SPARTA specifies nine space cyber *attack tactics* (i.e., an attacker's sub-objectives in different phases of attacks), with each supported by five to 18 techniques (some of which is further divided into one to nine sub-techniques); ATT&CK specifies 14 cyber attack tactics, with each supported by eight to 44 techniques (some of which is further divided into one to 17 sub-techniques).

**Definition VI.4** (attack capabilities). *We define the capabilities of an attacker $A$ as the set of attack techniques (as defined in ATT&CK and SPARTA) that are possessed by $A$, denoted by* $\mathsf{AT}_A$.

### 6.4.2 Mission Risk Analysis

#### 6.4.2.1 Modeling Attack Cascading Effects

We consider the following three kinds of *cascading effects* with respect to $e = (u, v) \in E_{\mathsf{infra}}$.

- Type-1 Cascading Effects: Compromising node (i.e., module) $u$ can cause the compromise of $v$. This can happen (e.g.) when $u$ is designed to send commands to $v$ via a cryptographically authenticated channel, because a compromised $u$ can send malicious commands to $v$ for execution.
- Type-2 Cascading Effects: Compromising arc (i.e., communication channel) $(u, v)$ can cause the compromise of $v$. This can happen when the attacker injects malicious commands to $v$ for execution because, for instance, when the channel $(u, v)$ is not cryptographically authenticated.
- Type-3 Cascading Effects: Compromising node $u$ can cause the compromise of $e = (u, v)$. This can happen (e.g.) when compromising $u$ gives $A$ the privilege to use $e$.

There could be other kinds of cascading effects, which are left to future studies.

#### 6.4.2.2 Modeling Space Cyber Risks

Risks are inherently associated with *uncertainty*, of which *certainty* is a special case. For instance, a space infrastructure may be attacked by attacker $A$ *potentially* (or *certainly* as a special case) with an attack technique, which can be captured via likelihood $\alpha \in [0, 1]$ (with $\alpha = 1$ being the special case of certainty). We treat likelihood $\alpha$ as a broader concept than *probability* because $\alpha$ in the real world may be *subjectively* estimated by domain experts without explicitly considering the relevant probabilistic structure, as shown by SPARTA NRS [139].

**Definition VI.5** (likelihood of mission disruption). *Let* $\mathbf{L}(\cdot)$ *be a function returning the likelihood that a node, arc, control flow, data flow, or mission is disrupted by space cyber attacks. We say mission* $j$*, which has* $m_j$ *control flows and* $m'_j$ *data flows specified via* $\left\{G^{(i,j)}_{\mathsf{mcf}} = \left(V^{(i,j)}_{\mathsf{mcf}}, E^{(i,j)}_{\mathsf{mcf}}\right)\right\}$ *and* $\left\{G^{(i',j)}_{\mathsf{mdf}} = \left(V^{(i',j)}_{\mathsf{mdf}}, E^{(i',j)}_{\mathsf{mdf}}\right)\right\}$ *where* $1 \leq i \leq m_j$, $1 \leq i' \leq m'_j$, *and* $1 \leq j \leq n$, *is disrupted with likelihood* $\eta \in [0,1]$, *denoted by* $\mathbf{L}(j) = \eta$, *if any of its control flow or data flow is disrupted at most with likelihood* $\eta$, *namely*

$$\mathbf{L}(j) \leftarrow \max\left\{\left\{\mathbf{L}\left(G^{(i,j)}_{\mathsf{mcf}}\right)\right\}_{1\leq i\leq m_j}, \left\{\mathbf{L}\left(G^{(i',j)}_{\mathsf{mdf}}\right)\right\}_{1\leq i'\leq m'_j}\right\}. \tag{VI.1}$$

Definition VI.5 captures the intuition of "weakest link" in cybersecurity, while noting that other aggregation functions than the $\max$ function in Eq.(VI.1) are possible. This definition offers this flexibility by not specifying how the disruption likelihood of a control (data) flow is derived or aggregated from the disruption likelihoods of the nodes and arcs in the control (data) flow, namely how $\mathbf{L}\left(G^{(i,j)}_{\mathsf{mcf}}\right)$ should be derived or aggregated from $\mathbf{L}(v)$ for $v \in V^{(i,j)}_{\mathsf{mcf}}$ and $\mathbf{L}(e)$ for $e \in E^{(i,j)}_{\mathsf{mcf}}$, and how $\mathbf{L}\left(G^{(i',j)}_{\mathsf{mdf}}\right)$ should be derived or aggregated from $\mathbf{L}(v)$ for $v \in V^{(i',j)}_{\mathsf{mdf}}$ and $\mathbf{L}(e)$ for $e \in E^{(i',j)}_{\mathsf{mdf}}$. In our case study, we also consider the aggregation function $\max$.

#### 6.4.2.3 Algorithm for Space Cyber Risk Analysis

We propose Algorithm 4 for computing space cyber risks to missions. Its input includes: cyber attack techniques $\mathsf{AT}_A$ that may be possessed by $A$ with likelihood $\mathsf{L}_{\mathsf{at}} \in (0,1]$ for $\mathsf{at} \in \mathsf{AT}$, which may be derived from or provided by cyber threat intelligence (CTI), while noting that $\mathsf{L}_{\mathsf{at}} = 0$ means the at does not need to be considered; the likelihood that at can compromise node $v$, denoted by $\beta(v, \mathsf{at}) \in [0, 1]$, or compromise arc $e$, denoted by $\beta(e, \mathsf{at}) \in [0, 1]$, which may be obtained via experiments or CTI; space infrastructure $G_{\mathsf{infra}}$; mission control flows and mission data flows for missions $1 \leq j \leq n$; a flag $\in \{0, 1\}$, which allows the analyst to determine the extent of cascading effects under consideration. It outputs $\mathbf{L}(j)$, the likelihood mission $j$ is disrupted for $1 \leq j \leq n$.

Algorithm 4 can be understood as follows. Lines 1-2 initialize $\mathbf{L}(v, \mathsf{at})$ and $\mathbf{L}(e, \mathsf{at})$, the likelihoods node $v \in V_{\mathsf{infra}}$ and arc $e \in V_{\mathsf{infra}}$ are compromised as a direct consequence of using attack technique at $\in \mathsf{AT}_A$ to attack them. Lines 3-4 initialize $\mathbf{L}(v)$ and $\mathbf{L}(e)$, the likelihoods $v \in V_{\mathsf{infra}}$ and $e \in V_{\mathsf{infra}}$ are compromised as a consequence of collective effect (i.e., multiple at's are applicable to them) and/or cascading effects.

Lines 5-11 compute the *direct* effect of each attack technique at $\in \mathsf{AT}_A$ on $v \in V_{\mathsf{infra}}$ and $e \in V_{\mathsf{infra}}$. The likelihood that a technique at compromises $v$, denoted by $\mathbf{L}(v, \mathsf{at})$, is the product of $\mathsf{L}_{\mathsf{at}}$, which is the likelihood that at is possessed by $A$, and $\beta(v, \mathsf{at})$, which is the likelihood that at compromises $v$. The likelihood $\mathbf{L}(e, \mathsf{at})$ that $e$ is compromised by a technique at is similarly computed.

Lines 12-13 compute the *joint* effect of direct attacks that employ the applicable at $\in \mathsf{AT}_A$ to attack node $v \in V_{\mathsf{infra}}$ via a suitable function $f$, where "suitable" (throughout this Dissertation) means a function is derived by practitioners via real-world experiments or data, or a function that is derived from certain (validated) assumptions (e.g., either the attacks are waged independently or waged under a sophisticated orchestration

**Algorithm 4:** Computing Space Cyber Risks to Missions

**Input:** a set $\mathsf{AT}_A$ of attack techniques possessed by attacker $A$, associated with likelihood of possession $\mathsf{L}_{\mathsf{at}} \in (0,1]$ for $\mathsf{at} \in \mathsf{AT}_A$; the likelihood that technique $\mathsf{at} \in \mathsf{AT}_A$ can compromise node $v \in V_{\mathsf{infra}}$, denoted by $\beta(v, \mathsf{at})$, or arc $e \in E_{\mathsf{infra}}$, denoted by $\beta(e, \mathsf{at})$; space infrastructure $G_{\mathsf{infra}} = (V_{\mathsf{infra}}, E_{\mathsf{infra}})$; a set of mission control flows $\left\{G^{(i,j)}_{\mathsf{mcf}} = \left(V^{(i,j)}_{\mathsf{mcf}}, E^{(i,j)}_{\mathsf{mcf}}\right)\right\}$ and a set of mission data flows $\left\{G^{(i',j)}_{\mathsf{mdf}} = \left(V^{(i',j)}_{\mathsf{mdf}}, E^{(i',j)}_{\mathsf{mdf}}\right)\right\}$ for $1 \le i \le m_j$, $1 \le i' \le m'_j$, and $1 \le j \le n$; $\mathsf{flag} \in \{0,1\}$

**Output:** mission disruption likelihood $\mathbf{L}(j)$ for $1 \le j \le n$

$\mathbf{L}(v, \mathsf{at}) \leftarrow 0$ for $v \in V_{\mathsf{infra}}$ and $\mathsf{at} \in \mathsf{AT}_A$<br>
$\mathbf{L}(e, \mathsf{at}) \leftarrow 0$ for $e \in E_{\mathsf{infra}}$ and $\mathsf{at} \in \mathsf{AT}_A$<br>
$\mathbf{L}(v) \leftarrow 0$ for $v \in V_{\mathsf{infra}}$<br>
$\mathbf{L}(e) \leftarrow 0$ for $e \in E_{\mathsf{infra}}$<br>
**for** *technique* $\mathsf{at} \in \mathsf{AT}_A$ **do** `// computing direct effects`<br>
    **for** $v \in V_{\mathsf{infra}}$ **do**<br>
        **if** $\mathsf{at}$ *compromises* $v$ *with likelihood* $\beta(v, \mathsf{at}) > 0$ **then**<br>
            $\mathbf{L}(v, \mathsf{at}) \leftarrow \beta(v, \mathsf{at}) \times \mathsf{L}_{\mathsf{at}}$<br>
    **for** $e \in E_{\mathsf{infra}}$ **do**<br>
        **if** $\mathsf{at}$ *compromises* $e$ *with likelihood* $\beta(e, \mathsf{at}) > 0$ **then**<br>
            $\mathbf{L}(e, \mathsf{at}) \leftarrow \beta(e, \mathsf{at}) \times \mathsf{L}_{\mathsf{at}}$<br>
**for** $v \in E_{\mathsf{infra}}$ **do** `// computing joint attack effects`<br>
    $\mathbf{L}(v) \leftarrow f(\{\mathbf{L}(v, \mathsf{at})\}_{\mathsf{at} \in \mathsf{AT}_A})$ `// using suitable` $f$<br>
**for** $e = (u, v) \in E_{\mathsf{infra}}$ **do** `// using suitable` $g$<br>
    $\mathbf{L}(e) \leftarrow g(\mathbf{L}(u), \{\mathbf{L}(e, \mathsf{at})\}_{\mathsf{at} \in \mathsf{AT}_A})$<br>
**if** $\mathsf{flag} = 1$ **then** `// Case 0 (`$\mathsf{flag} = 0$`) vs. 1 (`$\mathsf{flag} = 1$`)`<br>
    `delete` $\{v : \mathbf{L}(v) = 0$ `and` $\prod_{(u,v) \in E_{\mathsf{infra}}} \mathbf{L}(u) = 0\}$ `and their adjacent arcs from` $G_{\mathsf{infra}}$<br>
**repeat** `// computing cascading effects`<br>
    **for** $v \in V_{\mathsf{infra}}$ **do** `// using suitable` $h$<br>
        $\mathbf{L}(v) \leftarrow h(\{\mathbf{L}(u), \mathbf{L}(e = (u, v))\}_{e \in E_{\mathsf{infra}}}, \mathbf{L}(v))$<br>
        **for** $e' = (v, v') \in E_{\mathsf{infra}}$ **do** `// using suitable` $h'$<br>
            $\mathbf{L}(e') \leftarrow h'(\mathbf{L}(v), \{\mathbf{L}(e', \mathsf{at})\}_{\mathsf{at} \in \mathsf{AT}_A})$<br>
**until** $\{\mathbf{L}(v)\}_{v \in V_{\mathsf{infra}}}$ *converge (i.e., no longer updated)*;<br>
**for** $j = 1$ *to* $n$ **do**<br>
    `call a sub-routine (e.g., Algorithm 5) to compute mission disruption likelihood` $\mathbf{L}(j)$ `based on` $\{\mathbf{L}(v)\}_{v \in V_{\mathsf{infra}}}$, $\{\mathbf{L}(e)\}_{e \in E_{\mathsf{infra}}}$, $\left\{G^{(i,j)}_{\mathsf{mcf}} = \left(V^{(i,j)}_{\mathsf{mcf}}, E^{(i,j)}_{\mathsf{mcf}}\right)\right\}_{1 \le i \le m_j}$ `and` $\left\{G^{(i',j)}_{\mathsf{mdf}} = \left(V^{(i',j)}_{\mathsf{mdf}}, E^{(i',j)}_{\mathsf{mdf}}\right)\right\}_{1 \le i' \le m'_j}$<br>
**return** $\{\mathbf{L}(j)\}_{1 \le j \le n}$, $\{\mathbf{L}(v)\}_{v \in V_{\mathsf{infra}}}$, $\{\mathbf{L}(e)\}_{e \in E_{\mathsf{infra}}}$

that incurs a certain dependence structure that can be mathematically described [155]; in any case, a "suitable" function should obey the probability laws because likelihood is often interpreted or treated as probability in the real world. In the case that the successes of the $\mathsf{at}$'s are independent of each other, the likelihoods can be interpreted as independent probabilities and we can define function $f$ as:

$$\mathbf{L}(v) \leftarrow 1 - \prod_{\mathsf{at} \in \mathsf{AT}_A} (1 - \beta(v, \mathsf{at}) \times \mathsf{L}_{\mathsf{at}}) . \tag{VI.2}$$

In the case of dependence, the computation of $\mathbf{L}(v)$ would require further information about the dependence structure (see, e.g., [155] for a theoretical treatment of dependence in a related context); in the absence of such information, we suggest conducting experiments to empirically determine $\mathbf{L}(v)$, or asking domain experts to estimate them as shown by the SPARTA NRS [139]. Note that Eq.(VI.2) accommodates the worst-case scenario in the sense that we do not consider the possibility that an attack attempt is detected and further thwarted by the defender; incorporating such defense capabilities is an important matter that is left for future studies.

Lines 14-15 compute the *joint* effect of direct attacks waging the applicable $\mathsf{at} \in \mathsf{AT}_A$ against arc $e \in V_{\mathsf{infra}}$, via an appropriate function $g$. In the case that the successes of these events are independent of each other, the likelihoods can be interpreted as independent probabilities and we can define function $g$ as:

$$\mathbf{L}(e) \leftarrow 1 - \prod_{\mathsf{at} \in \mathsf{AT}_A} (1 - \beta(e, \mathsf{at}) \times \mathsf{L}_{\mathsf{at}}) . \tag{VI.3}$$

In the case of dependence, $\mathbf{L}(e)$ can be treated in the same fashion as how $\mathbf{L}(v)$ would be treated. Similarly, Eq.(VI.3) also accommodates the worst-case scenarios that the attacks are not detected by defenses.

Line 16 uses a flag to determine whether to consider cascading effects on the nodes or arcs that cannot be directly attacked by the techniques $\mathsf{at} \in \mathsf{AT}_A$. Specifically, $\mathsf{flag} = 0$, also dubbed Case 0, means these nodes and their adjacent arcs can be compromised as a consequence of cascading effects; otherwise (i.e. $\mathsf{flag} = 1$, also dubbed Case 1), these nodes and their adjacent arcs can be deleted from $G_{\mathsf{infra}}$ as they need no further consideration. To be more precise, the nodes at stake are the $v$'s in the following set:

$$\bar{V}_{\mathsf{infra}} \leftarrow V_{\mathsf{infra}} - \{v : \exists \mathsf{at} \in \mathsf{AT}_A \text{ such that } \beta(v, \mathsf{at}) > 0\}. \tag{VI.4}$$

As we will show in our case study, contrasting Case 0 vs. 1 will lead to deep insights into the impact of cascading effects. Note that $\{v : \mathbf{L}(v) = 0 \text{ and} \prod_{(u,v)\in E_{\mathsf{infra}}} \mathbf{L}(u) = 0\}$ means the set of nodes that cannot be compromised by the techniques in $\mathsf{AT}_A$ and their parent nodes cannot be compromised by techniques in $\mathsf{AT}_A$.

Lines 18-23 compute the cascading effects of attacks via an appropriate function $h$. Similar to functions $f$ and $g$, the form of function $h$ relies on whether the relevant events are independent of each other or not. In the case of independence while assuming compromising $u$ can cause the compromise of $v$ (i.e., Type-1 cascading effects) and compromising $e = (u, v)$ can cause the compromise of $v$ (i.e., Type-2 cascading effects), the likelihoods can be interpreted as independent probabilities and we can

define $h$ as:

$$\mathbf{L}(v) \leftarrow \mathbf{L}(v) + (1 - \mathbf{L}(v))(1 - (1 - \mathbf{L}(u))(1 - \mathbf{L}(e))). \qquad \text{(VI.5)}$$

We also leave the treatment of dependence to future studies. Lines 21-22 update the likelihood that arc $e' = (v, v')$ is compromised as the likelihood $\mathbf{L}(v)$ is compromised has just been updated (i.e., Type-3 cascading effects). Under a similar independence assumption, we define $h'$ as:

$$\mathbf{L}(e') \leftarrow \mathbf{L}(e') + (1 - \mathbf{L}(e'))\mathbf{L}(v). \qquad \text{(VI.6)}$$

Line 23 deserves special mention. While it is intuitive that the loop will eventually stop because *all vulnerable nodes are eventually compromised with a respective likelihood that depends on the structure of* $G_{\mathsf{infra}}$ *in the absence of defense*, e.g., a node $v$ with in-degree 0 is compromised with likelihood $\mathbf{L}(v)$ of direct attacks without being affected by any cascading effect. A rigorous proof of this property is left to future studies, but could be obtained by leveraging the theoretical results reported in [79, 175], which rigorously prove that a large family of cybersecurity dynamics is globally convergent, hinting Algorithm 4 always converges.

Lines 24-26 call a sub-routine to compute the likelihood that mission $j$ is disrupted, where $1 \leq j \leq n$. There can be multiple methods for this purpose, such as the one described in Algorithm 5, which essentially uses the $\max$ function to aggregate attack effects (i.e., corresponding to, as mentioned above, the intuition of "weakest link" in cybersecurity).

The computational complexity of Algorithm 4 is estimated as follows. The first loop (Lines 5-11) incurs a complexity of $O(|\mathsf{AT}_A| \times |V_{\mathsf{infra}}|^2)$, assuming there are at most a constant number of multi-arcs between any pair of nodes (which would be true in practice). This is also the upper bound of the number of calls to function $f$ (Lines 12-13), of the number of calls to function $g$ (Lines 14-15), and of Lines 16-17 which only incurs $O(|V_{\mathsf{infra}}|^2)$. However, it is difficult to estimate how many iterations it takes the cascading effects to converge (Lines 18-23) because the convergence result, even if theoretically proven (as shown in [79, 175]), is asymptotic. In our experiments, they converge quickly. The last loop (Lines 24-26) incurs a complexity of $n$ times the complexity of Algorithm 5.

#### 6.4.2.4 A Sub-Routine for Computing Mission Disruption Likelihood

Algorithm 5 is a specific method for computing mission disruption likelihood $\mathbf{L}(j)$, which uses the $\max$ aggregation function corresponding to the cybersecurity notion of "weakest link." For a given mission control flow, its disruption likelihood is defined as the maximum likelihood among the $\mathbf{L}(v)$'s for $v \in V_{\mathsf{mcf}}$ and the $\mathbf{L}(e)$'s for $e \in E_{\mathsf{mcf}}$ (Lines 1-2). For a given mission data flow, its disruption likelihood is similarly computed (Lines 3-4). The mission's disruption likelihood is the maximum among the control flow disruption likelihoods and the data flow disruption likelihoods (Line 5).

The complexity of Algorithm 5 is $O\left((m+m')|V_{\textsf{infra}}|^2\right)$, where $m = \max\{m_1, \ldots, m_n\}$ and $m' = \max\{m'_1, \ldots, m'_n\}$, while noting that $G^{(i,j)}_{\textsf{mcf}} \subseteq G_{\textsf{infra}}$ and $G^{(i,j)}_{\textsf{mdf}} \subseteq G_{\textsf{infra}}$ and assuming there is a constant number of multi-arcs between any pair of nodes.

---

**Algorithm 5:** Sub-routine for computing $\mathbf{L}(j)$, the likelihood that mission $j$ is disrupted by attacker $A$

---

**Input:** $\{\mathbf{L}(v)\}_{v\in V_{\textsf{infra}}}$, $\{\mathbf{L}(e)\}_{e\in E_{\textsf{infra}}}$, $\left\{G^{(i,j)}_{\textsf{mcf}} = \left(V^{(i,j)}_{\textsf{mcf}}, E^{(i,j)}_{\textsf{mcf}}\right)\right\}_{1\le i\le m_j}$ and $\left\{G^{(i',j)}_{\textsf{mdf}} = \left(V^{(i',j)}_{\textsf{mdf}}, E^{(i',j)}_{\textsf{mdf}}\right)\right\}_{1\le i'\le m'_j}$

**Output:** $\mathbf{L}(j)$, the likelihood mission $j$ is disrupted

**for** $i = 1$ *to* $m_j$ **do**

  compute the likelihood that the $i$th control flow of mission $j$'s is disrupted as $\mathbf{L}\left(G^{(i,j)}_{\textsf{mcf}}\right) \leftarrow \max\left\{\max_{v\in V^{(i,j)}_{\textsf{mcf}}} \mathbf{L}(v), \max_{e\in E^{(i,j)}_{\textsf{mcf}}} \mathbf{L}(e)\right\}$

**for** $i' = 1$ *to* $m'_j$ **do**

  compute the likelihood that the $i'$th data flow of mission $j$'s is disrupted as $\mathbf{L}\left(G^{(i',j)}_{\textsf{mdf}}\right) \leftarrow \max\left\{\max_{v\in V^{(i',j)}_{\textsf{mdf}}} \mathbf{L}(v), \max_{e\in E^{(i',j)}_{\textsf{mdf}}} \mathbf{L}(e)\right\}$ compute the likelihood mission $j$ is disrupted via Eq.(VI.1) or $\mathbf{L}(j) \leftarrow \max\left\{\left\{\mathbf{L}\left(G^{(i,j)}_{\textsf{mcf}}\right)\right\}_{1\le i\le m_j}, \left\{\mathbf{L}\left(G^{(i',j)}_{\textsf{mdf}}\right)\right\}_{1\le i'\le m'_j}\right\}$

**return** $\mathbf{L}(j)$

---

### 6.4.3 Mission Hardening

Let $\tau \in [0, 1]$ denote the tolerable degree of mission disruption. If the output of Algorithm 4 is greater than $\tau$, namely $\mathbf{L}(j) > \tau$ for some $1 \le j \le n$, then *security controls* (i.e., countermeasures) need to be identified and employed to reduce the mission disruption likelihood to a tolerable level (i.e., below $\tau$). This triggers the need for a *mission hardening* algorithm, which is presented as Algorithm 6, for identifying security controls, denoted by sc's, that can be employed to harden missions.

For this purpose, we leverage the security controls specified in NIST SP 800-53 [41] because they are widely used.

Algorithm 6 determines whether mission hardening is required, and if so, how. It can be understood as follows. Line 1 keeps the input $\mathsf{AT}_A$ in $\mathsf{AT}^*_A$ because $\mathsf{AT}_A$ will be dynamically updated. Hardening is required when there is some mission $j$ ($1 \leq j \leq n$) that is disrupted with a likelihood $\mathbf{L}(j) > \tau$, as shown in Line 3. In the case of hardening (Lines 6-20), the algorithm proceeds as follows. The loop in Lines 6-7 identify the $\mathsf{at}$'s that must be mitigated because they would immediately disrupt a mission via some *nodes* even without considering the joint or cascading attack effects. The loop in Lines 8-9 identify the $\mathsf{at}$'s that must be mitigated because they would immediately disrupt a mission via some *arcs* even without considering the joint or cascading attack effects. After mitigating these $\mathsf{at}$'s, Line 10 calls Algorithm 4 to determine whether some mission is still disrupted, where $\mathsf{flag} = 0$ because Algorithm 6 will solely determine which $\mathsf{at}$'s to mitigate (while accommodating the worst-case scenarios that the nodes and arcs that cannot be directly attacked by $\mathsf{at} \in \mathsf{AT}_A$ are affected by attack cascading effects). If so, the loop in Lines 12-17 further mitigates the remaining $\mathsf{at}$'s that contributes to the cascading effects. The loop in Lines 18-19 identifies the NIST security controls that can mitigate the $\mathsf{at} \in \mathsf{AT}^*_A - \mathsf{AT}_A$ as identified by the algorithm.

The computational complexity of Algorithm 6 is analyzed as follows. The loop in Lines 6-7 incurs an $O(|\mathsf{AT}_A| \times |V_{\mathsf{infra}}|)$ complexity. The loop in Lines 8-9 incurs an $O(|\mathsf{AT}_A| \times |V_{\mathsf{infra}}|^2)$ complexity, while assuming the number of multi-arcs between a pair of nodes is upper bounded by a constant (which is often true in practice). The loop

**Algorithm 6:** Hardening Missions $j$ $(1 \leq j \leq n)$

**Input:** tolerable mission disruption likelihood threshold $\tau \in [0,1]$; a set $\mathsf{AT}_A$ of attack techniques possessed by attacker $A$, associated with likelihood of possession $\mathsf{L}_{\mathsf{at}} \in (0,1]$ for $\mathsf{at} \in \mathsf{AT}_A$; the likelihood that attack technique $\mathsf{at} \in \mathsf{AT}_A$ can compromise node $v \in V_{\mathsf{infra}}$, denoted by $\beta(v, \mathsf{at})$, or arc $e \in E_{\mathsf{infra}}$, denoted by $\beta(e, \mathsf{at})$; space infrastructure $G_{\mathsf{infra}} = (V_{\mathsf{infra}}, E_{\mathsf{infra}})$; a set of mission control flows $\left\{G_{\mathsf{mcf}}^{(i,j)} = \left(V_{\mathsf{mcf}}^{(i,j)}, E_{\mathsf{mcf}}^{(i,j)}\right)\right\}$ and a set of mission data flows $\left\{G_{\mathsf{mdf}}^{(i',j)} = \left(V_{\mathsf{mdf}}^{(i',j)}, E_{\mathsf{mdf}}^{(i',j)}\right)\right\}$ for $1 \leq i \leq m_j$, $1 \leq i' \leq m'_j$, and $1 \leq j \leq n$; $\{\mathbf{L}(j)\}_{1 \leq j \leq n}$, $\{\mathbf{L}(v)\}_{v \in V_{\mathsf{infra}}}$, $\{\mathbf{L}(e)\}_{e \in E_{\mathsf{infra}}}$ (output of Algorithm 4)

**Output:** security controls for mitigating risks to missions

$\mathsf{AT}_A^* \leftarrow \mathsf{AT}_A$
$J \leftarrow \{j : (1 \leq j \leq n) \text{ and } (\mathbf{L}(j) > \tau)\}$
**if** $J = \emptyset$ **then**
    **return** *"Hardening not necessary"*
**else**
    **for** $v \in V_{\mathsf{infra}}$ **do** `// using updated` $G_{\mathsf{infra}}$ `and` $\mathsf{AT}_A$
        If $\exists \mathsf{at} \in \mathsf{AT}_A$ such that $\beta(v, \mathsf{at}) \times \mathsf{L}(\mathsf{at}) > \tau$ $\mathsf{AT}_A \leftarrow \mathsf{AT}_A - \{\mathsf{at}\}$; $V_{\mathsf{infra}} \leftarrow V_{\mathsf{infra}} - \{v\}$
    **for** $e \in E_{\mathsf{infra}}$ **do** `// using updated` $G_{\mathsf{infra}}$ `and` $\mathsf{AT}_A$
        If $\exists \mathsf{at} \in \mathsf{AT}_A$ such that $\beta(e, \mathsf{at}) \times \mathsf{L}(\mathsf{at}) > \tau$ $\mathsf{AT}_A \leftarrow \mathsf{AT}_A - \{\mathsf{at}\}$; $E_{\mathsf{infra}} \leftarrow E_{\mathsf{infra}} - \{e\}$
    call Algorithm 4 with flag $= 0$ and the updated $G_{\mathsf{infra}}$ and $\mathsf{AT}_A$ to re-compute $\mathbf{L}(j)$ with cascading effects
    $J \leftarrow \{j : (1 \leq j \leq n) \text{ and } (\mathbf{L}(j) > \tau)\}$
    **while** $J \neq \emptyset$ **do**
        **for** $e \in V_{\mathsf{infra}}$ *such that* $\mathbf{L}(e = (u,v)) > \tau$ **do** `// using updated` $G_{\mathsf{infra}}$ `and` $\mathsf{AT}_A$
            $\mathsf{AT}_A \leftarrow \mathsf{AT}_A - \{\mathsf{at} : \beta(u, \mathsf{at}) > 0\}$
            delete $u$ and its adjacent arcs from $G_{\mathsf{infra}}$;
            call Algorithm 4 with flag $= 0$ and updated $G_{\mathsf{infra}}$ and $\mathsf{AT}_A$ to re-compute $\mathbf{L}(j)$ to accommodate cascading effects;
            $J \leftarrow \{j : (1 \leq j \leq n) \text{ and } (\mathbf{L}(j) > \tau)\}$
    **for** $\mathsf{at} \in \mathsf{AT}_A^* - \mathsf{AT}_A(v)$ **do**
        use sub-routine $\Delta$ to identify security control to thwart $\mathsf{at}$, denoted by $\mathsf{sc}(\mathsf{at})$ at the relevant nodes and arcs
    **return** *security controls* $\{\mathsf{sc}(\mathsf{at})\}_{\mathsf{at} \in \mathsf{AT}_A^* - \mathsf{AT}_A}$

in Lines 12 repeats at most $|\mathsf{AT}_A|$ times, with the inner loop in Lines 13-17 for at most $O(|V_{\mathsf{infra}}|^2)$ calls to Algorithm 4, leading to a total of at most $O(|\mathsf{AT}_A| \times |V_{\mathsf{infra}}|^2)$ calls to Algorithm 4. The loop in Lines 18-19 incurs $O(|\mathsf{AT}_A|)$ calls to the sub-routine $\Delta$, which could be implemented with a simple table-querying algorithm (reflecting our manual process in our case study). Thus, the complexity of the algorithm is $O(|\mathsf{AT}_A| \times |V_{\mathsf{infra}}|^2)$ calls to Algorithm 4.

### 6.4.4 Targeted Users and Use Cases

Our framework can be used by the following users: (i) space infrastructure designers, who conceptually identify the system and threat models; (ii) space infrastructure developers, who engineer countermeasures against space cyber threats that can impact space missions; (iii) space infrastructure defenders, who manage and operate cyber defenses during the operation phase for the space infrastructure; (iv) space cyber risk analysis and mitigation team, who plan space missions and assure their success; (v) researchers who want to improve the framework; (vi) educators who want to teach space cyber risk analysis and mitigation in their cybersecurity curricula; and (vii) trainers who want to train professionals whose job duties involve space cyber risk analysis and mitigation.

Example use cases of our framework include: (i) identifying and mitigating *potential* space cyber risks to space infrastructures and missions; (ii) identifying and mitigating *actual* space cyber risks to real-world space infrastructure and missions, such as those imposed by identified vulnerabilities; (iii) simulating or emulating space cyber risks in a testbed for a hypothetic or real-world space infrastructure. In any use case, one can always set $\mathsf{L}(\mathsf{at}) = 1$ for $\mathsf{at} \in \mathsf{AT}_A$ to accommodate the worst-case scenario.

Our framework can leverage the input to SPARTA's NRS [139] as follows. NRS's base risk scores, which are provided by domain experts as described in Line 3 of the NRS algorithm [22], can serve the purpose of our $\mathsf{L}_{\mathsf{at}}$'s as described in Algorithm 4. Our framework goes much beyond NRS because our Algorithms 4-6 consider a space infrastructure as a whole, while NRS deals with individual space systems or modules without considering the cascading effects of successful attacks.

## 6.5 Case Study

To demonstrate the usefulness of the framework, we now conduct a case study based on our SATCOM (satellite communications) testbed. Note that SATCOM is representative as our study in Chapter III shows that 58 real-world space cyber attacks (out of the 108 reported therein) target SATCOM missions. Our experiments include re-implementing three (out of the 108) attacks in our testbed and mitigating them via the security controls identified by Algorithm 6.

### 6.5.1 Our SATCOM Testbed

Figure 6.2 highlights our testbed, which includes four segments (i.e., space, link, ground, and user). Its components and modules are elaborated below.

Figure 6.3 describes the $G_{\mathsf{infra}}$ corresponding to our testbed at the module-level, with $|V_{\mathsf{infra}}| = 19$ modules in total. The modules are named as follows: SM.NAME

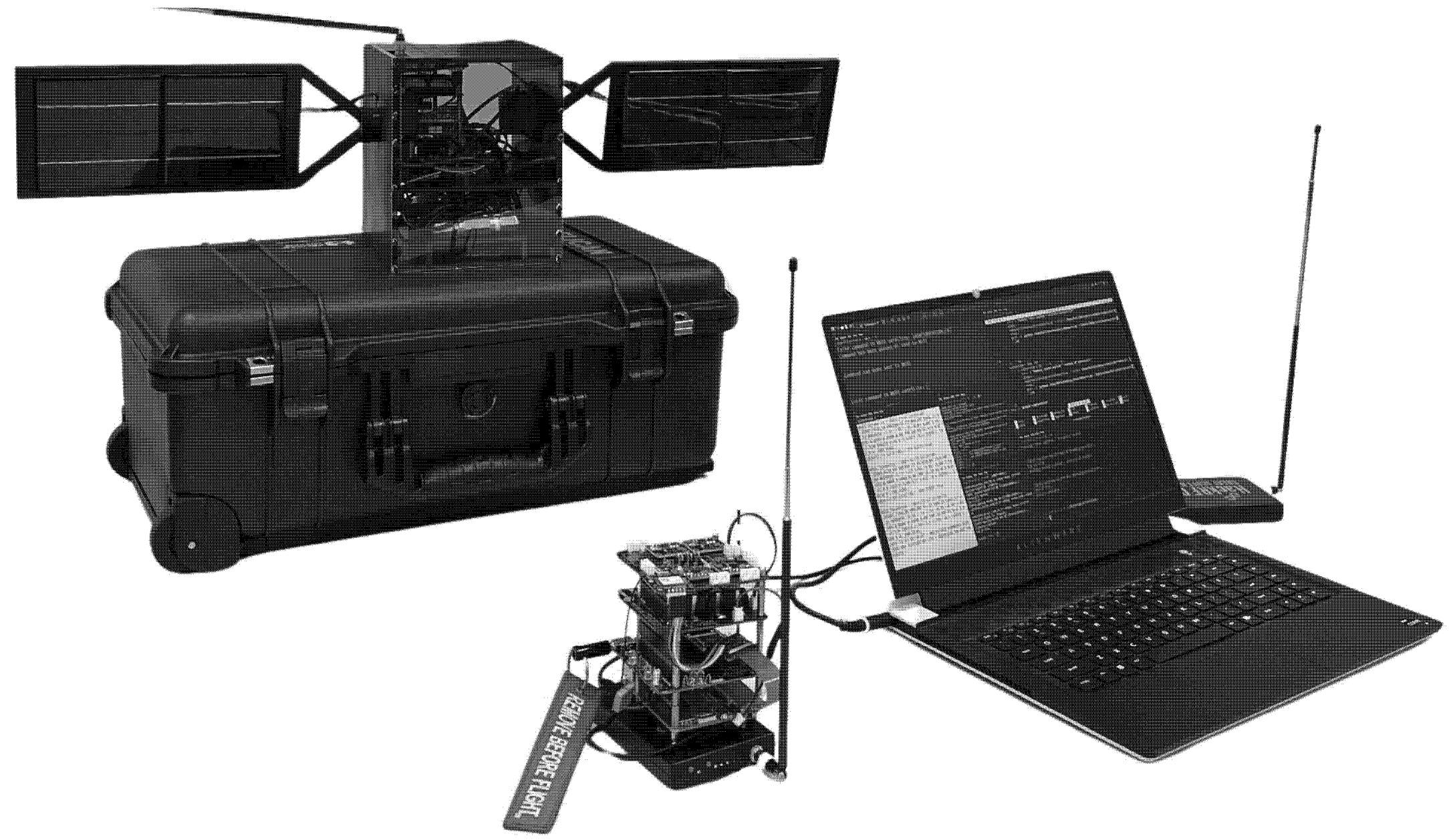

Figure 6.2: Our SATCOM space infrastructure testbed.

denotes a module in the Space segment, GM.NAME denotes a module in the Ground segment, and UM.NAME denotes a module in the User segment. The space segment consists of two components: bus system (six modules) and payload (one module); the payload or application mission is to enable two users to communicate with each other via the satellite, as described in [150]. The ground segment consists of four components: ground station (two modules), mission control (three modules), data processing (two modules), and remote terminal (two modules). The user segment has one component: user (three modules). The link segment is indicated by bold arrows. The modules indicated with an '*' are involved in the aforementioned 58 real-world space cyber attacks; these nine modules are also involved in the mission that is used in our case study. Among the 19 modules, 15 are fully implemented and the other four are emulated (highlighted in light blue color) because these four involve proprietary hardware and software that we do not have access to at the time of writing.

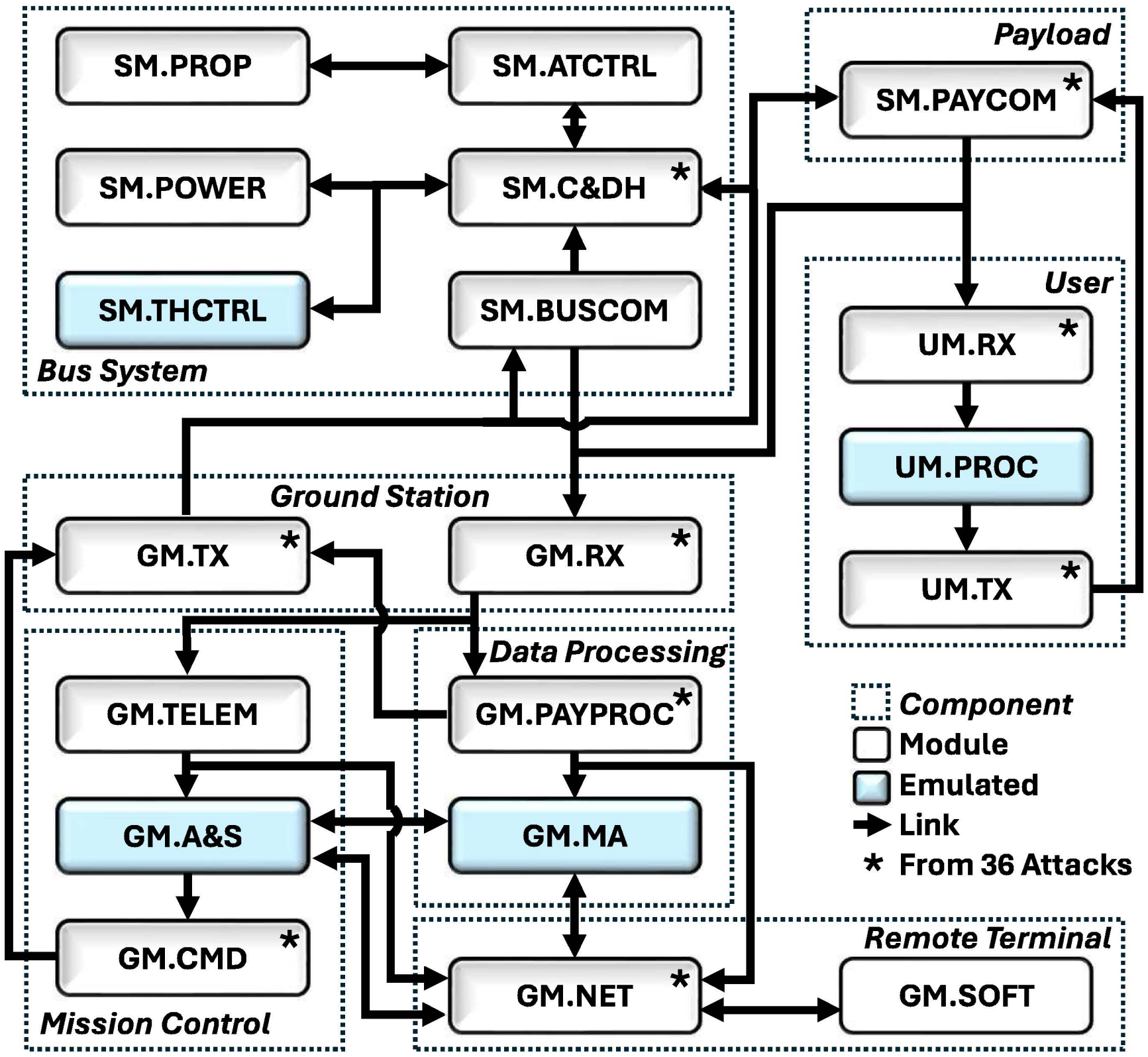

Figure 6.3: $G_{\text{infra}}$ of our space testbed, with $|V_{\text{infra}}| = 19$ modules.

The 15 fully implemented modules are briefly described as follows. In the ground segment, we employ Ball Aerospace's Comprehensive Open-architecture Solution for Mission Operations Systems (COSMOS) ground software and a Debian OS on an Intel i7 processor to implement GM.TELEM (Telemetry), GM.PAYPROC (Payload Processing), and GM.CMD (Command). We employ Netgate's pfSense appliance and an OpenVPN service to implement GM.NET (Network Access) and GM.SOFT (Software Access). In the Space segment, we employ NASA's core Flight System (cFS) flight software and a Debian OS on an ARM Cortex-A76 processor to implement SM.C&DH (Command & Data Handling), while implementing SM.ATCTRL (Attitude Control) via custom firmware and an ARM Cortex-M0+ processor on a RP2040 microcontroller. We implement SM.PROP (Propulsion) with an electric relay, solenoid valve, $CO_2$ canister, and exhaust ports. We implement SM.POWER (Electrical Power) with electric relays, servomotors, solar panel arrays, Texas Instruments' INA219 current and voltage sensors, and batteries. In the Ground, Space, and User segments, we use GNU Radio libraries and Great Scott Gadgets' HackRFs to implement GM.TX (Ground Transmission), GM.RX (Ground Reception), UM.TX (User Transmission), UM.RX (User Reception), SM.PAYCOM (Payload Communications), and SM.BUSCOM (Bus Communications).

The other four modules are emulated as follows. In the Ground segment, we emulate GM.A&S (Analysis & Support) and GM.MA (Mission Analysis) with Bash and Python scripts for input/output interactions. In the Space segment, we emulate SM.THCTRL (Thermal Control) by producing ambient temperature, barometric pressure, and relative humidity measurements with Bosch Sensortec's BME280 sensor. In the User segment, we emulate UM.PROC (User Processing) with Bash and python scripts.

### 6.5.2 Implementing the SATCOM Mission

The testbed currently supports the SATCOM mission, meaning $n = 1$. As highlighted in Figure 6.4, the mission has $m_1 = 3$ control flows and $m'_1 = 2$ data flows; orange nodes belong to the control flows, purple nodes belong to the data flows, yellow nodes belong to both control flows and data flows, orange arcs belong to the control flows, and purple arcs belong to the data flows.

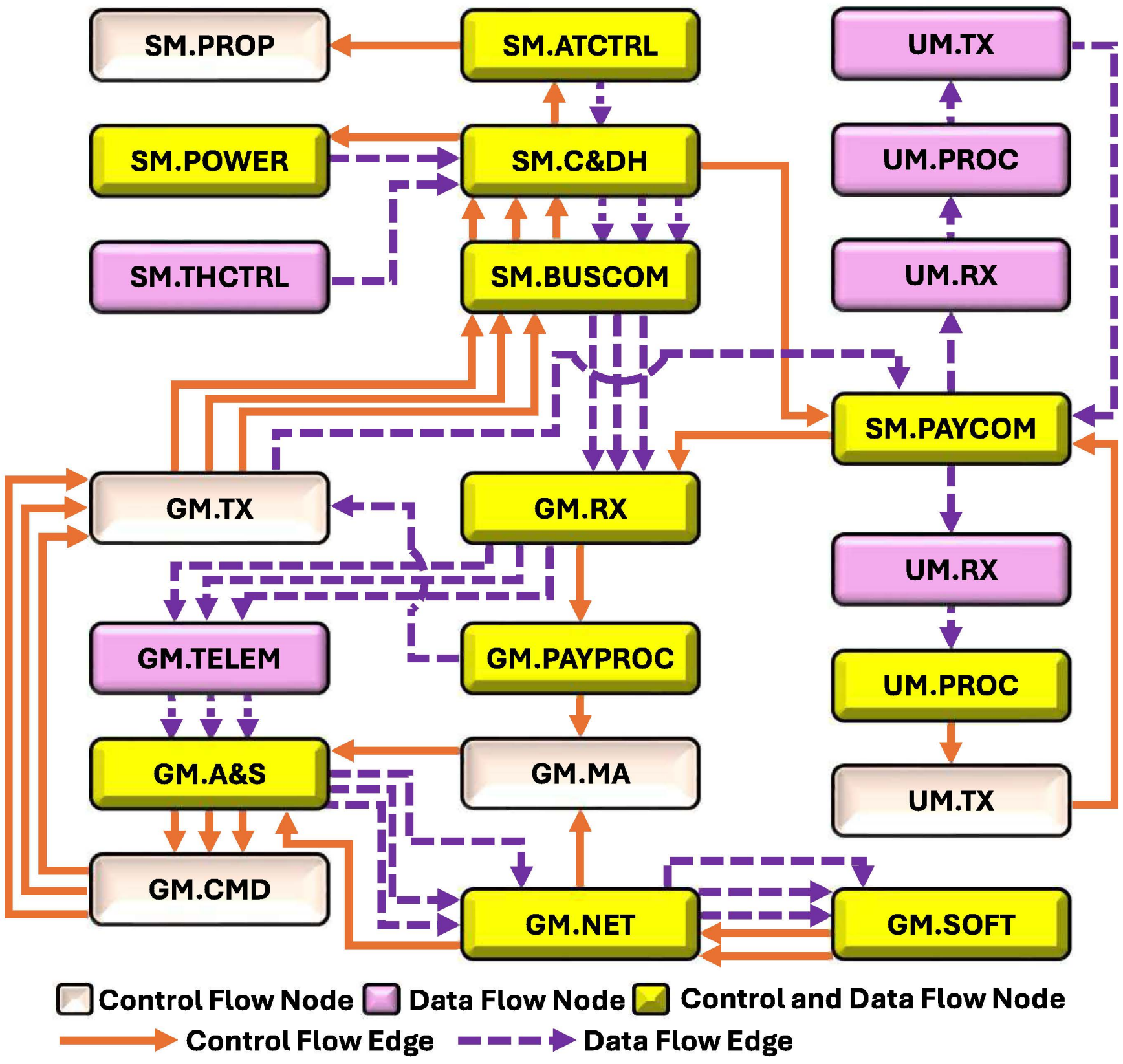

Figure 6.4: $\left(\bigcup_{i=1}^{3} G_{\text{mcf}}^{(i,1)}\right) \cup \left(\bigcup_{i'=1}^{2} G_{\text{mdf}}^{(i',1)}\right)$ of the SATCOM mission with $n = 1$, $m_1 = 3$ control flows and $m'_1 = 2$ data flows.

#### 6.5.2.1 Bus Management Control Flow $G_{\text{mcf}}^{(1,1)}$

As highlighted in Figure 6.5 depicts, this control flow allows for commanding the modules of the bus component of the satellite. More specifically, this control flow is for purposes that include: (a) maneuvering the satellite to modify the pitch, roll, or yaw of the satellite to ensure optimal communications antenna position, or to maintain / correct / change the satellite's orbit to meet the SATCOM's mission requirements; and (b) managing the electrical power distribution of the satellite, such as powering off certain components to preserve power or controlling the direction of the solar panel arrays.

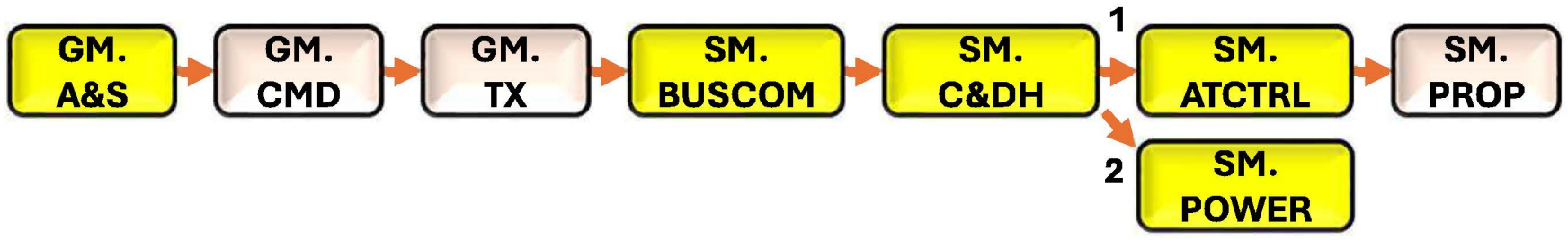

Figure 6.5: Bus Management Mission Control Flow $G_{\mathsf{mcf}}^{(1,1)}$.

For purpose (a), the control flow begins at the Analysis & Support module GM.A&S of the Mission Control component and ends at the Propulsion module SM.PROP of the Bus component. It progresses as follows: (i) GM.A&S determines whether a change in the satellite's pitch, roll, yaw, or orbit is required, and if so, sends the required state change (of the Propulsion module SM.PROP) to the Command module GM.CMD; (ii) GM.CMD interprets and encodes the change requirement into the Consultative Committee for Space Data Systems (CCSDS) protocol, and sends the command data packet to the Transmission module GM.TX; (iii) GM.TX modulates the command data packet for Radio Frequency (RF) transfer to the Communications module SM.BUSCOM of the Bus component; (iv) SM.BUSCOM demodulates the RF signal to retrieve the command data packet and forwards it to the Command & Data Handling module SM.C&DH; (v) SM.C&DH parses, validates, and translates the command data packet to command directives for the Attitude Control module SM.ATCTRL; (vi) SM.ATCTRL parses and executes the command directives to send the corresponding electric signals to the Propulsion module SM.PROP; and (vii) SM.PROP receives the electric signals and actuates the solenoid vales to control propellant discharge to achieve the desired satellite maneuver.

For purpose (b), GM.A&S determines a change in power distribution or solar panel direction is required. Eventually, the Command & Data Handling module SM.C&DH pushes the power re-configuration command onto the flight software bus to make the Power module SM.POWER to retrieve the command and execute it to manage the satellite's electrical power.

#### 6.5.2.2 Remote Management Control Flow $G_{\text{mcf}}^{(2,1)}$

As depicted in Figure 6.6, this control flow allows operators to remotely control the ground segment when away from the main Mission Control and Data Processing facilities. More specifically, it is for purposes that include: (a) for operators to establish, modify, or delete mission analysis parameters to ensure SATCOM mission success, such as directing the collection of channel saturation levels; and (b) for operators to establish, modify, or delete infrastructure analysis parameters to ensure optimal satellite health, such as setting threshold levels for triggering sensor alerts.

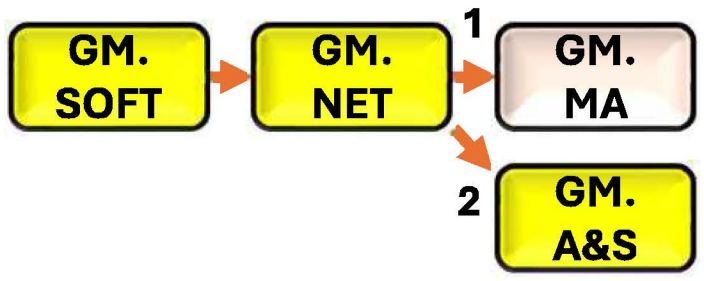

Figure 6.6: Remote Management Mission Control Flow $G_{\text{mcf}}^{(2,1)}$.

For purpose (a), the control flow begins at the Software Access module GM.SOFT and ends at the Mission Analysis module GM.MA. It progresses as follows: (i) the GM.SOFT module enables an operator to compose and transmit configuration directives through a client interface that is connected to the GM.NET module via a VPN service that traverses commercial networks; (ii) the Network Access module GM.NET transmits the remote operator's configuration directives to the GM.MA module on a local LAN; and (iii) the GM.MA module receives and executes the new configuration directives to modify its analysis parameters for the SATCOM mission.

For purpose (b), the control flow also begins at GM.SOFT but ends at the Analysis & Support module GM.A&S. Eventually, the GM.A&S module receives the configuration directives from the Network Access module GM.NET and executes it to modify its analysis parameters for the SATCOM infrastructure.

#### 6.5.2.3 Subscription Management Control Flow $G^{(3,1)}_{\mathsf{mcf}}$

As depicted in Figure 6.7, this control flow is for managing the channel subscription configuration of the satellite; more specifically, it allows SATCOM users to establish, modify, or delete their channel subscription, such as increasing the bandwidth of current subscription. It contains 12 nodes, beginning at the User Processing module UM.PROC and ending with the Payload Communications module SM.PAYCOM. It progresses as follows: (i) UM.PROC receives a users' subscription change request and forwards it to the Transmission module UM.TX; (ii) UM.TX sends the change request to SM.PAYCOM via RF; (iii) SM.PAYCOM acts as a repeater, forwarding the request to the Ground Reception module GM.RX via RF; (iv) GM.RX transmits the request to the Ground Payload Processing module GM.PAYPROC; (v) GM.PAYPROC attaches the context of the users' current subscription to the change request and forwards it to the Mission Analysis module GM.MA; (vi) GM.MA assesses if the change is allowed within the mission context, and if so, sends the request to the Analysis & Support module GM.A&S; (vii) GM.A&M assesses if the change is allowed, and if so, sends the request to the Command module GM.CMD; (viii) GM.CMD converts the change request into standard CCSDS commands and forwards the commands to the Transmission module GM.TX; (ix) GM.TX modulates the command data packet and sends the signal to the Communications module SM.BUSCOM via RF; (x) SM.BUSCOM demodulates the RF signal and forwards the command data packet to the Command & Data Handling module SM.C&DH; (xi) SM.C&DH parses and validates the command data packet and sends the command to the Payload Communications module SM.PAYCOM; and (xii) SM.PAYCOM executes the commands to modify its channel subscription hardware and software configuration to meet the users' requirements.

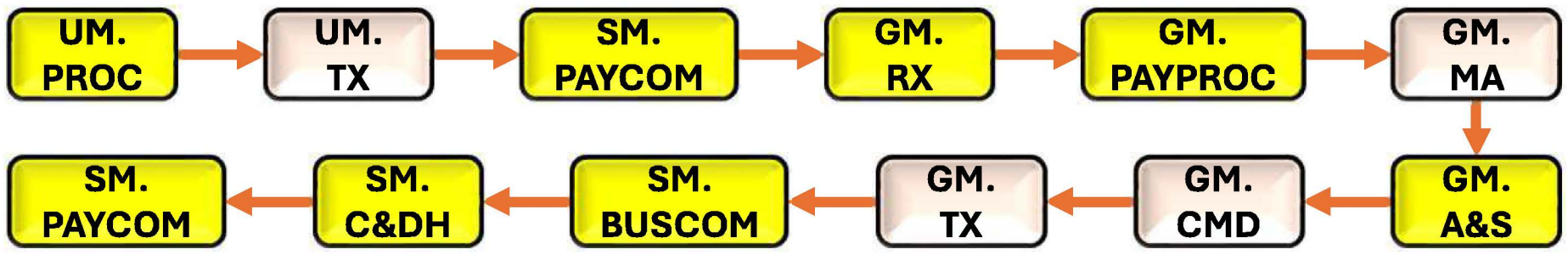

Figure 6.7: Subscription Mgmt Mission Control Flow $G_{\mathsf{mcf}}^{(3,1)}$.

#### 6.5.2.4 Telemetry Reporting and Analysis Data Flow $G_{\mathsf{mdf}}^{(1,1)}$

As depicted in Figure 6.8, allows for transmitting satellite health and status data to operators in the Ground segment. More specifically, it serves three purposes: (a) transmitting sensor data about the pitch, roll, yaw of the satellite, as well as its propellant status (e.g., positions of relays, solenoid valves, and rates of propellant releases) to satellite operators at the Analysis & Support module GM.A&S or the Software Access module GM.SOFT; (b) transmitting sensor data about the satellite's electric power (e.g., currents and voltages) and the servomotor angle position for the solar panel ar-rays; and (c) transmitting sensor data about the satellite's thermal environment, such as the temperature of the flight computer.

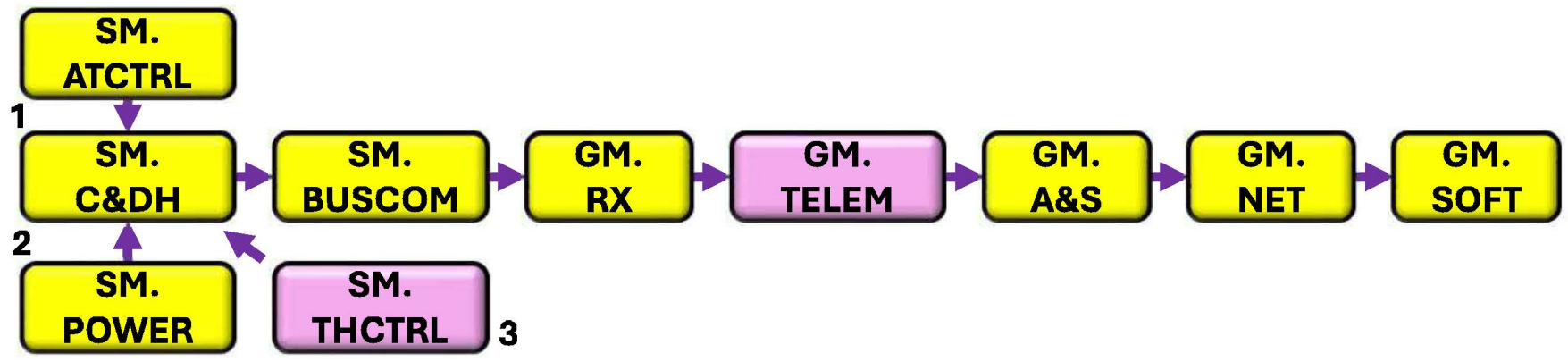

Figure 6.8: Telemetry Mission Data Flow $G_{\mathsf{mdf}}^{(1,1)}$.

For purpose (a), the data flow begins at the Attitude Control module SM.ATCTRL and ends at the Analysis & Support module GM.A&S. It progresses as follows: (i) SM.ATCTRL pushes its sensor data onto the flight software bus; (ii) the Command & Data Handling module SM.C&DH periodically retrieves the sensor data, packs them as an archive, and sends the archive to the Communication module SM.BUSCOM; (iii) SM.BUSCOM modulates the archive and sends the data signal via RF to the Ground Reception module GM.RX; (iv) GM.RX receives and demodulates the RF signal into the data archive and forwards it to the Telemetry module GM.TELEM; (v) GM.TELEM unpacks the data archive and populates the sensor data on its dashboards to monitor real-time satellite health and then forwards the sensor data to the Analysis & Support module GM.A&S; (vi) GM.A&S analyzes the sensor data to determine trends in satellite health and forwards the data to the Network Access module GM.NET if there is a remote operator; (vii) GM.NET transmits the sensor data through its VPN service to the Software Access module GM.SOFT; and (viii) GM.SOFT displays the sensor data in a client interface for the remote operator to examine.

For purpose (b), it begins at the Power module SM.POWER and ends at the Software Access module GM.SOFT. Specifically, SM.POWER pushes its sensor data onto the flight software bus where the Command & Data Handling module SM.C&DH periodically retrieves the sensor data. From there, the process for the remaining seven nodes is the same as for purpose (a).

For purpose (c), it begins at the Thermal Control module SM.THCTRL and ends at the Software Access module GM.SOFT. Specifically, SM.THCTRL pushes its sensor data onto the flight software bus. From there, the process for the remaining seven nodes is the same as for purpose (a).

#### 6.5.2.5 Channel Operations Data Flow $G_{\text{mdf}}^{(2,1)}$

As depicted in Figure 6.9, this data flow is for transmitting data from its source to end-users. Specifically, this control flow serves two purposes: (a) transmitting communications data from one user to another user via a satellite to accomplish beyond-line-of-sight voice and data communications; and (b) transmitting communication data from the Ground segment to end-users (e.g., satellite television broadcasting).

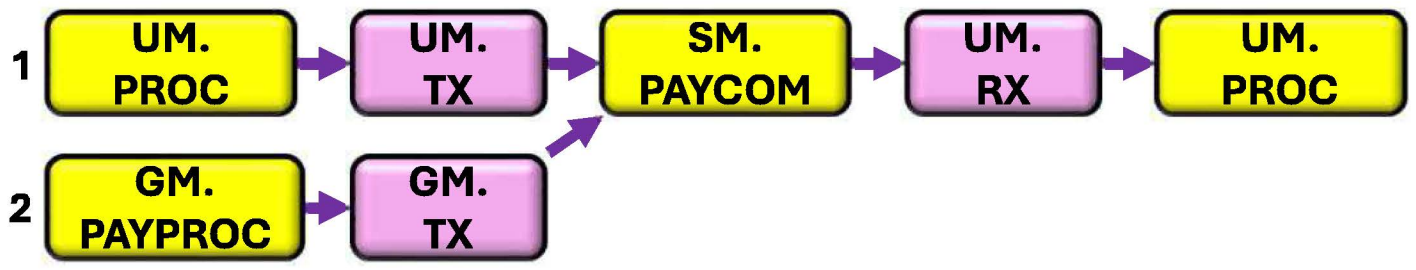

Figure 6.9: Channel Operations Mission Data Flow $G_{\text{mdf}}^{(2,1)}$.

For purpose (a), the control flow begins at one user's Processing module (i.e., UM.PROC) and ends at another user's UM.PROC. It progresses as follows: (i) the sending user's UM.PROC module receives the voice or data communications and packs the byte stream according to the CCSDS packet structure; (ii) the sending user's Transmission module UM.TX modulates the data packets and transmits the data signal via RF to the Bus Communications module SM.BUSCOM; (iii) SM.BUSCOM receives the data signal and acts as a repeater by re-transmitting the data signal to the receiving user's Reception module UM.RX; (iv) the receiving user's UM.RX demodulates the data signal and sends the resulting data to the receiving user's Processing module UM.PROC; and (v) the receiving user's UM.PROC presents the communications data to the end-user via a client terminal interface.

For purpose (b), the control flow begins at the Ground Payload Processing module GM.PAYPROC and ends at the User Processing module UM.PROC. Specifically, the GM.PAYPROC module packages the communications data according to the CCSDS standard and the Ground Transmission module GM.TX module modulates the packets into a data signal. From there, the remaining process is the same as for purpose (a).

### 6.5.3 Specifying Space Cyber Attacks

We consider three (out of the 108) real-world space cyber attacks reported in Chapter III, by leveraging our testbed to show that they can succeed *before* employing our mission hardening and they are largely thwarted by our mission hardening Algorithm 6 to render $\mathbf{L}(j) < \tau$. These three attacks collectively use 10 attack techniques, meaning $|\mathsf{AT}_A| = 10$.

#### 6.5.3.1 Attacker Capabilities

The 10 attack techniques are $\mathsf{AT}_A = \{$T1210, T1199, T1595, T1592, T1566.001, EX-0012, EX-0009.03, IA-0007.02, IA-0008.01, REC-0005.02$\}$, with associated likelihood of possession by $A$, namely $\mathsf{L}_{\mathsf{at}} = \{0.23, 0.38, 0.38, 0.15, 0.25, 0.24, 0.23, 0.27, 0.23, 0.23\}$. The 10 attack techniques are elaborated as follows: (i) *Downlink Intercept* (SPARTA REC-0005.02), whereby an attacker eavesdrops on data signal broadcasts from a satellite; (ii) *Rogue Ground Station* (SPARTA IA-0008.01), whereby an attacker injects their own ground station into a space infrastructure; (iii) *Modify On-Board Values* (SPARTA EX-0012), whereby an attacker modifies either stored data or the configuration settings of modules in the space segment to alter or disrupt satellite behavior; (iv) *Gather Victim Host Information* (ATT&CK T1592), whereby an attacker identifies important information about a victim host, such as operating system version; (v) *Spearphishing Attachment* (ATT&CK T1566.001), whereby an attacker convinces a victim to click on a malicious file; (vi) *Active Scanning* (ATT&CK T1595), whereby an attacker identifies important information about a victim network, such as open ports of connected hosts; (vii) *Exploitation of Remote Services* (ATT&CK T1210), whereby an attacker gains remote access through (e.g.) a misconfigured service; (viii) *Trusted Relationship* (ATT&CK T1199), whereby an attacker leverages the established trust relationship between modules to gain access; (ix) *Malicious Commanding* (SPARTA IA-0007.02), whereby an attacker transmits a satellite command through a valid ground station; and (x) *Exploit Code Flaw: Known Vulnerability* (SPARTA EX-0009.03), whereby an attacker exploits an unpatched software.

To determine $\beta(v, \mathsf{at})$ and $\beta(e, \mathsf{at})$, we leverage our domain knowledge in the same fashion as the SPARTA NRS [139]. Figure 6.10 highlights which $v \in V_{\mathsf{infra}}$ and $e \in E_{\mathsf{infra}}$ can be directly attacked by which $\mathsf{at} \in \mathsf{AT}_A$. Likelihood $\beta(*, \mathsf{at})$ is annotated with each $v$ and $e$, while noting $\beta(v, *) = 0$ and $\beta(e, *) = 0$ for all other $v$'s and $e$'s. Specifically, we have $\beta$(GM.A&S, T1210) $= 0.11$, $\beta$(GM.CMD, T1199) $= 0.33$, $\beta$(GM.TX, T1199) $= 0.11$, $\beta$(GM.NET, T1595) $= 0.40$, $\beta$(GM.SOFT, T1592) $= 0.11$, $\beta$(GM.SOFT, T1566.001) $= 0.07$, $\beta$(SM.C&DH, EX-0012) $= 0.20$, $\beta$(SM.C&DH, EX-0009.03) $= 0.17$, $\beta$(SM.PAYCOM, IA-0007.02) $= 0.45$, $\beta$(GM.TX$\rightarrow$SM.BUSCOM, IA-0008.01) $= 0.33$, $\beta$(SM.BUSCOM$\rightarrow$GM.RX, REC-0005.02) $= 0.33$, $\beta$(SM.PAYCOM $\rightarrow$UM.RX, REC-0005.02) $= 0.33$, and $\beta$(SM.PAYCOM$\rightarrow$UM.RX, IA-0008.01) $= 0.33$. Note that the annotation $(0)$ on arc $e$ means $\beta(e, \mathsf{at}) = 0$ for any $\mathsf{at} \in \mathsf{AT}_A$, meaning that $e$ cannot be directly attacked by any of the 10 techniques but is still relevant because $e$ can be compromised via cascading effects. Similarly, the annotation $(0)$ on node $v$ (e.g., UM.RX) means that $\beta(v, \mathsf{at}) = 0$ for any $\mathsf{at} \in \mathsf{AT}_A$, namely that $v$ cannot be attacked by any of the 10 techniques but can be compromised via cascading effects.

#### 6.5.3.2 Attack 1: RF-based Attack

This attack corresponds to the 2007 space cyber attack, originally reported in [43], that hijacked a Czech satellite TV channel to transmit its own audio and video. As highlighted by the dashed brown arcs in Figure 6.10, this attack employs two attack techniques and progresses as follows: (i) the attacker targets the arc SM.PAYCOM$\rightarrow$UM.RX, where the attacker intercepts legitimate downlink broadcasts (SPARTA REC-0005.02) of an Israeli satellite TV channel to identify the properties of the legitimate data signal; and (ii) the attacker targets the same arc to generate the attacker's own broadcast content via its rogue ground station (IA-0008.01) to overpower the legitimate signal.

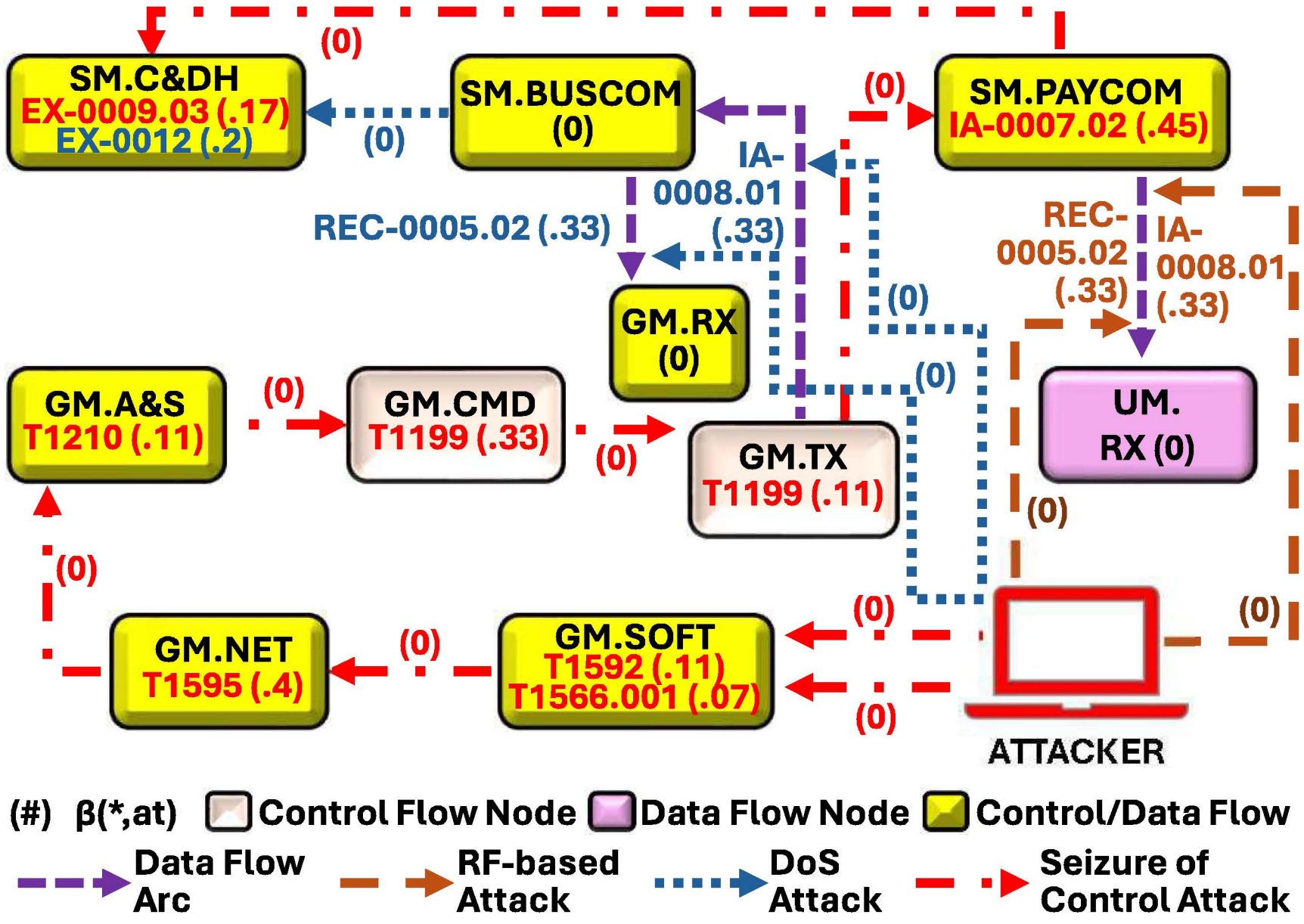

Figure 6.10: The 10 attack techniques in $AT_A$ employed by the three attacks against modules and links.

We repeat this attack in our testbed as follows. (i) We equip our attacking terminal with a HackRF, which is programmed to scan through Multi-Use Radio Service (MURS) frequencies, and position the terminal in the vicinity of the legitimate user terminal to intercept downlink signals (SPARTA REC-0005.02); upon receipt of a valid signal, we analyze the signal to understand the modulation, data packing, and transfer protocols used, while assuming the communications data is not encrypted (which is not uncommon in the space domain [152, 169]). (ii) We construct a competing signal and broadcast it (SPARTA IA-0008.01) at a gain level appropriate to overpower the legitimate satellite (SM.PAYCOM), while not over-saturating the victim user terminal (UM.RX).

#### 6.5.3.3 Attack 2: DoS Attack

This attack corresponds to the 1998 incident, originally reported in [43], that caused the on-board processor of the PANAMSAT Galaxy 4 SATCOM satellite to malfunction, leading to a four-day disruption of 90% of US pagers. As highlighted by the dotted blue arcs in Figure 6.10, this attack employs three attack techniques and progresses as follows: (i) the attacker targets the arc SM.BUSCOM→GM.RX, by intercepting and identifying the legitimate data signals (SPARTA REC-0005.02) from the Galaxy 4 satellite; (ii) the attacker uses its rogue ground station (SPARTA IA-0008.01) to send a malicious command to the SM.BUSCOM module; and (iii) SM.BUSCOM automatically forwards the malicious command to the SM.C&DH module, which parses the command and executes it, corrupting its on-board configuration (SPARTA EX-0012) and leading to a halt of the flight computer.

We repeat the DoS Attack in our testbed as follows. (i) We intercept the downlink (SPARTA REC-0005.02) at SM.BUSCOM→GM.RX. (ii) We craft a malicious flight command to trigger a memory access violation in the flight software and use the attacker terminal's HackRF as a rogue ground station (SPARTA IA-0008.01) to transmit the signal to SM.BUSCOM. (iii) SM.BUSCOM automatically places the malicious command on the flight software bus, where SM.C&DH retrieves, parses, and executes the malicious command (SPARTA EX-0012) to cause (e.g.) a buffer overflow, resulting in a crash of the flight software.

#### 6.5.3.4 Attack 3: Seizure of Control Attack

This attack corresponds to the 2008 space cyber attack, originally reported in [43], that used a malware to exploit an unpatched module in the Ground segment at the Johnson Space Center; the malware further leverages an uplink to gain control of the International Space Station (ISS) in the Space segment. As highlighted by the dashed and dotted red arcs in Figure 6.10, this attack employs seven attack techniques and progresses as follows: (i) the attacker employs ATT&CK technique T1592 to target the GM.SOFT module, where the attacker gains important information about the client computer used (e.g., operating system version); (ii) the attacker convinces the GM.SOFT user to execute a malicious file, such as malicious email attachment (ATT&CK T1566.001) to entice the user; (iii) the malware allows the attacker to scan (ATT&CK T1595) GM.NET for additional hosts with open services; (iv) the attacker uses the scan results to exploit a remote service (ATT&CK T1210) in GM.A&S, such as an overly permissive Samba service; (v) the attacker laterally moves to GM.CMD by exploiting its trust relationship (ATT&CK T1199) with GM.A&S and prepositions the malware within a transmission request; (vi) the attacker continues to laterally move to GM.TX by again exploiting its trust relationship (ATT&CK T1199) with GM.CMD, embedding the malware in a command packet archive; (vii) the attacker conducts a malicious command transmission (SPARTA IA-0007.02) from GM.TX to SM.PAYCOM to spread the malware to the ISS; and (viii) SM.PAYCOM pushes the command packet archive to the flight software bus where SM.C&DH unpacks, parses, and executes the command, triggering the malware to exploit a known vulnerability (SPARTA EX-0009.03) at SM.C&DH, such as a software misconfiguration, to gain control of SM.C&DH's services.

We repeat this attack in our testbed as follows. (i) We gain information about the operating system of the Software Access module GM.SOFT (ATT&CK T1592) by enticing the space operator using the terminal to visit a malicious website. (ii) We employ MSFvenom to embed a reverse shell in a space calculator utility to entice the space operator to download and execute our malware (ATT&CK 1566.001). (iii) With remote access to GM.SOFT, we employ Nmap to scan the Network Access module GM.NET through the VPN connection. (iv) We discover that the Analysis & Support module (GM.A&S) runs a custom web server that is vulnerable to command injection and exploit this vulnerability (ATT&CK T1210) to gain access to GM.A&S. (v) We design a beacon-based Command and Control (C2) bash script that piggybacks off the established Bus Management Mission Control Flow (Figure 6.5) and the Telemetry Mission Data Flow (Figure 6.8) and embed it in a command to use the script execution function in the flight software and place the command as a file in a directory on GM.A&S, where the Command module GM.CMD trusts (ATT&CK T1199) and pulls files from GM.A&S' shared directory to construct satellite commands. (vi) Similarly, the Ground Transmission module GM.TX automatically receives the constructed command because it trusts GM.CMD (ATT&CK T1199). (vii) We accomplish our malicious command transmission (SPARTA IA-0007.02) because the Payload Communications module SM.PAYCOM automatically receives the signal broadcast from GM.TX. (viii) SM.PAYCOM pushes our command onto the flight software bus, where the Command & Data Handling module SM.C&DH retrieves and executes our command, which exploits the flight software's lack of input validation (SPARTA EX-0009.03), leading to that our C2 script implants itself in the SM.C&DH module with root privileges. Our C2 script beacons back to GM.A&S every time the satellite is within range of the Ground Station.

### 6.5.4 Analyzing Mission Risks

We use Algorithms 4 and 5 to assess the SATCOM mission disruption likelihood, with the attack capabilities described in Section 6.5.3.1, $G_{\mathsf{infra}}$ as described in Figure 6.3, $G^{(i,1)}_{\mathsf{mcf}}$ for $1 \leq i \leq m'_1 = 3$ and $G^{(i',1)}_{\mathsf{mdf}}$ for $1 \leq i' \leq m'_1 = 2$ as described in Figures 6.5-6.9, respectively.

We run Algorithm 4 on an Alienware x15 R2 with an Intel i9-12900H processor and 32GB RAM. We measure the execution time until the update on each element of the vector $(\mathbf{L}(v))_{v \in V_{\mathsf{infra}}}$ is less than or equal to $\epsilon = 1^{-10}$.

- Case 0 ($\mathsf{flag} = 0$): The nodes and arcs that cannot be attacked by any $\mathsf{at} \in \mathsf{AT}_A$ are vulnerable to cascading effects. In this case, Lines 16-17 are *not* executed and the results show that $\mathbf{L}(v) > \tau$ for all 19 modules in $V_{\mathsf{infra}}$ and $\mathbf{L}(e) \to 1$ for $e \in E_{\mathsf{infra}}$. This is expected because the lack of defense renders every vulnerable node and arc eventually compromised. The measured execution time is 90 milliseconds, while noting that it only takes 42 iterations for the computation in the **repeat** loop (Line 18) to converge.
- Case 1 ($\mathsf{flag} = 1$): The nodes and arcs that cannot be attacked by any $\mathsf{at} \in \mathsf{AT}_A$ are considered not vulnerable to cascading effects, meaning that they all can be disregarded in the subsequent space cyber risk analysis. In this case, Lines 16-17 are executed. The results show all 10 mission-required modules in the resulting $V_{\mathsf{infra}}$ have $\mathbf{L}(v) > \tau$ and $\mathbf{L}(e) \to 1$ for the arcs in the resulting $E_{\mathsf{infra}}$. This is also expected because the lack of defense renders every vulnerable node and arc eventually compromised. The measured execution time is also 90 milliseconds, while it only takes 36 iterations for the computation in the **repeat** loop (Line 18) to converge. Note that for $e = (u, v) \in G_{\mathsf{infra}}$, $\mathbf{L}(u) \leq \mathbf{L}(e)$ is possible because $e$, in addition to be affected by the compromise of $u$, is also subject to attacks that directly target $e$ without compromising $u$.

To see the impact of cascading effects, we observe that in Case 0, there are 16 (out of the 19) modules with $\mathbf{L}(v) < \tau$ and 35 (out of the 36) arcs with $\mathbf{L}(e) < \tau$ *before* considering cascading effects (Lines 18-23); however, cascading effects result in $\mathbf{L}(v) > \tau$ for all 19 nodes and $\mathbf{L}(e) \to 1$ for all 36 arcs. In Case 1, there are seven (out of the 10) nodes with $\mathbf{L}(v) < \tau$ and 13 (out of the 14) arcs with $\mathbf{L}(e) < \tau$ *before* considering cascading effects (Lines 18-23); cascading effects result in $\mathbf{L}(v) > \tau$ for all 10 nodes and $\mathbf{L}(e) \to 1$ for all 14 arcs. This further implies that $A$ using $G_{\mathsf{infra}}$ to guide its attacks (i.e., effectively Case 0) causes more damage than using the $G_{\mathsf{mcf}}$'s and/or $G_{\mathsf{mdf}}$'s to guide its attacks, because $G_{\mathsf{infra}}$ offers more opportunities in leveraging the cascading effects to compromise nodes and arcs with eventual impact on the $G_{\mathsf{mcf}}$'s and/or $G_{\mathsf{mdf}}$'s.

**Insight 9.** *Cascading effects offer attacker flexibilities in disrupting missions, especially when attackers leverage the entire infrastructure to wage attacks (in contrast to only leveraging the target control flows and data flows).*

Insight 9, while sounding intuitive, has an important implication: defenders should secure the modules that are not part of any mission control / data flows because attackers can pivot from them to attack the control / data flows of targeted missions.

### 6.5.5 Hardening Space Missions

As discussed above and summarized in Figure 6.11, Attack 1 (Section 6.5.3.2) uses two techniques to disrupt mission data flow $G_{\mathsf{mdf}}^{(2,1)}$; Attack 2 (Section 6.5.3.3) uses three techniques to disrupt $G_{\mathsf{mcf}}^{(1,1)}$, $G_{\mathsf{mcf}}^{(3,1)}$ and $G_{\mathsf{mdf}}^{(1,1)}$; Attack 3 (Section 6.5.3.4) uses seven techniques to disrupt $G_{\mathsf{mcf}}^{(1,1)}$, $G_{\mathsf{mcf}}^{(2,1)}$, $G_{\mathsf{mcf}}^{(3,1)}$, $G_{\mathsf{mdf}}^{(1,1)}$ and $G_{\mathsf{mdf}}^{(2,1)}$. In what follows we show how Algorithm 6, using the same parameters as in Section 6.5.4, to identify some of the 10 the attack techniques used by the three attacks for mitigation, rendering the mission disruption likelihood to below $\tau = 0.1$. Also as summarized in Figure 6.11, in Case 0, Algorithm 6 identifies eight (out of the 10) at's to mitigate, reducing the mission disruption likelihood to $\mathbf{L}(j) = 0.04 < \tau = 0.1$; in Case 1, the algorithm identifies five (out of the 10) at's to mitigate, reducing the mission disruption likelihood $\mathbf{L}(j) = 0.08 < \tau = 0.1$. We observe that the five at's identified in Case 1 are a subset of the eight at's in Case 0; it is an outstanding open problem to prove (or disprove) this is universally true.

We manually identify security controls to mitigate the eight at's. Also as summarized in Figure 6.11, we only need four security controls to adequately mitigate the eight at's and render $\mathbf{L}(j = 1) < \tau$.

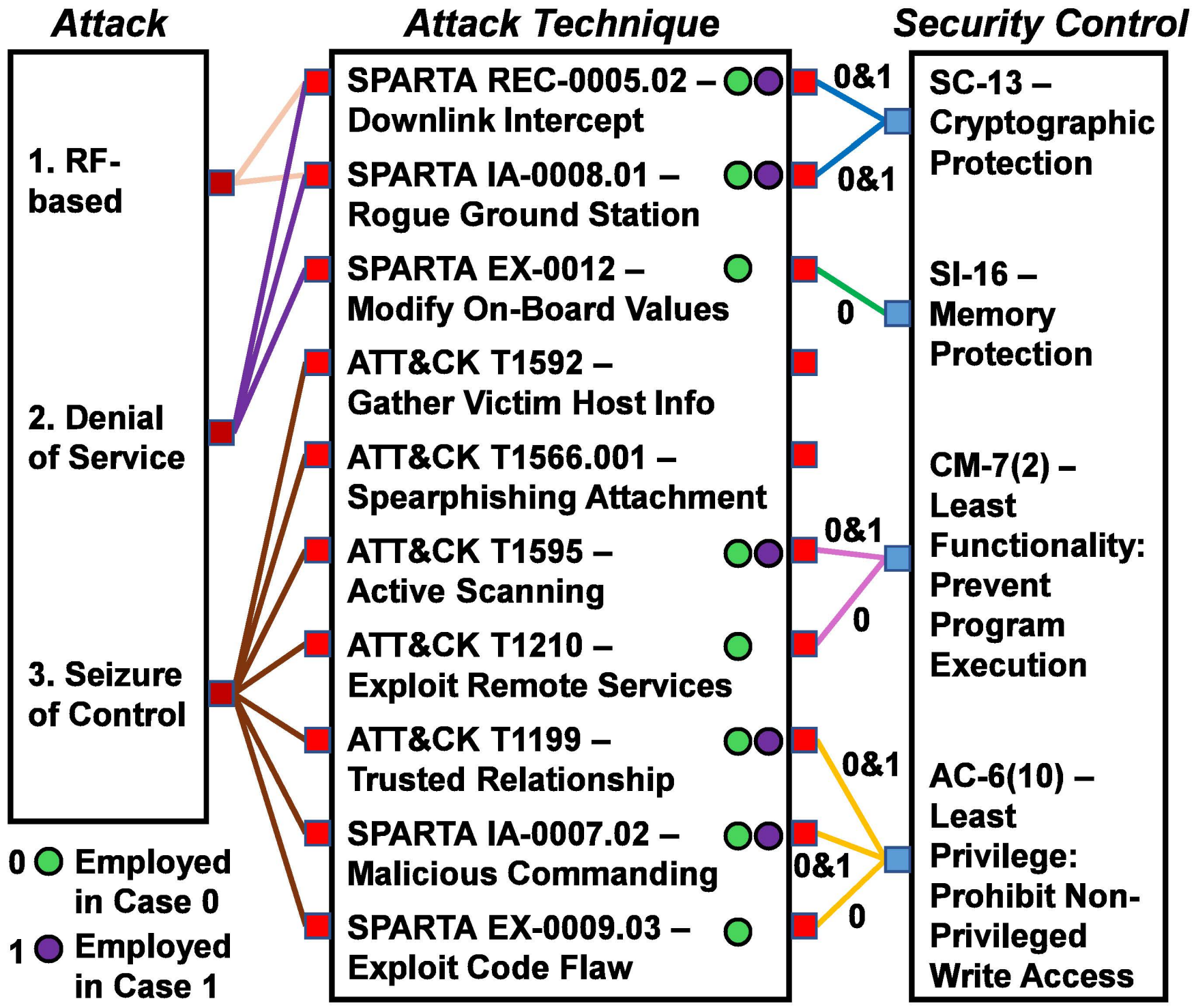

Figure 6.11: The three attacks, the 10 attack techniques they use, and the four security controls Algorithm 6 identifies.

Our experiment for showing how the mitigation completely thwarts Attack 1, namely the RF-based Attack, proceeds as follows. In Case 0, we implement security control SC-13 (Cryptographic Protection), a python-based implementation of AES-256 together with HMAC/SHA-256, to mitigate technique REC-0005.02 and protect the arc SM.PAYCOM→UM.RX from both eavesdropping and manipulation attacks. After employing this mitigation, we re-wage Attack 1 to intercept and inspect the signal transmitted over arc SM.PAYCOM→UM.RX, only finding that the signal is cryptographically protected, meaning REC-0005.02 is thwarted. This mitigation also thwarts

IA-0008.01 (injecting malicious data / signal over arc SM.PAYCOM→UM.RX), as shown in our experiment that our injection attempt fails because the injected data cannot pass the message authentication verification. In Case 1, the same experiment and result apply when using security control SC-13 to mitigate REC-0005.02 and IA-0008.01.

Our experiment for showing how the mitigation largely thwarts Attack 2, namely the DoS Attack, proceeds as follows. In Case 0, we implement security control SC-13 to protect the arc SM.BUSCOM→ GM.RX against data eavesdropping technique REC-0005.02 and the arc GM.TX→SM.BUSCOM against the data injection technique IA-0008.01. Our experiment to eavesdrop the communication over SM.BUSCOM→GM.RX failed because we cannot decrypt the demodulated signal without using the key. Our experiment to inject commands over the arc SM.BUSCOM→GM.RX also fails because the injected data cannot pass the message authentication verification and is thus discarded. We also implement security control SI-16 (Memory Protection) at the SM.C&DH module to run an on-orbit intrusion protection system (IPS) via allowlists and denylists (e.g., allowing commands that conform to the CCSDS Telecommand protocol, while denying commands that attempt to access the underlying operating system). Our attempt to apply technique EX-0012 (triggering the buffer overflow) at SM.C&DH fails because security control SI-16 detects and blocks the attempt. Thus, Attack 2 is completely thwarted. In Case 1, we only need to employ security control SC-13 to protect SM.BUSCOM→GM.RX against REC-0005.02 and protect GM.TX→SM.BUSCOM against IA-0008.01 (i.e., we do not need to employ SI-16 at SM.C&DH because the employed SC-13 has already closed all known routes to gaining access to SM.C&DH.

Our experiment for showing how the mitigation largely thwarts Attack 3 (the Seizure of Control Attack) proceeds as follows. In Case 0, we implement CM-7(2) (Least Functionality: Prevent Program Execution) via an application denylist utility to inspect operating system processes and file writes at the GM.NET, GM.A&S, GM.CMD, GM.TX, SM.PAYCOM, and SM.C&DH modules. Our attempt to execute T1595 (using Nmap to conduct active scanning) at GM.NET fails because of the mitigation. Our attempt to execute T1210 (spawning Python and Bash terminals through a command injection vulnerability) at GM.A&S is instantly blocked by the security control. We also implement AC-6(10) (Least Privilege: Prohibit Non-Privileged Write Access) at GM.CMD, GM.TX, SM.PAYCOM, and SM.C&DH with immutable file settings to prevent command files from tampering. Our attempt to execute T1199 at GM.CMD and GM.TX (manipulating their data storage with a malicious satellite command file) fails because AC-6(10) blocks the file write operation. Our attempting to locally execute IA-0007.02 (using a python script to inject a malicious flight command) at SM.PAYCOM fails because of the mitigation. Our attempt to execute EX-0009.03 (executing a malicious flight command at SM.C&DH by using a python script to place the malicious command on the software bus) also fails because of AC-6(10). That is, the mitigation largely thwarts Attack 3. In Case 1, the same experiment and results apply except that we only need to employ security control CM-7(2) at GM.NET to prevent T1595, and employ security control AC-6(10) at GM.CMD to prevent T1199 and at SM.PAYCOM to prevent IA-0007.02, because these employments already close all known routes for attackers to gain access from GM.SOFT to GM.A&S, GM.TX, and SM.C&DH.

As highlighted in Figure 6.11, in Case 0, two attack techniques (T1592 and T1566.001) are not mitigated, explaining the resulting *residual* risk of $\mathbf{L}(j) = 0.04$; in Case 1, five techniques (EX-0012, T1592, T1566.001, T1210, and EX-0009.03) are not mitigated, Explaining the resulting residual risk of $\mathbf{L}(j) = 0.08$.

**Insight 10.** *Our framework can effectively harden missions by mitigating* some*, but not necessarily all (for cost-effectiveness purposes), attack techniques that are possibly possessed by the attacker. Moreover, we observe that attack cascading effects might demand the mitigation of more attack techniques (e.g.,* $8 - 5 = 3$ *in our case study, highlight the price we pay).*

Insight 10 is important because it is significant to mitigate fewer than $|\mathsf{AT}_A|$ attack techniques (i.e., incurring a lower cost). This cost-effectiveness is amplified when CTI does not provide accurate estimation of $|\mathsf{AT}_A|$, meaning the defender or analyst may be forced to consider a set $|\mathsf{AT}_A|$ that is much larger than necessary. Similar to what was mentioned above, this insight is derived from an analysis with a specific system model (including the modules, links, mission control flows, and mission data flows in a particular space infrastructure) and threat models (including the attack techniques used by attackers). This means that the insight, like any model-driven study, is valid under the premise that the models correctly describe the attackers as well as their attacks. It is an outstanding open problem to investigate and quantify the uncertainties associated with the system and threat models as well as their consequences (e.g., whether the insight still holds when the threat model is not accurate).

## 6.6 Applying the Desired Properties to Assess our Framework

We assess our formal framework with respect to the ideal properties of space cyber risk analysis and mitigation tools presented in Chapter V. Our assessment is summarized in Figure 6.12, which also contains the results of the assessment of NRS and CTAP from Chapter V for comparison purposes, while recalling that the details of the assessments of NRS and CTAP may be found in Chapter V. Overall, our framework substantially advances the status quo regarding usefulness in terms of MISSION-CENTRICITY, robustness in terms of OBJECTIVITY, RIGOROUSNESS and DYNAMICS, and usability in terms of AUTOMATION.

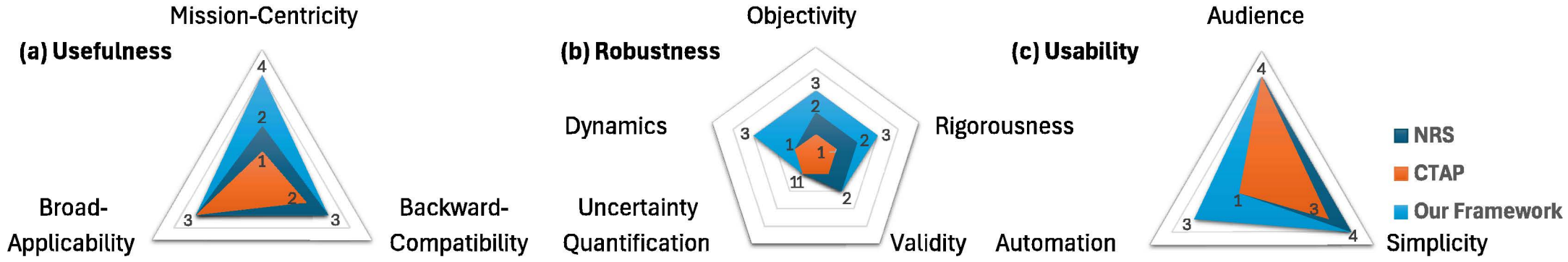

Figure 6.12: Scores for our framework, NRS, and CTAP with respect to the desired properties in terms of (a) usefulness, (b) robustness, and (c) usability.

Specifically, in terms of usefulness properties, our framework satisfies MISSION-CENTRICITY to a larger degree than NRS and CTAP by providing formal definitions of missions via mission control flows and mission data flows to holistically approach space infrastructure risk, as opposed to NRS which looks at risk from an individual attack technique perspective. It somewhat satisfies BACKWARDS-COMPATIBILITY, for example, by accommodating NRS, as discussed in Chapter VI. It somewhat satisfies BROAD-APPLICABILITY to at least the same level as NRS and CTAP because it can accommodate all space missions and has a modular design to plug in sub-routines or other custom functions, as required by the practitioner.

In terms of robustness properties, it somewhat satisfies OBJECTIVITY, RIGOROUSNESS and DYNAMICS because most of its computations are performed by a machine rather than a human, formal and explicit definitions and algorithms are provided, and cascading effects are accounted for. It scores lower on VALIDITY and UNCERTAINTY QUANTIFICATION, but at least as high as NRS and CTAP, because it has currently only been validated through case studies, and lack explicit uncertainty quantification.

In terms of usability properties, our framework somewhat achieves them because we accommodate at least as much for AUDIENCE as NRS; we use common input parameters (e.g., CTI) that provide SIMPLICITY; and 12 out of 13 steps in our algorithms are completed by a machine while additional risk analysis and mitigation steps need to be incorporated into the framework and automated, i.e., somewhat satisfying AUTOMATION.

## 6.7 Chapter Summary

We have presented a novel framework for space cyber risk analysis and mitigation. The framework includes a formal model of space infrastructures, their missions, and cyber attacks against them. We leveraged our space testbed to conduct experiments to demonstrate the usefulness of the framework, by re-implementing three real-world space cyber attacks, using our framework (more specifically its mission hardening algorithm) to identify security countermeasures, and employing our countermeasures to effectively mitigate the space cyber risks.

# CHAPTER VII

## CONCLUSION AND FUTURE RESEARCH DIRECTIONS

### 7.1 Dissertation Contributions

The contributions of this Dissertation are summarized as follows:

1. It presents an innovative framework for characterizing real-world space cyber attacks, including a principled *extrapolation* methodology and three metrics. Its usefulness is demonstrated by its capability in analyzing 108 real-world space cyber attacks, leading to a number of insights, such as: (i) space cyber attacks of average sophistication can be successful; (ii) proper security of the link segment between the space and user segments (e.g., using cryptography) could have thwarted nearly half of the 108 space cyber attacks; and (iii) traditional IT security controls against cyber social engineering attacks could have thwarted 32% (20/63) of the cyber attacks that compromise the ground segment.

2. The study presents a characterization of the state-of-the-practice in space cyber risk analysis and mitigation, namely NRS, as well as the first algorithmic descrip-tion of NRS, which has been incorporated into SPARTA, since version 1.6. It characterizes NRS' strengths and weaknesses, via two real-world space cyber attacks, leading to the insight: both the attack against the Terra satellite and the Turla attack could have been prevented by applying NRS at satellite design time.

3. The Dissertation introduces a set of desired properties that should be satisfied by any competent space cyber risk analysis and mitigation tool. It demonstrates the usefulness of the proposed properties by applying them to assess NRS and CTAP, leading to the useful insight: the current space cyber risk analysis and mitigation tools, namely NRS and CTAP, are far from competent or ideal.

4. Our study presents a novel and actionable framework to analyze and mitigate space cyber risks, which include: *explicit* modeling of space cyber attack cascading effects, and two core algorithms for *mission risk analysis* and *mission hardening*.It demonstrates the actionability and usefulness of the framework by implementing the algorithms and applying them in our space testbed to manage space cyber risks, leading to useful insights, such as: cascading effects offer attacker flexibilities in disrupting missions, especially when attackers leverage the entire infrastructure to wage attacks (in contrast to only leveraging the target control flows and data flows).

## 7.2 Dissertation Limitations

### 7.2.1 Chapter III Limitations

Chapter III, which analyzes real-world space cyber attacks, has three limitations. First, it addresses the missing data problem in a *manual*, rather than automated, fashion. Since we leverage our domain expertise to manually extrapolate attack tactics, attack techniques, space cyber kill chains, and space cyber attack campaign, the process is inevitably subjective even though we strive to be as objective as possible. Future research should investigate automated and objective methods for this process.

Second, the metrics can be refined. For example, the attack consequence metrics, except those associated with Link Segments, are geared toward *availability* because the raw dataset lacks enough information about what kinds of confidential data are processed in these space infrastructures; as a matter of fact, only 3 out of the 108 attack descriptions contain such information to some extent. Future research should refine the metrics to accommodate other metrics, such as data-specific *confidentiality* and *integrity*. For this purpose, one source of inspiration would be the CVSS scores of vulnerabilities [107] because their exploitation causes attacks. However, one cannot simply adopt the CVSS scores until after figuring out what software vulnerabilities enabled space cyber attacks, as vulnerabilities are not documented in raw incident reports.

Third, it assumes the measurements of the “building-block” metrics are given as input. While reasonable because of the focus of the present study, it is important to investigate how to obtain these measurements, which may require a community effort. In our case study, we use our own subjective measurements.

### 7.2.2 Chapter IV Limitations

Chapter IV, which analyzes NRS has the following three limitations. First, it would be ideal if we could automate NRS, by replacing the subjective steps with objective algorithms. However, this is a non-trivial task because it requires deep cyber-security knowledge, which does not appear to be possessed even by the current Large Language Models.

Second, the study lacks analysis of how NRS base scores are constructed because we did not have access to the necessary details. This is worthwhile to investigate for future studies because it would a aid in validating the quality of the base scores and it would significantly aid practitioners in tailoring risk scores for their own space infrastructures.

Third, the study does not analyze the use of ATT&CK in cooperation with NRS and SPARTA which is orthogonal to this study, but nevertheless important in the context of space cyber risk analysis and mitigation tools because the case studies in Chapters III, IV and VI show that ATT&CK techniques are integral to characterizing space cyber attacks that involve the ground segment.

### 7.2.3 Chapter V Limitations

Chapter V, which proposes a set of desired properties, has the following three limitations. First, it does not present objective methods to assess the quality of these properties themselves. For instance, are they equally desirable, or some are more desirable than others?

Second, the desired properties are necessarily at a high level to ensure their own broad applicability. It may be helpful, especially for practitioners, to investigate appropriate sub-properties that may be oriented toward specific space missions (e.g., SATCOM), or even specific mission control or data flows. The investigation could also approach from an infrastructure perspective to define sub-properties related to specific segments, components, or modules.

Third, the study provides a first steps toward tackling challenges imposed by zero-day vulnerabilities and interdependence as an example to illustrate the usefulness of the desired properties. Both of these challenges warrant deeper analysis where the desired properties remain helpful for tackling them.

### 7.2.4 Chapter VI Limitations

Chapter VI, which introduces a new framework to analyze and mitigate space cyber risks, has eight limitations. First, it currently focuses on dealing with likelihood of compromises without considering impact of compromises (effectively assuming the impacts on all modules are the same). Thus, it is important to study the extension to accommodate non-constant impacts.

Second, Algorithm 6 is by no means optimal, meaning that it may mitigate more attack techniques than necessary. Thus, it is important to design optimal hardening algorithm.

Third, when dealing with aggregation of attack effects, we make simplifying assumptions. This is shown by Eq.(VI.1), which aggregates attack effects over different control flows and data flows, and by Lines 2 and 4 in Algorithm 5, which aggregate attack effects over the nodes and arcs in a control or data flow, we use the simple $\max$ function. To (in)validate this choice of aggregation function, we plan to design experiments to identify the correct form of aggregation functions (if applicable). Our analysis shows that the experimental are non-trivial to conceive and execute, warranting separate studies.

Fourth, when dealing with cascading effects of successful attacks, we also make simplifying assumptions, especially the independence assumptions behind Eqs.VI.2-VI.6. It is an outstanding and challenging open problem to cope with the dependence that might exist, because there are different kinds of dependence structures. To our knowledge, this is a completely unexplored territory and we are not aware of any experimental studies on coping with this problem.

Fifth, we have presented some examples of "suitable" functions $f, g, h$ and $h'$ based on simplifying assumptions. We anticipate that real-world situations are more complex or complicated and potentially invalidate these simplifying assumptions. Therefore, it is an outstanding open problem to conduct experimental attack-defense research in real testbeds to collect data, conduct analysis, and identify these functions. Moreover, we anticipate that each of these functions will have different forms in different scenarios, such as different attack strategies or attack orchestration methods. This is an area that has not been investigated or understood in cybersecurity in general (i.e., not just in space cybersecurity). Nevertheless, these functions should obey probability laws because the derivation of a "composite" likelihood from "atomic" likelihoods (i.e., input) are often treated or interpreted as probabilistic calculations, even though the "atomic" likelihoods may not be defined with clear probabilistic structures; clearly describing such structures is another outstanding open problem, especially in complex or complicated cyber attack scenarios.

Sixth, the framework currently can deal with the so-called *unknown unknowns* to a certain degree, namely by assuming, for instance, that an attacker may possess zero-day attack capabilities that are not known to the defender, with likelihoods, respectively; these likelihoods can be seen as subjectively estimated probabilities based on analyst's experience and/or cyber threat intelligence. A future research direction is to enhance the framework with *uncertainty quantification*, which is not addressed in the present version, namely to quantify the uncertainty that may be associated with the results output by the algorithms (e.g., uncertainty associated with the cyber attack consequences resulting from the analysis), perhaps by first considering the uncertainty associated with the "atomic" or input likelihoods to the algorithms.

Seventh, the case study focuses on $n = 1$ mission, despite that it has $m'_1 = 3$ control flows and $n'_1 = 2$ data flows. This restriction is imposed by the scale of our current space testbed, which we plan to expand in the future, permitting us to investigate more sophisticated mission scenarios. Still, we manually created the concrete model, including $G_{\mathsf{infra}}$, $G_{\mathsf{mcf}}$, and $G_{\mathsf{mdf}}$, based on our domain knowledge. It is an outstanding open problem to automatically extract these graph structures either from their specifications and/or code.

Eighth, the case study considers two extreme cases: attack cascading effects apply to all (Case 0) vs. none (Case 1) of the modules that are not directly affected by the attack techniques possibly possessed by the attackers. We observe that there are many cases in between, which need to be investigated by analyzing the code and the security configurations / policies at modules (if applicable).

## 7.3 Future Research Directions

### 7.3.1 Future Research Related to Analyzing Real-World Space Cyber Risks

In relation to Chapter III, future research needs to investigate automated and objective methods for the process of extrapolating missing data. Future research could extend the metrics to accommodate other metrics, such as data-oriented *confidentiality* and *integrity*. For this purpose, one good source of inspiration would be the CVSS scores of vulnerabilities [142] because their exploitation likely caused attacks. However, one cannot simply adopt the CVSS scores because it is not clear what software vulnerabilities enabled the cyber attacks against space systems as this is not documented in the raw incident dataset. Moreover, are there other metrics that should be defined? Additional future research should address the limitations and challenges of this study discussed earlier. This includes the problem of how to conduct measurements, which is anticipated to require a community-wide effort to solve. One approach is to conduct expert surveys to solicit practitioners' perspectives, but this approach may be overly subjective. Another approach is to leverage full-fledged space testbeds to conduct experiments, collect pertinent data, and measure metrics.

### 7.3.2 Future Research Related to NRS

In relation to Chapter IV, future research should address the limitations mentioned above. For instance, it is important to refine the design of NRS, especially replacing the subjective sub-tasks in Algorithm 3 with objective processes, or to propose a completely new NRS. Along this line, open problems include: How can the base risk scores be made more objective and transparent? How can practitioners tailor space cyber risk scores objectively? How can an NRS support a broad range of audiences while maintaining adequate simplicity to remain intuitive? How can we design automated tools to help practitioners determine the criticality level and impact of each node in a mission graph (i.e., space infrastructure unit)?

### 7.3.3 Future Research Related to the Ideal Properties

In relation to Chapter V, future research should address the limitations discussed above. For instance, it is important to further explore the desired properties that should be exhibited by any competent space cyber risk management tool. For example: Can the properties be refined, and are there other desired properties than what are specified above? What are other challenges than the two discussed above (i.e., dealing with zero-day vulnerabilities and interdependence)? Towards fulfilling the desired properties, there are many open problems, such as: How can we develop automated space cyber risk management tools (i.e., fulfilling AUTOMATION)? How can we automatically extract a mission graph from the mission specification (i.e., fulfilling MISSION-CENTRICITY)? How can the criticality and impact of space system units in a mis-sion graph be automatically computed from space system specification, environmental conditions, cyber threat intelligence, and mission requirements (i.e., fulfilling BROAD-APPLICABILITY, OBJECTIVITY, RIGOROUSNESS, and VALIDITY)?

### 7.3.4 Future Research Related to Mitigating Space Cyber Risks

In relation to Chapter VI, future research should address the limitations discussed above. For instance, how should we incorporate non-constant impacts into our mission risk analysis and mission hardening algorithms? How can we optimize these algorithms? Further, experiments in robust high-fidelity testbeds and analysis of empirical data can validate or invalidate the simplifying assumptions we make, such as the independence between attacks, the validity of the simplified functions we use for aggregating attack effects, and the effectiveness of the algorithms to mission-specific requirements, processes, and hardware/software stacks. We anticipate that dependence is pertinent, especially for orchestrated cyber attacks. This opens the door to an exciting research field, such as the identification and characterization of the dependence structures and the treatment of them. Further, how should we quantify the uncertainty associated with our assumptions and model parameters, as well as the cascading uncertainty incurred by the process of derivations? How should we design experiments in high-fidelity testbeds to validate and/or refine our framework? What can we learn from case studies of more sophisticated mission scenarios, such as $n > 1$ missions and a spectrum of cyber attack cascading effects?

### 7.3.5 Future Research on Space Cybersecurity Dynamics

The present Dissertation follows the *Cybersecurity Dynamics* way of thinking [160, 164, 165], which provides a systematic framework for modeling, quantifying, reasoning, and enhancing cybersecurity from a holistic perspective. This framework has three pillars: cybersecurity metrics, cybersecurity data analytics, and first principle modeling. The present Dissertation can be seen as the first endeavor towards understanding the manifestation of Cybersecurity Dynamics in the context of Space Infrastructures, leading to what we call *Space Cybersecurity Dynamics*. It is an outstanding research problem to characterize the appropriate entry and intersection points where space cybersecurity dynamics can merge or mesh with the existing space system engineering metrics and validation techniques, as well as to explicitly define where space cybersecurity dynamics can revolutionize current space system engineering approaches. This is an outstanding research direction because the cybersecurity community and the system engineering community have their own, established way of thinking. One potential starting point is NRS and SPARTA. We envision that this new field will have the following three pillars.

**Space Cybersecurity Metrics**. While space cybersecurity dynamics would inherit most, if not all, cybersecurity metrics in other contexts, such as those metrics investigated in [11,12,15,20,92,114,166], we envision that space cybersecurity would demand its unique metrics. For instance, this Dissertation, or more specifically Chapters III-VI, already introduce some new metrics that do not appear to be immediately applicable to other settings (e.g., IT networks) without non-trivial adaptations. One direction of particular interest is to define metrics to quantify humans' susceptibility to cyber social engineering attacks, because Chapter III shows that many space cyber attacks start with cyber social engineering attacks in the ground segment. Along this line, there are some very significant progresses in modeling and quantifying cyber social engineering attacks [13, 80–83, 94, 95, 124, 125]. One exciting open problem is to investigate whether space infrastructures offer more opportunities to cyber social engineering attacks than what are already known, and if so, how to design new solutions to effectively thwart space-tailored cyber social engineering attacks.

**Space Cybersecurity Data Analytics**. This pillar focuses on data- and experiments-driven cybersecurity research. One direction with substantial results is cyber threats forecasting (similar to weather forecasting) [36,37,115,116,146,156,157], which could be adapted to forecast the evolution of space cyber attacks and incidents. This capability would support quantitative space cyber risk mitigation, via approaches that also include *space cyber insurance* (in a fashion similar to cyber insurance [173]). Another direction is to investigate whether space cyber attacks would be more or less than, or as evasive as, their counterparts in IT networks, which have seen sophisticated attacks that evolve to evade existing machine learning-based defenses [59, 63–67, 96–98]. Yet another exciting research direction is to investigate the evolution of vulnerabilities in space infrastructures and systems and how to detect them. For these purposes, we anticipate that the progresses in software vulnerability detection in non-space contexts (e.g., [69–75, 131, 176, 177]) can be adapted to the space domain.

**Space Cybersecurity First-Principle Modeling**. This pillar focuses on theoretical studies, especially on the cascading effect studies in space infrastructures. Chapter VI makes a significant step in modeling space cyber attack cascading effects. More systematic studies can benefit from the many cybersecurity first-principle models that are introduced in the non-space contexts (e.g., [19,48,49,68,79,86,155,158,159,161–163, 167,174,175]). One particular open problem is to investigate space cybersecurity first-principle models for satellite constellations, which exhibit more unique properties than terrestrial IT networks because satellites are orbiting. Another particular open problem is to tackle the matter of dependence in these theoretical models, where dependence can be incurred by the use of common hardware and software in space infrastructures. This is one notoriously difficult problem despite significant progresses that have been made in the non-space contexts [11,12,19,87,155,158]. One promising approach is to first understand the unique dependence characteristics exhibited by space infrastructures.